\pdfoutput=1
\documentclass[usenatbib]{mn2e}
\usepackage{apjfonts}
\usepackage{amssymb}
\usepackage{amsmath}
\usepackage{ctable}
\usepackage{fixltx2e} 

\newcommand{\be}{\begin{equation}}
\newcommand{\ee}{\end{equation}}

\newcommand{\msun}{M_{\sun}}

\newcommand{\qeos}{q_{\rm eos}}

\newcommand{\papertwo}{Paper {\small II}}

\newcommand{\scaleup}{}

\newcommand{\breaker}{}

\newcommand\plotone[1]
 {\centering \leavevmode \includegraphics[width={0.99\columnwidth}]{#1}}

\newcommand{\plotside}[1]
 {\centering \leavevmode \includegraphics[width={0.95\textwidth}]{#1}}
\newcommand{\plotsideup}[1]
 {\centering \leavevmode \includegraphics[width={0.97\textwidth}]{#1}}
\newcommand{\plotsidesmall}[1]
 {\centering \leavevmode \includegraphics[width={0.90\textwidth}]{#1}}
\newcommand{\plotsidesmallrotate}[1]
 {\centering \leavevmode \includegraphics[angle=-90,width={0.90\textwidth}]{#1}}
\newcommand{\plotsidesmallest}[1]
 {\centering \leavevmode \includegraphics[width={0.75\textwidth}]{#1}}
\newcommand{\acknowledgments}{\begin{small}\section*{Acknowledgments}\end{small}}
\newcommand\altaffilmark[1]{$^{#1}$}
\newcommand\altaffiltext[1]{$^{#1}$}
\title[How Do Massive Black Holes Get Their Gas?]{How Do Massive Black Holes Get Their Gas?}

\author[Hopkins and Quataert]{
\parbox[t]{\textwidth}{ 
Philip F. Hopkins\altaffilmark{1}\thanks{E-mail:phopkins@astro.berkeley.edu} \&
Eliot Quataert\altaffilmark{1}} 
\vspace*{6pt} \\
\altaffiltext{1}{Department of Astronomy and Theoretical Astrophysics
  Center, University of California
  Berkeley, Berkeley, CA 94720} \\
}

\date{Submitted to MNRAS, Dec.\ 14, 2009}
\begin{document}
\maketitle
\label{firstpage}
\begin{abstract}
  We use multi-scale smoothed particle hydrodynamic simulations to
  study the inflow of gas from galactic scales ($\sim 10 \,$kpc) down
  to $\lesssim 0.1\,$pc, at which point the gas begins
  to resemble a traditional, Keplerian  accretion disk. The key
  ingredients of the simulations are gas, stars, black holes (BHs),
  self-gravity, star formation, and stellar feedback (via a subgrid
  model); BH feedback is
  not included. We use $\sim 100$ simulations to survey a large
  parameter space of galaxy
  properties and subgrid models for the interstellar medium
  physics.
  We generate initial conditions for our simulations of galactic
  nuclei
  ($\lesssim 300$ pc) using galaxy scale simulations, including both
  major galaxy mergers and isolated bar-(un)stable disk galaxies. 
  For sufficiently gas-rich, disk-dominated systems, we find that
  a series of gravitational instabilities generates large
  accretion rates of up to $\sim1-10\,\msun\,{\rm
    yr^{-1}}$ onto the BH (i.e., at $\lesssim 0.1$ pc); this is
  comparable to what is needed to fuel the most luminous quasars. The
  BH  accretion rate  is highly time variable for a given set of conditions in the  galaxy at $\sim$ kpc.
  At radii $\gtrsim 10$ pc, our simulations resemble the ``bars within
  bars'' model of Shlosman et al, but we show that the gas can have a
  diverse array of morphologies, including spirals, rings, clumps, and
  bars; the duty cycle of these features is modest,
  complicating attempts to correlate BH accretion with the 
  morphology of gas in galactic nuclei. At $\sim 1-10$ pc, the
  gravitational potential becomes
  dominated by the BH and bar-like modes are no longer present.
  However, we show that the gas can become unstable to a standing,
  eccentric disk or a single-armed spiral mode ($m=1$), in which the
  stars and gas precess at different rates, driving the gas to
  sub-pc scales (again for sufficiently gas-rich, disk-dominated
  systems). A proper treatment of this mode
  requires including star formation and the self-gravity of both the
  stars {\it and} gas
  (which has not been the case in many previous calculations).
  Our simulations predict a correlation between the BH accretion rate
  and the star formation rate at different galactic radii. We find
  that  nuclear star formation is more tightly coupled to AGN activity than
  the global star formation rate of a galaxy, but a reasonable
  correlation remains even for the latter. 
\end{abstract}

\begin{keywords}
  galaxies: active --- quasars: general --- galaxies: evolution ---
  cosmology: theory
\end{keywords}

\section{Introduction}
\label{sec:intro}

The inflow of gas into the central parts of galaxies plays a critical
role in galaxy formation, ultimately generating phenomena as diverse
as bulges and spheroidal galaxies, starbursts and ultra-luminous
infrared galaxies (ULIRGs), nuclear stellar clusters, and accretion
onto super-massive black holes (BHs). The discovery, in the past
decade, of tight correlations between black hole mass and host
spheroid properties including mass
\citep{KormendyRichstone95,magorrian}, velocity dispersion
\citep{FM00,Gebhardt00}, and binding energy or potential well depth
\citep{hopkins:bhfp.obs,aller:mbh.esph} implies that these phenomena
are tightly coupled.

It has long been realized that bright, high-Eddington ratio accretion
(i.e., a quasar) dominates the accumulation of mass in the
supermassive BH population \citep{soltan82,salucci:bhmf,shankar:bhmf}.
In order to explain the existence of black holes with masses $\sim
10^{9} M_\odot$, the amount of gas required is comparable to that
contained in entire large galaxies. Given the short lifetime of the
quasar phase $\lesssim\,10^{8}\,{\rm yr}$ \citep{martini04}, the
processes of interest must deliver a galaxy's worth of gas to the
inner regions of a galaxy on a relatively short timescale.

There is also compelling evidence that quasar activity is preceded
and/or accompanied by a period of intense star formation in galactic
nuclei \citep{sanders88:quasars,sanders88:warm.ulirgs,
  dasyra:pg.qso.dynamics,kauffmann:qso.hosts}. The observed properties
of bulges at $z \sim 0$ independently require that dissipative
processes (gas inflow) dominate the formation and structure of the
inner $\sim$kpc
\citep{ostriker80,carlberg:phase.space,gunn87,kormendy:dissipation,
  hernquist:phasespace}. 
\citet{hopkins:cusps.ell,hopkins:cores,hopkins:cusps.fp} showed that
this inner dissipational component can constitute a large fraction
$\sim5-30\%$ of the galaxy's mass, with stellar (and at some point
probably gas) surface densities reaching $\sim 10^{11-12}\,\msun\,{\rm
  kpc^{-2}}$.

On large (galactic) scales, several viable processes for initiating
such inflows are well-known. Major galaxy-galaxy mergers produce
strong non-axisymmetric disturbances of the constituent galaxies; such
disturbances may also be produced in some minor mergers and/or
globally self-gravitating isolated galactic disks. Observationally,
major mergers are associated with enhancements in star formation in
ULIRGs, sub-millimeter galaxies, and pairs more generally
\citep[e.g.][]{sanders96:ulirgs.mergers, schweizer98,jogee:review,
  dasyra:mass.ratio.conditions,woods:tidal.triggering,
  veilleux:ir.bright.qso.hosts.merging}. Numerical simulations of
mergers have shown that when such events occur in gas-rich galaxies,
resonant tidal torques lead to rapid inflow of gas into the central
$\sim\,$kpc \citep{hernquist.89,barnes.hernquist.91,
  barneshernquist96}.  The resulting high gas densities trigger
starbursts \citep{mihos:starbursts.94, mihos:starbursts.96}, and are
presumed to feed rapid black hole growth. Feedback from the starburst
and a central active galactic nucleus (AGN) may also be important,
both for regulating the BH's growth
\citep{dimatteo:msigma,hopkins:lifetimes.letter,
debuhr:momentum.feedback,
johansson:bh.scalings.in.remergers} and for shutting down
future star formation (\citealt{springel:red.galaxies,
johansson:mixed.morph.mbh.sims}; see, however,
\citealt{debuhr:momentum.feedback}).

However, the physics of how gas is transported from $\sim 1$ kpc to
much smaller scales remains uncertain (e.g.,
\citealt{goodman:qso.disk.selfgrav}). Typically, once gas reaches
sub-kpc scales, the large-scale torques produced by a merger and/or
large-scale bar/spiral become less efficient. In the case of stellar bars or spiral waves,
there can even be a ``hard'' barrier to further inflow in the form of
an inner Linblad resonance, if the system has a non-trivial bulge.
In mergers, the coalescence of the two systems generates perturbations
on all scales, and so allows gas to move through the resonances, but
the perturbations relax rapidly on small scales, often before gas can
inflow.

Local viscous stresses -- which are believed to dominate angular
momentum transport near the central BH (e.g.,
\citealt{balbus.hawley.review.1998}) -- are inefficient at radii
$\gtrsim 0.01-0.1$ pc (e.g.,
\citealt{shlosman:inefficient.viscosities,
  goodman:qso.disk.selfgrav,thompson:rad.pressure}).  It is in
principle possible that some molecular clouds could be scattered onto
very low angular momentum orbits, but even the optimistic fueling
rates from this process are generally  insufficient to produce luminous
quasars \citep[see e.g.][]{hopkins:seyferts,kawakatu:disk.bhar.model,
  nayakshin:forced.stochastic.accretion.model}.  As a consequence,
many models invoke some form of gravitational torques (``bars
within bars''; \citealt{shlosman:bars.within.bars}) to continue
transport to smaller radii. As gas is driven into the central kpc by
large-scale torques, it will cool rapidly into a disky structure; if
this gas reservoir is massive enough, the gas will be self-gravitating
and thus again vulnerable to global instabilities (e.g., the
well-known bar and/or spiral wave instabilities) that can drive some of the gas to yet
smaller radii.

To date, numerical simulations have seen the formation of such
secondary bars in some circumstances, such as in adaptive mesh
refinement (AMR) simulations of galaxy formation
\citep{wise2007:protogalaxy.collapse,levine2008:nuclear.zoom,
  escala:nuclear.gas.transport.to.msigma} or particle-splitting
smoothed particle hydrodynamics (SPH) simulations of some idealized
systems \citep{escala:bh.mgr.idealized,mayer:bh.binary.sph.zoom.sim}.
These studies have served as a critical ``proof of concept.'' 
However, these examples have generally been limited by computational
expense to studying a single system at one instant in its evolution,
and thus it is difficult to assess how the sub-pc dynamics depends on
the large parameter space of possible inflow conditions from large
radii and galaxy structural parameters.

Alternatively, some simulations simply take an assumed small-scale
structure and/or fixed inflow rate as an initial/boundary condition,
and study the resulting gas dynamics at small radii \citep[e.g.][]{
  schartmann:2009.stellar.fb.effects.on.torus,dotti:bh.binary.inspiral,
  wada:starburst.torus.model,wada:torus.mol.gas.hydro.sims}.  These
studies have greatly informed our understanding of nuclear obscuration
on small scales (the ``torus''), and the role of stellar feedback in
determining the structure and dynamics of the gas at these radii; it
is, however, unclear how to relate this small-scale dynamics to the
larger-scale properties of the host galaxy.  This is critical for
understanding black hole growth and nuclear star formation in the
broader context of galaxy formation.

Observationally, a long standing puzzle has been that many systems,
especially those with weaker inflows on large scales (e.g.\
bar or spiral wave-unstable disks with some bulge, as opposed to major mergers),
exhibit no secondary instabilities at $\sim 0.1-1$ kpc -- in several
cases, torques clearly reverse sign inside these radii
\citep{block:obs.bar.torque.maps,garcia.burillo:torques.in.agn.nuclei.obs.maps.no.inflow}.
Whether this is generic, or the consequence of a low duty cycle, or
the result of the large-scale inflows simply being too weak in these
cases, is not clear. Moreover, even among systems that do show nuclear
asymmetries, and that clearly exhibit enhanced star formation and
luminous AGN, the observed features at smaller radii are very often
{\it not} traditional bars. Rather, they exhibit a diverse morphology,
with spirals quite common, along with nuclear rings, barred rings,
occasional one or three-armed modes, and some clumpy/irregular
structures \citep{martini:seyfert.host.morph,
peletier:seyfert.morph.imaging,knapen:seyfert.morphology,
laine:nested.bars.in.seyferts,kanpen:nuclear.region.in.bars.vs.host.prop,
greene:2009.sigma.gas.in.agn}.

Even if secondary bars or spirals are present at intermediate radii $\sim 10-100$
pc, it has long been recognized that they will cease to be important
at yet smaller scales, when the potential becomes quasi-Keplerian and
the global self-gravity of the gas less important; this occurs as one
approaches the BH radius of influence, which is $\sim10\,$pc in
typical $\sim L_{\ast}$ galaxies
\citep{athanassoula:bar.orbits,athanassoula:bar.vs.cmc,
  shlosman:bars.within.bars,heller:secondary.bar.instability,
  begelman:direct.bh.collapse.w.turbulence}.  Indeed, in previous
simulations and most analytic calculations, the ``bars-within-bars''
model appears to break down at these scales \citep[see e.g.][and
references therein]{jogee:review}.  However, local angular momentum
transport is still very inefficient at $\sim 10$ pc, and the gas is
still locally self-gravitating, and so should be able to form stars
rapidly (e.g., \citealt{thompson:rad.pressure}). Understanding the
physics of inflow through these last few pc, especially in a
consistent model that connects to gas on galactic scales
($\sim10\,$kpc), remains one of the key open questions in our
understanding of massive BH growth. 

In this paper, we present a suite of multi-scale hydrodynamic
simulations that follow gravitational torques and gas inflow from the
kpc scales of galaxy-wide events through to $<0.1\,$pc where the
material begins to form a standard thin accretion disk. These
simulations include gas cooling, star formation, and self-gravity;
feedback from supernovae and stellar winds is crudely accounted for via a subgrid
model.  In order to isolate the physics of angular momentum transport,
we do not include BH feedback in our calculations.  We systematically
survey a large range of galaxy properties (e.g., gas fraction and
bulge to disk ratio) and gas thermodynamics, in order to understand
how these influence the dynamics, inflow rates, and observational
properties of gas on small scales in galactic nuclei ($\sim 0.1-100$
pc).  Our focus in this paper is on the results of most observational
interest: what absolute inflow rates, star formation rates, and
gas/stellar surface density profiles result from secondary
gravitational instabilities? What is their effective duty cycle? And
what range of observational morphologies are predicted?  In a future
paper (\papertwo) we will present a more detailed comparison between
our numerical results and analytic models of inflow and angular
momentum transport induced by non-axisymmetric instabilities in
galactic nuclei.

The remainder of this paper is organized as follows.  In
\S~\ref{sec:sims} we describe our simulation methodology, which
consists of two levels of ``re-simulations'' using initial
conditions motivated by galactic-scale simulations.  In
\S~\ref{sec:results:transport} we present an overview of our results
and show how a series of gravitational processes leads to gas
transport from galactic scales to sub-pc scales. In \S
\ref{sec:results:duty} we quantify the resulting inflow rates and gas properties as a function of time and radius in the simulations.  \S \ref{sec:stability} summarizes the conditions required for global gravitational instability and significant gas inflow.  In \S~\ref{sec:results:sf} we show how the physics of accretion induced by gravitational instabilities leads to a correlation (with
significant scatter) between star formation at different radii and BH
accretion; we also compare these results to observations. In
\S~\ref{sec:discussion} we summarize our results and discuss a number
of their implications and several additional observational
tests. Further numerical details and tests of our methodology are
discussed in \S~\ref{sec:numerical}. In \S~\ref{sec:ism} we show how
the subgrid model of the ISM we use influences our results.

\vspace{-0.6cm}
\breaker
\section{Methodology}
\label{sec:sims}

We use a suite of hydrodynamic simulations to study the physics of gas
inflow from $\sim 10\,$kpc to $\sim0.1\,$pc in galactic nuclei.  In
order to probe the very large range in spatial and mass scales, we
carry out a series of ``re-simulations.'' First, we simulate the
dynamics on galaxy scales.  Specifically, we use representative
examples of gas-rich galaxy-galaxy merger simulations and isolated,
moderately bar-unstable disk simulations. These are well-resolved down
to $\sim100-500\,$pc.  We use the conditions at these radii (at
several times) as the initial conditions for intermediate-scale
re-simulations of the sub-kpc dynamics. In these re-simulations,
the smaller volume is simulated at higher resolution, allowing us to
resolve the subsequent dynamics down to $\sim10\,$pc scales -- these
re-simulations approximate the nearly instantaneous behavior of the
gas on sub-kpc scales in response to the conditions at $\sim$kpc set
by galaxy-scale dynamics. We then repeat our re-simulation method
to follow the dynamics down to sub-pc scales where the gas begins to
form a standard accretion disk.

Our re-simulations are not intended to provide an exact realization of the small-scale dynamics of the larger-scale simulation that motivated the initial conditions of each re-simulation (in the manner of particle-splitting or adaptive-mesh refinement techniques). 
Rather, our goal is to identify the dominant mechanism(s) of angular momentum transport in galactic nuclei and what parameters they depend on.  This approach clearly has limitations, especially at the outer boundaries of the simulations; however, it also has a major advantage. By  not requiring the conditions at small radii to be uniquely set by a larger-scale ``parent'' simulation, we can run a series of simulations with otherwise identical conditions (on that scale) but systematically vary one parameter (e.g., gas fraction or ISM model) over a large dynamic range.  This allows us to identify the physics and galaxy properties that have the biggest effect on gas inflow in galactic nuclei.   As we will show, the diversity of behaviors seen in the simulations, and desire to marginalize over the uncertain ISM  physics, makes such a parameter survey critical.

This methodology is discussed in more detail below. First, we describe
the physics in our simulations, in particular our treatment of gas
cooling, star formation, and feedback from supernovae and young stars
(\S~\ref{sec:sims:gas}).  We then summarize the galaxy-scale
simulations that are used to motivate the initial conditions for
subsequent re-simulations (\S~\ref{sec:sims:kpc}). The
intermediate-scale re-simulations, and the methodology used to
construct their initial conditions, are discussed in
\S~\ref{sec:sims:100pc}.  Finally, we discuss the nuclear-scale
resimulations, which are themselves motivated by the
intermediate-scale resimulations (\S~\ref{sec:sims:10pc}).

\vspace{-0.3cm}
\subsection{Gas Physics, Star Formation, and Stellar Feedback}
\label{sec:sims:gas}

\begin{figure*}
    \centering
    \scaleup
    \plotside{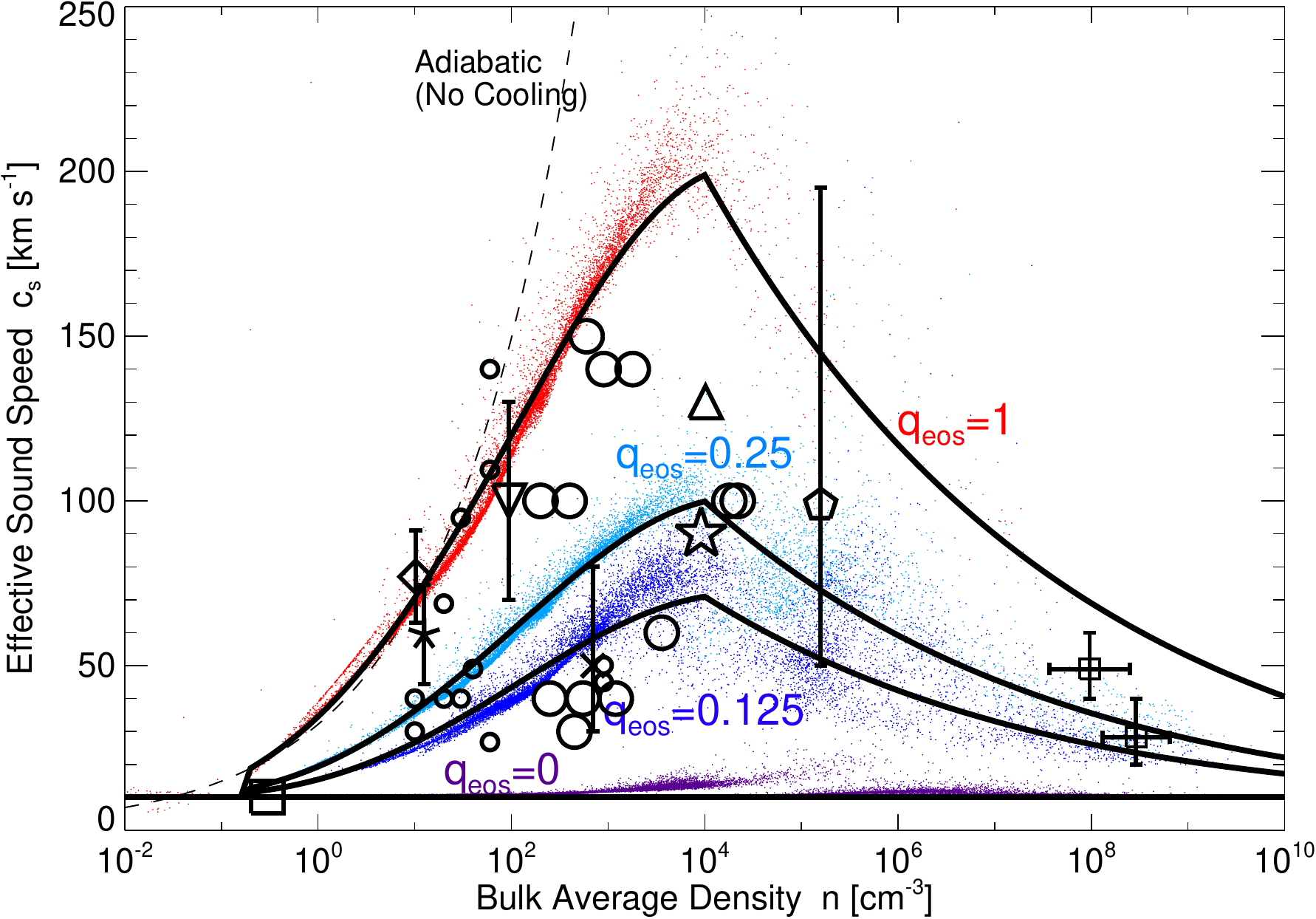}
    \caption{Effective equation of state of the ISM in our
      simulations; this accounts for the effects of stellar feedback
      that are not resolved in our calculations. We plot the effective
      sound speed $c_{s}$ (i.e., the turbulent speed generated by
      feedback) versus average ISM density $n$.  The $\qeos=0$
      case is an isothermal ``floor'' at $c_{s}=10\,{\rm km\,s^{-1}}$.
      The $\qeos=1$ line is the ``maximal feedback'' model in
      \citet{springel:multiphase}, in which the ISM is multi-phase
      above a minimum $n$ and all supernova energy goes into
      pressurizing the ``diffuse'' medium. 
      Intermediate $\qeos$ interpolate between the two
      (eq.~\ref{eqn:qeos.interpol}).  For each $\qeos$, we show an analytic
      curve for the equilibrium $c_{s}(n)$, and simulation results
      (colored points).  We compile measurements of the turbulent or
      non-thermal velocities in observed systems (black points): 
      local ULIRGs  \citep[][circles;
      large/small for the inner/outer observed radii in
      each]{downes.solomon:ulirgs}, the nuclei of merging 
      LIRGs \citep[][star]{bryant.scoville:ulirgs.co}, the core of 
      M82  \citep[][$\times$]{westmoquette:m82.sb.core.gasdynamics}, 
      the central 400 pc of NGC 6240 \citep[][triangle]{tacconi:ngc6240.gasdynamics}, 
      nuclear clumps in NGC 6240 \citep[][pentagon]{iono:ngc6240.nuclear.gas.huge.turbulence}, and
     the maser disk of NGC 3079 \citep[][small
      squares]{kondratko:3079.acc.disk.maser}.  The canonical spiral disk 
      velocity dispersion is also shown (large square).  At high-redshifts, we
      show recent observations of ``normal'' star-forming disks
      from \citet[][$z\sim2$;
      diamond]{forsterschreiber:z2.disk.turbulence} and
      \citet[][$z=3-4$;
      asterisk]{lemoinebusserolle:z3.disk.turbulence}, and luminous
      sub-millimeter galaxies from \citet[][inverted
      triangle]{tacconi:smg.maximal.sb.sizes}.  The observations favor
      a median $\qeos\sim0.125-0.3$, which we adopt as our ``fiducial''
      models. 
      \label{fig:qeos}}
\end{figure*}

The simulations were performed with the parallel TreeSPH code {\small
  GADGET-3} \citep{springel:gadget}, based on a fully conservative
formulation of smoothed particle hydrodynamics (SPH), which conserves
energy and entropy simultaneously even when smoothing lengths evolve
adaptively \citep[see e.g.,][]{springel:entropy,hernquist:sph.cautions,oshea:sph.tests}.
The detailed numerical methodology is described in
\citet{springel:gadget}, \citet{springel:multiphase}, and
\citet{springel:models}.

The simulations include supermassive black holes (BHs) as additional
collisionless particles at the centers of all progenitor galaxies.  In
our calculations the BH's only dynamical role is via its gravitational
influence on the smallest scales $\sim 1-10$ pc.  In particular, to
cleanly isolate the physics of gas inflow, we do {\it not} include the
subgrid models for BH accretion and feedback that have been used in
previous works (e.g., \citealt{springel:models}).  During a galaxy
merger, the BHs in each galactic nucleus are assumed to coalesce and
form a single BH at the center of mass of the system once they 
are within a single SPH smoothing length of one another and are moving 
at a relative speed lower than both the local gas sound speed and 
relative escape velocities. 

In our models, stars form from the gas using a
prescription motivated by the observed \citet{kennicutt98} relation.
Specifically, we use a star formation rate per unit volume $\dot
\rho_{\ast} \propto \rho^{3/2}$ with the normalization chosen so that
a Milky-way like galaxy has a total star formation rate of about
$1\,M_{\sun} \, {\rm yr^{-1}}$.  

The precise slope, normalization, and scatter of the Schmidt-Kennicutt relation, 
and even whether or not such a relation is generally applicable, are somewhat 
uncertain on the smallest spatial scales we model here. This is especially true when 
the dynamical times become short relative to the main-sequence stellar lifetime 
($t_{\rm dyn}\sim10^{5}-10^{6}\,$yr in the smallest regions simulated).   Nonetheless, there is some observational and physical motivation for the ``standard'' parameters we have adopted,
even at high surface densities.   
For the densest star forming galaxies, observational studies favor a logarithmic slope $\simeq 1.7$ for the relation between $\dot \Sigma_*$ and $\Sigma_g$ \citep{bouche:z2.kennicutt}, not that different from what our model implements.  
In addition, \citet{tan:mol.cloud.formation.times} and \citet{krumholz:sf.eff.in.clouds} 
show that local observations imply a constant 
star formation efficiency in units of the dynamical time (i.e.\ $\dot{\rho}_{\ast}\propto \rho^{1.5}$) 
at all densities observed, $n\sim 10^{1-6}\,{\rm cm^{-3}}$ -- the highest gas densities in these studies are 
comparable to the highest gas densities in our simulations ($\sim10^{8}\,\msun$ of gas inside $\sim10$\,pc).   Finally, \citet{davies:sfr.properties.in.torus} \&\ \citet{hicks:obs.torus.properties} 
estimate the star formation rate (SFR) and gas surface densities in AGN on exactly the small  
scales of interest here ($\sim 1-10\,$pc); they find a SFR-density relation 
continuous with that implied at ``normal'' galaxy densities.   

To understand the possible impact of uncertainties in the Schmidt-Kennicutt relation on our conclusions, 
we have adjusted the slope ${\rm d} \ln\dot \rho_{\ast}/{\rm d} \ln \rho$ adopted in our simulations between $1.0-2.0$ in a small set of test runs, fixing the star formation rate at MW-like surface densities of
$\simeq 10^{9}\,\msun\,{\rm kpc^{-2}}$ where the observational
constraints are tight. This amounts to varying the absolute star formation efficiency 
on the smallest resolved scales by a factor of $\gtrsim100$; 
qualitatively, this could presumably mimic a wide variety of 
different physics associated with stellar feedback and star formation. 
This variation can, unsurprisingly, have a dramatic affect on the 
quasi-equilibrium gas densities at small radii, which are set by gas inflow balancing star formation. 
However, even over this large range of star formation efficiencies, the qualitative behavior of the 
angular momentum transport and gas inflow does not change dramatically; the gas dynamics in 
a low-star formation efficiency run is similar to that in a run with much higher 
initial gas content but also higher star formation efficiency.   
As a result, although the absolute star formation efficiency is clearly somewhat 
uncertain, we do not believe that this qualitatively affects our conclusions.  

The largest uncertainties in our modeling stem from the treatment of
the interstellar medium (ISM) gas physics and the impact of stellar
feedback on the ISM.  Our simulations are relatively coarse and
average over many star-forming clumps, HII regions, supernova
remnants, etc.  As a result, the simulations use a sub-resolution
model of a multiphase ISM in which the gas has a sound speed much
larger than its true thermal velocity \citep{springel:multiphase}.
Our assumption is that the large-scale gravitational torques produced
by bars, spiral waves, and other non-axisymmetric features, will not
depend critically on the small-scale structure of the ISM; although we
believe that this is qualitatively correct, more detailed calculations
will be required to ultimately assess this assumption.  The key role
of stellar feedback in this model is to suppress the runaway
fragmentation and clumping of gas on small scales.  In reality, this
likely occurs via turbulence generated by stellar feedback and via the
disruption of star clusters and molecular clouds (e.g.,
\citealt{murray:molcloud.disrupt.by.rad.pressure}).  In our model, all
of this physics is ``accounted for'' by the large effective sound
speed, which increases the Jeans and Toomre masses, thus suppressing
the formation of small-scale structure.

Figure~\ref{fig:qeos} shows the range of effective sound speeds
$c_{s}$ in our calculations as a function of the ISM density $n$,
compared to a number of observational constraints (large symbols).
The solid lines in Figure \ref{fig:qeos} represent analytic
approximations to the equation of state, while the small colored
points are representative results from simulations that also include
adiabatic cooling/heating and shock heating.  We adopt a
parameterization of the sound speed as a function of density -- i.e.\
the effective equation of state for the ISM gas -- following
\citet{springel:models,springel:spiral.in.merger,robertson:disk.formation,
robertson:msigma.evolution}.
With this model, we can interpolate freely between two extremes using
a parameter $\qeos$. At one extreme, the gas has an effective sound
speed of $10 \, {\rm km\,s^{-1}}$, motivated by, e.g., the observed
turbulent velocity in atomic gas in nearby spirals or the sound speed
of low density photo-ionized gas; this is the ``no-feedback'' case
with $\qeos=0$.  The opposite extreme, $\qeos=1$, represents the
``maximal feedback'' sub-resolution model of \citet{springel:models},
based on the multiphase ISM model of \citet{mckee.ostriker:ism}; in
this case, $100\%$ of the energy from supernovae is assumed to stir up
the ISM.  This equation of state is substantially stiffer, with
effective sound speeds as high as $\sim200\,{\rm km\,s^{-1}}$.
Note that at the highest densities, $c_{s}$ begins to decline in all
of the models (albeit slowly), as the efficiency of star formation
asymptotes but cooling rates continue to increase.

Varying $\qeos$ between these two extremes amounts to varying the
effective sound speed of the ISM, with the interpolation
\begin{equation}
c_{s} = \sqrt{\qeos\,c_{s}^{2}[q=1] + (1-\qeos)\,c_{s}^{2}[q=0]}\ .
\label{eqn:qeos.interpol}
\end{equation}
The resulting sound speeds for $\qeos=0.125$ and $0.25$ are shown in
Figure~\ref{fig:qeos}; these correspond to more moderate values of
$c_s \sim30-100\,{\rm km\,s^{-1}}$ for the densities of interest.

Figure \ref{fig:qeos} compares these models to observations of the
turbulent (non-thermal) velocities in atomic and molecular gas in a
number of systems (large symbols).  At low mean densities,
$n\lesssim0.3-1\,{\rm cm^{-3}}$, the turbulent velocity in nearby
spirals is $\sim10 \, {\rm km\,s^{-1}}$.
\citet{downes.solomon:ulirgs} present a detailed study of a number of
luminous local starbursts that have significantly higher mean
densities; they decompose the observed molecular line profiles into
bulk (e.g., rotation) and turbulent motions.  We plot their
determination of the mean density and turbulent velocities in each
system at several radii.  We also show the results of similar
observations of the core of M82
\citep{westmoquette:m82.sb.core.gasdynamics}, additional nearby
luminous infrared galaxies \citep{bryant.scoville:ulirgs.co}, NGC 6240
\citep{tacconi:ngc6240.gasdynamics,iono:ngc6240.nuclear.gas.huge.turbulence},
and luminous starbursts at high redshift, $z\sim2-3$
\citep{forsterschreiber:z2.disk.turbulence,tacconi:smg.maximal.sb.sizes,lemoinebusserolle:z3.disk.turbulence};
at the highest densities, we also show the random velocities observed
in the nuclear maser disk in the nearby Seyfert 2 galaxy NGC3079
\citep{kondratko:3079.acc.disk.maser}.

The observational results in Figure \ref{fig:qeos} favor models with
$\qeos\approx0.1-0.3$, albeit with significant scatter.  We thus take
these values of $\qeos$ as our ``standard'' choices, although we have
carried out numerical experiments over the entire range $\qeos=0-1$.
Note that the observations clearly do {\em not} support a simple
no-feedback ($\qeos=0$) model.  Within the range $\qeos \approx
0.1-0.3$, our results on AGN fueling are not particularly sensitive to
the precise value of $\qeos$.  Moreover, the functional form $c_s(n)$
is also not crucial: simulations using a constant $c_s \simeq 50 \,
{\rm km s^{-1}}$ yield similar results.
However, our simulations with $\qeos=0$ and $\qeos=1$ predict results
that are inconsistent with observations of galactic nuclei -- thus,
our results themselves favor $\qeos \approx 0.1-0.3$ (see
Appendix~\ref{sec:ism}).

For the gas densities of interest in this paper, the precise form of the cooling law does not significantly affect our conclusions.   This is because the cooling time is almost always 
much shorter than the local dynamical time (typical $t_{\rm cool}\sim 10^{-6}-10^{-4}\,t_{\rm dyn}$).  As a result the "sound speed" of the gas is nearly always pinned to the subgrid `turbulent' value discussed above (this is why the numerical points in Fig. \ref{fig:qeos} are so close to the analytic models).   
This is true even when the gas is optically thick to the infrared radiation produced by 
dust, as can readily occur in the central $\sim 100$ pc:   the cooling time (diffusion time)
is still much less than the dynamical time in the optically thick limit for the radii that we resolve
(e.g., \citealt{thompson:rad.pressure}). 
We have, in fact, experimented with alternative cooling rate 
prescriptions: including or excluding metal-line cooling, uniformly increasing 
or decreasing the cooling rate by a factor of $\approx3$, and in the most 
extreme case, assuming instantaneous gas cooling (any gas parcel above the 
cooling floor is assumed to immediately radiate the excess energy in 
a single timestep). We do not see any significant changes in our results 
with these variations, simply because the gas always cools rapidly in our calculations 
\citep[in contrast, in regimes such as 
the $\alpha$-disk where the cooling time is comparable to the dynamical 
time, the details of the cooling function can have a significant effect; see][]{gammie:2001.cooling.in.keplerian.disks,
nayakshin:sfr.in.clumps.vs.coolingrate,
cossins:2009.grav.instab.vs.coolingrate}. 
If the effective minimum $c_{s}$ comes from turbulent velocities, 
then the ÒeffectiveÓ cooling time around this ßoor should be given by the
turbulent decay time, which can be comparable to the dynamical time 
\citep{begelman:direct.bh.collapse.w.turbulence}; this is not included in our calculations.   
In the presence of such an effective cooling time, local gravitational instability 
may lead to tightly wound spirals as opposed to fragmentation into star-forming
clumps. These could be important for angular momentum transport at some radii.

To conclude our discussion of the ISM physics in our simulations, it is important to reiterate that the key role of the sub-resolution sound speed $c_s$ is that it determines the local Jeans and Toomre
criteria, and thus the physical scale on which gravitational physics
dominates.
The Jeans mass for a disk of surface density $\Sigma$ and sound speed
$c_s$ is given by $M_{\rm J}=(\pi\,c_{s}^4)/(4\,G^{2}\,\Sigma)$. For
the outer regions of a galactic disk with
$\Sigma\sim10^{8}-10^{9}\,\msun\,{\rm kpc^{-2}}$ and
$c_{s}\sim10\,{\rm km\,s^{-1}}$, $M_{J}\sim10^{6}\,\msun$, comparable
to that of a molecular cloud; the corresponding Jeans length is tens
of pc, comparable to that of massive molecular cloud complexes in
galaxies.  Thus our sub-resolution model is effectively averaging over
discrete molecular clouds and star clusters in galaxies.  Large-scale
inflows can increase the surface density in the central regions of
galaxies, but $c_{s}$ also rises.  In our models with $\qeos \simeq
0.1-0.3$, the Jeans mass remains roughly similar down to $\sim $pc
scales, but as a result the Jeans length is significantly smaller in
galactic nuclei where the ambient density is much higher.
These physical mass and size-scales motivate the numerical resolution
in our simulations; in all cases, we ensure that the resolution is
sufficient to formally resolve the Jeans mass and length.  Higher
resolution simulations may be numerically achievable, but can provide
only minimal gains in the ``reality'' of the simulation without a
corresponding increase in the sophistication of the ISM model.

\vspace{-0.2cm}
\subsection{Large Scale Galaxy Mergers and Bars: 100\,kpc to 100\,pc}
\label{sec:sims:kpc}

\begin{footnotesize}
  \ctable[ caption={{\normalsize Galaxy-Scale
      Simulations}\label{tbl:galaxy}},center,star ]{lcccccccccc}{
    \tnote[ ]{Parameters describing representative examples of our
      galaxy-scale simulations of galaxy-galaxy mergers and unstable
      isolated disks.  The top row gives the range spanned in each
      parameter across our suite of simulations. Subsequent rows
      pertain to specific examples at chosen times in their evolution.
      {\bf (1)} Simulation name/ID.  {\bf (2)} Minimum smoothing
      length (in pc).  {\bf (3)} Equation of state parameter
      (Figure~\ref{fig:qeos}).  {\bf (4)} Gas fraction $\equiv M_{\rm
        gas}/(M_{\rm gas}+M_{\ast,\, \rm disk})$ of the disky/cold
      component (inside the given radius).  {\bf (5)} Scale height of
      the disky/cold component within the given radius (median
      $|z|/R$; the plane of the disk is defined by the total angular
      momentum vector).  {\bf (6)} Bulge-to-total mass ratio inside a
      given radius (bulge here includes all spherical components:
      stellar bulge, dark matter halo, and black hole).  {\bf (7)}
      Maximum surface density of the disky/cold component (gas plus
      stars) -- for an exponential profile, the surface density is
      nearly constant at radii below the disk scale length. Otherwise,
      averaged over a couple times our minimum smoothing length.  {\bf
        (8)} Average surface density inside $300$\,pc for the bulge
      component.  {\bf (9)} Total mass enclosed inside a given radius.
      All simulations include black holes, but these are dynamically
      unimportant on these scales.
      \\
    } 
\tnote[a]{For each of the two simulations here, we have also run three ultra high-resolution 
simulations which also act as a moderate-resolution intermediate-scale 
simulations ($\sim10^{7}$ particles). They are identical in initial conditions to the 
standard merger and isolated disk run here, with initial gas fraction equal to, 
one-half, and one-quarter that shown (six simulations in total). 
}
    }{ \hline\hline \multicolumn{1}{c}{Simulation} &
    \multicolumn{1}{c}{$\epsilon$} & \multicolumn{1}{c}{$\qeos$} &
    \multicolumn{1}{c}{$f_{\rm gas}$} & \multicolumn{1}{c}{$h/R$} &
    \multicolumn{2}{c}{$B/T(<R)$} & \multicolumn{1}{c}{$\Sigma_{\rm
        d}(0)$} & \multicolumn{1}{c}{$\Sigma_{\rm b}(0)$} &
    \multicolumn{2}{c}{$M_{\rm tot}(<R)$  $[M_{\sun}]$} \\
    \multicolumn{1}{c}{Name} & \multicolumn{1}{c}{[pc]} &
    \multicolumn{1}{c}{ } & \multicolumn{1}{c}{($500$\,pc)} &
    \multicolumn{1}{c}{($500$\,pc)} & \multicolumn{1}{c}{$(300$\,pc)}
    & \multicolumn{1}{c}{$(1$\,kpc)} &
    \multicolumn{2}{c}{$[M_{\sun}\,{\rm kpc^{-2}}]$} &
    \multicolumn{1}{c}{$(300$\,pc)} &
    \multicolumn{1}{c}{$(1$\,kpc)} \\
    \multicolumn{1}{c}{{\bf (1)}} & \multicolumn{1}{c}{{\bf (2)}} &
    \multicolumn{1}{c}{{\bf (3)}} & \multicolumn{1}{c}{{\bf (4)}} &
    \multicolumn{1}{c}{{\bf (5)}} & \multicolumn{2}{c}{{\bf (6)}} &
    \multicolumn{1}{c}{{\bf (7)}} & \multicolumn{1}{c}{{\bf (8)}} &
    \multicolumn{2}{c}{{\bf (9)}} \\
    \hline
\multicolumn{11}{c}{Merger \&\ Isolated Galaxy Simulations: Parameter Studies} \\
\hline
{\bf ---} & 20,50,100,150 & 0.0-1.0 & 0.05-1.0 & 
0.05-0.35 & 0.1-0.9 & 0.01-0.5 & 1.0e9-1.0e11 & 1.0e9-1.0e12 & 1.0e8-1.0e11 & 1.0e8-1.0e11 \\ 
\hline
\multicolumn{11}{c}{Typical Gas-Rich $\sim L_{\ast}$ Merger: Initial Conditions} \\
\hline
{\bf b3ex(ic)} & 10\tmark[a],\,50 & 0.25 & 0.80 & 0.25 & 0.8 & 0.4 & 2.5e8 & 2.5e8 & 1.4e8 & 1.9e9 \\ 
\hline
\multicolumn{11}{c}{Typical Gas-Rich $\sim L_{\ast}$ Merger: At Coalescence} \\
\hline
{\bf b3ex(co)} & 10\tmark[a],\,50 & 0.25 & 0.45 & 0.33 & 0.15 & 0.25 & 2.0e10 & 2.6e9 & 4.4e9 & 2.5e10 \\ 
\hline
\multicolumn{11}{c}{Typical Gas-Rich $\sim L_{\ast}$ Merger: $10^{8}$\,yr Post-Coalescence} \\
\hline
{\bf b3ex(po)} & 10\tmark[a],\,50 & 0.25 & 0.08 & 0.37 & 0.10 & 0.15 & 2.3e11 & 9.1e9 & 2.8e10 & 4.6e10 \\ 
\hline
\multicolumn{11}{c}{Bar-Unstable $\sim L_{\ast}$ Disk: Initial Conditions} \\
\hline
{\bf barex(ic)} & 10\tmark[a],\,50 & 0.25 & 0.40 & 0.15 & 0.55 & 0.35 & 1.7e8 & 2.4e8 & 1.5e8 & 1.9e9 \\ 
\hline
\multicolumn{11}{c}{Bar-Unstable $\sim L_{\ast}$ Disk: At Peak of Inflow} \\
\hline
{\bf barex(pk)} & 10\tmark[a],\,50 & 0.25 & 0.48 & 0.20 & 0.13 & 0.23 & 1.1e9 & 2.5e8 & 5.2e8 & 3.8e9 \\ 
\hline
\multicolumn{11}{c}{Bar-Unstable $\sim L_{\ast}$ Disk: After Bar Relaxation} \\
\hline
{\bf barex(re)} & 10\tmark[a],\,50 & 0.25 & 0.04 & 0.11 & 0.16 & 0.18 & 1.2e10 & 5.5e9 & 5.6e9 & 1.5e10 \\ 
\hline\hline\\
}
\end{footnotesize}

Our galaxy-scale simulations motivate the initial conditions chosen
for the smaller-scale re-simulation calculations described in \S \ref{sec:sims:100pc} \&
\ref{sec:sims:10pc}.  The galaxy-scale simulations include isolated
disks (both globally stable and bar unstable) and galaxy-galaxy
mergers.  We will ultimately focus on a few representative examples,
but we chose those having surveyed a large parameter space.  These
simulations and the methodology used for building the initial galaxies
are described in more detail in a series of papers \citep[see
e.g.][]{dimatteo:msigma,robertson:msigma.evolution,cox:kinematics,
  younger:minor.mergers}. We briefly review the key points here.

For each simulation, we generate one or two stable, isolated disk
galaxies, each with an extended dark matter halo with a
\citet{hernquist:profile} profile, an exponential disk of gas and
stars, and an optional stellar bulge.  The initial systems are chosen
to be consistent with the observed baryonic Tully-Fisher relation and
estimated halo-galaxy mass scaling laws \citep[][and references
therein]{belldejong:tf,kormendyfreeman:scaling, mandelbaum:mhalo}.
The galaxies have total masses $M_{\rm vir}=V_{\rm
  vir}^{3}/(10GH[z])$ for an initial redshift $z$, with the baryonic
disk having a mass fraction $m_{\rm d}$ (typically $m_{\rm d} \simeq
0.041$) relative to the total mass. The system has an initial bulge-to-total 
baryonic mass ratio $B/T$, and the disk has initial gas fraction $f_{\rm gas}$. The
dark matter halos are assigned a concentration parameter scaled as in
\citet{robertson:msigma.evolution} for the galaxy mass and redshift
following \citet{bullock:concentrations}.
Disk scale lengths are set in accordance with the above scaling laws. 

In previous papers (referenced above), a large suite of these
simulations have been presented, with several hundred simulations of
varying equation of state, numerical resolution, merger orbital
parameters, structural properties (e.g.\ profile shapes, initial
bulge-to-disk ratios, and scale lengths), initial gas fractions, and
halo concentrations.  In this suite, galaxies have baryonic masses
$\sim10^{8}-10^{13}\,\msun$ and gas fractions $f_{\rm gas}=0-1$;
mergers spanning mass ratios from 1:1 to 1:20, and isolated disks have
Toomre Q parameters from $0.1-10$.

In this work, we focus on galaxies with baryonic masses $\sim
10^{11}\,\msun$.  Based on the survey above, we select a
representative simulation of a gas-rich major merger and one of an
isolated disk, to provide the basis for our subsequent re-simulations.
Some of the salient parameters of these simulations are given in
Table~\ref{tbl:galaxy}.  The merger is equal-mass, with
$10^{11}\,\msun$ galaxies that have gas fractions of $\sim40\%$ at the
time of the merger/coalescence.  The orbit is a moderately tilted
prograde, parabolic case \citep[orbit {e}
in][]{cox:kinematics}. Together this makes for a fairly violent, gas
rich major merger, representative of many of our other gas-rich major
merger simulations at both low and high redshifts.  The isolated
system is a $10^{11}\,\msun$ disk with $f_{\rm gas}=0.4$, $B/T=0.3$,
scale length $h=3.2\,{\rm kpc}$, and $m_{d}=0.041$; it has a Toomre Q
of order one and develops a moderate bar (amplitude $\sim15\%$), but
the gas encounters an inner Linblad resonance at $\sim$1-2\,kpc.

Small variations in the orbits or the structural properties of the galaxies
will change the details of the tidal and bar features on large scales.  However, the precise details of these large-scale
simulations are not important for our study of the dynamics on small
scales (see Appendix~\ref{sec:numerical}).  Rather, the small-scale
dynamics depends on global parameters such as the gas mass channeled
into the central region, relative to the pre-existing bulge, disk, and
black hole mass. In these respects, we have chosen the simulations
summarized in Table \ref{tbl:galaxy} to be representative of a broad
class of gas-rich systems.

Our galaxy-scale simulations have spatial resolution -- gravitational
softening length and minimum adaptive SPH smoothing length -- of
$50\,$pc. In the suite described above, the resolution scales with
galaxy mass and is $\sim50-100\,$pc for $M_{\ast} \sim 10^{11} \,
\msun$ systems, but in a subset of higher-resolution cases is as small
as $20\,$pc. In \citet{hopkins:cusps.mergers} and
\citet{hopkins:cusps.ell} we have demonstrated that this resolution is
sufficient to properly resolve not only the mass fractions but also
the spatial extent of the ``starburst'' formed from gas which loses
angular momentum in a merger or via a strong bar instability.
However, to assess how much of this gas can ultimately fuel a central
BH requires that we determine the dynamics on even smaller spatial
scales.

\subsection{Intermediate Scales: Re-Simulating from 1 kpc to 10\,pc}
\label{sec:sims:100pc}

\begin{footnotesize}
\ctable[
  caption={{\normalsize Intermediate-Scale Resimulations ($\sim 10-1000$  pc)}\label{tbl:intermediate}},center,star
  ]{lcccccccccc}{
\tnote[ ]{Parameters describing our re-simulations of the 
$0.01-1$ kpc regions from galaxy scale simulations. 
Parameters separated by commas denote simulations with otherwise identical initial 
conditions, re-run with the specified parameter varied. 
{\bf (1)} Simulation name/ID.
{\bf (2)} Minimum smoothing length (in pc).
{\bf (3)} Equation of state parameter (Figure~\ref{fig:qeos}). 
{\bf (4)} Initial gas fraction of the disky/cold component (inside the given radius). 
{\bf (5)} Initial scale height of the disk component (inside the given radius). 
{\bf (6)} Initial bulge-to-total mass ratio inside a given radius (again, bulge refers to all spherical 
components).
{\bf (7)} Initial maximum surface density of the disky/cold component (gas plus stars). 
{\bf (8)} Initial average surface density inside $300$\,pc for the bulge component
{\bf (9)} Initial total mass enclosed inside a given radius. 
All simulations include BHs and dark matter, but these are dynamically 
unimportant on these scales.\\ 
}
\tnote[a]{A series of 7 runs testing different 
means of constructing initial conditions, described in Appendix~\ref{sec:numerical}. 
}
\tnote[b]{
Isothermal equation of state, but with a large $c_{s}=50\,{\rm km\,s^{-1}}$ 
cooling ``floor.''
}
\tnote[c]{
Cooling allowed down to $100\,K$, i.e.\ $c_{s}=1\,{\rm km\,s^{-1}}$. 
}
\tnote[d]{
Somewhat larger-scale simulation (between 
``galaxy scale'' and standard ``intermediate scale''). Instead of $B/T(<R)$ and $M_{\rm tot}(<R)$ 
being evaluated at $100$\,pc and $300\,$pc, they are here evaluated at 
$500\,$pc and $1\,$kpc, respectively. 
}
\tnote[e]{
Series where the gas disk profile is allowed to vary independent of the stellar disk profile. 
The gas has exponential, power-law, and truncated power-law profiles, with varying concentrations 
with respect to the disk (for example including an extended gas ``reservoir'' at a distance $\sim2$ times 
the regular nuclear stellar disk length, with surface density profile $\Sigma\propto R^{-1}$). 
}
\tnote[f]{
Very high-resolution simulations which also act as a moderate-resolution nuclear-scale 
simulations ($2\times10^{7}$ particles; gas particle mass $\approx500\,\msun$). 
A series of 6 galaxy-scale runs with very high ($\sim10\,$pc) resolution, used as 
moderate-resolution intermediate-scale simulations, are also described in the text. 
}
}{
\hline\hline
\multicolumn{1}{c}{Simulation} &
\multicolumn{1}{c}{$\epsilon$} &
\multicolumn{1}{c}{$\qeos$} & 
\multicolumn{1}{c}{$f_{\rm gas}$} & 
\multicolumn{1}{c}{$h/R$} & 
\multicolumn{2}{c}{$B/T(<R)$} & 
\multicolumn{1}{c}{$\Sigma_{\rm d}(0)$} & 
\multicolumn{1}{c}{$\Sigma_{\rm b}(0)$} & 
\multicolumn{2}{c}{$M_{\rm tot}(<R)$  $[M_{\sun}]$} \\
\multicolumn{1}{c}{Name} &
\multicolumn{1}{c}{[pc]} &
\multicolumn{1}{c}{ } & 
\multicolumn{1}{c}{($500$\,pc)} & 
\multicolumn{1}{c}{($500$\,pc)} & 
\multicolumn{1}{c}{$100$\,pc} & 
\multicolumn{1}{c}{$300$\,pc} & 
\multicolumn{2}{c}{$[M_{\sun}\,{\rm kpc^{-2}}]$} & 
\multicolumn{1}{c}{$100$\,pc} & 
\multicolumn{1}{c}{$300$\,pc} \\
\multicolumn{1}{c}{{\bf (1)}} &
\multicolumn{1}{c}{{\bf (2)}} &
\multicolumn{1}{c}{{\bf (3)}} &
\multicolumn{1}{c}{{\bf (4)}} &
\multicolumn{1}{c}{{\bf (5)}} &
\multicolumn{2}{c}{{\bf (6)}} &
\multicolumn{1}{c}{{\bf (7)}} &
\multicolumn{1}{c}{{\bf (8)}} &
\multicolumn{2}{c}{{\bf (9)}} \\
\hline
{\bf If9b5}\tmark[a] & 1.0 & 0\tmark[b],0.175,0.25 & 0.90 & 0.30 & 0.5 & 0.15 & 1.0e10 & 1.1e10 & 5.4e8 & 2.9e9 \\ 
{\bf If9b5thin} & 1.0 & 0.125,0.25 & 0.90 & 0.08,0.16 & 0.4 & 0.2 & 1.0e10 & 1.1e10 & 5.4e8 & 2.9e9 \\ 
{\bf If9b5res} & 0.3,1,3,10 & 0.125 & 0.90 & 0.30 & 0.5 & 0.15 & 1.0e10 & 1.1e10 & 5.4e8 & 2.9e9 \\ 
{\bf If9b5q} & 1.0 & 0,0\tmark[c],0.125,0.25,0.5,1 & 0.90 & 0.30 & 0.5 & 0.15 & 1.0e10 & 1.1e10 & 5.4e8 & 2.9e9 \\ 
{\bf Ilowresq} & 3.0 & 0,0.25,1 & 0.95 & 0.27 & 0.0 & 0.0 & 6.0e10 & 0.0 & 1.2e9 & 4.9e9 \\ 
{\bf If1b1late} & 1.0 & 0.125 & 0.096 & 0.25 & 0.06 & 0.03 & 1.0e11 & 1.1e10 & 3.4e9 & 2.2e10 \\ 
{\bf If1b0late} & 1.0 & 0.25 & 0.091 & 0.28 & 0.002 & 0.003 & 6.0e10 & 5.0e8 & 1.3e9 & 5.3e9 \\ 
{\bf If1b0lateL}\tmark[d] & 2.0 & 0.25 & 0.091 & 0.28 & 0.005 & 0.01 & 6.0e10 & 5.0e8 & 7.4e9 & 8.5e9 \\ 
{\bf If3b3mid} & 1.0 & 0.125 & 0.34 & 0.25 & 0.3 & 0.15 & 3.1e10 & 2.0e10 & 1.3e9 & 7.2e9 \\ 
{\bf If3b3midRg}\tmark[e] & 1.0 & 0.125 & 0.45,0.20,0.05 & 0.3,0.5 & 0.26 & 0.12 & 3.6e10 & 2.0e10 & 1.5e9 & 9.2e9 \\ 
{\bf If1b3Lmid} & 1.0 & 0.25 & 0.10 & 0.30 & 0.07 & 0.10 & 6.4e10 & 1.6e9 & 1.4e9 & 5.7e9 \\ 
{\bf If1b3LmidL}\tmark[d] & 2.0 & 0.25 & 0.10 & 0.30 & 0.15 & 0.25 & 6.4e10 & 1.6e9 & 8.4e9 & 1.1e10 \\ 
{\bf If5b4mbul} & 1.0 & 0.25 & 0.50 & 0.24 & 0.40 & 0.55 & 7.4e8 & 1.9e8 & 0.3e8 & 1.3e8 \\ 
{\bf If5b8mbul} & 1.0 & 0.25 & 0.50 & 0.24 & 0.80 & 0.90 & 7.4e8 & 3.1e9 & 1.2e8 & 5.3e8 \\ 
{\bf If9b1lowm} & 1.0 & 0.25 & 0.90 & 0.15 & 0.03 & 0.06 & 6.4e9 & 1.2e7 & 1.3e8 & 5.4e8 \\ 
{\bf If3b9dsk} & 1.0 & 0.25 & 0.32 & 0.22 & 0.95 & 0.80 & 1.6e9 & 1.1e11 & 2.1e9 & 6.2e9 \\ 
{\bf If3b9dskL}\tmark[d] & 2.0 & 0.25 & 0.32 & 0.22 & 0.71 & 0.62 & 1.6e9 & 1.1e11 & 8.0e9 & 9.3e9 \\ 
{\bf IfXb2gas} & 1.0 & 0.25 & 0.16,0.32,0.50 & 0.26 & 0.16 & 0.08 & 3.6e10 & 1.0e10 & 1.3e9 & 7.7e9 \\ 
{\bf Inf28b2} & 1.0 & 0.20 & 0.20,0.80 & 0.25 & 0.51 & 0.28 & 6.3e9 & 3.0e8 & 3.6e8 & 1.7e9 \\ 
{\bf Inf28b4} & 1.0 & 0.20 & 0.20,0.80 & 0.25 & 0.67 & 0.43 & 3.3e9 & 3.0e8 & 2.7e8 & 1.1e9 \\ 
{\bf Inf28b6} & 1.0 & 0.20 & 0.20,0.80 & 0.25 & 0.78 & 0.56 & 3.3e9 & 6.0e8 & 4.0e8 & 1.4e9 \\ 
{\bf Inf28b8} & 1.0 & 0.20 & 0.20.0.80 & 0.25 & 0.92 & 0.81 & 1.0e9 & 6.0e8 & 3.4e8 & 9.5e8 \\ 
{\bf Inf2b9} & 1.0 & 0.20 & 0.20 & 0.25 & 0.96 & 0.89 & 5.2e8 & 6.0e8 & 3.2e8 & 8.6e8 \\ 
{\bf Inf28b2h}\tmark[f] & 0.3 & 0.125 & 0.20,0.80 & 0.25 & 0.38  & 0.21 & 5.5e9 & 5.0e7 & 2.3e8 & 1.1e9 \\ 
{\bf Inf8b2hr}\tmark[f] & 0.3 & 0.20 & 0.80 & 0.25 & 0.17  & 0.11 & 5.5e9 & 2.5e7 & 2.2e8 & 1.1e9 \\ 
{\bf Inf28b9h}\tmark[f] & 0.3 & 0.125 & 0.20,0.80 & 0.25 & 0.81  & 0.75 & 5.5e9 & 5.0e10 & 2.1e9 & 1.1e10 \\ 
\hline\hline\\
}
\end{footnotesize}

In order to follow the behavior of gas inflow on smaller scales, we
re-simulate the central regions of interest at higher resolution, in a
series of progressively smaller-scale runs.  We begin by selecting a
number of representative outputs from the galaxy-scale simulations
described above, near the peak of activity.  We select several
snapshots in the gas-rich merger at key epochs: early close
encounters of the two galaxies, just at nuclear coalescence (which is
the peak of star formation in the nuclear region), and at the ``end''
of the merger (roughly $\sim10^{8}\,$yr after the final coalescence).
We also select snapshots typical of isolated, moderately bar-unstable
systems, at times where a bar and some inflow has developed; for
comparison, we also consider a fully stable (pre-bar) galaxy disk.  In
each case, we focus on the central $0.1-1$ kpc region, which includes
the majority of the gas that has been driven in from larger scales.
Some of the representative properties of these snapshots, at these
scales and times, are outlined in Table~\ref{tbl:galaxy}.

Our approach to re-simulating the nuclear region is to use the
larger-scale simulations to motivate the initial conditions of a
smaller scale calculation (a ``zoom-in'' or ``re-simulation'').  We do so by de-composing the
potential, density, and velocity distributions of the gas, stars, and
dark matter at a given time in the larger-scale simulation using the
basis expansion proposed in
\citet{hernquist:potential.expansion}. This allows us to not only
re-construct a smoothed density profile, but also to include the
asymmetric structures from the larger-scale simulation (if desired)
and to define where the potential is noise-dominated.\footnote{As one
  continues the expansion to include arbitrarily small-scale modes,
  the best-fit mode amplitude will eventually yield an amplitude
  consistent with the shot noise in the simulation, roughly at the
  scale of the median inter-particle spacing; we discard higher mode
  numbers as they are particle-noise dominated.}  From these stellar, gas, and dark matter distributions, we re-populate the gas and stars in the central regions (the scale we wish to re-simulate; generally out to an outer radius of $\sim1-2\,$kpc) and use this as the initial condition for a new simulation that we run for several local dynamical times.  To be conservative, we typically initialize only a small amount of gas in the inner parts of the re-simulation\footnote{To avoid numerical effects from a step-function cutoff in the mass profile at small radii, we typically truncate the gas mass profile with a $\Sigma_{\rm gas}\propto (R/R_{0})^{2}$ power law inside of a radius $R_{0}\sim3-5\,\epsilon$ in the parent simulation  ($\epsilon$ is the minimum smoothing length). 
Gas within this radius ($<1\%$ of the re-simulated mass)
is initialized with circular orbits.}, since the larger-scale simulation from which the initial condition is drawn has little information about the gas properties on small scales; in Appendix ~\ref{sec:numerical} we show that the subsequent dynamics does not depend significantly on these details of the initial conditions.  

We have carried out a total of $\sim50$ simulations at these
intermediate scales, which together span a wide range in the key
parameters: the equation of state of the gas and the relative mass
fraction in a pre-existing bulge, gas disk, and stellar disk.
Table~\ref{tbl:intermediate} summarizes the key properties of physical
importance in several of these simulations (some numerical studies and
surveys of initial condition, which turn out to have little effect on
the key results, are not listed). 
Because our general approach is to systematically survey the initial 
conditions, we do not identify every simulation in Table~\ref{tbl:intermediate} 
with an exact snapshot from Table~\ref{tbl:galaxy}; rather, they should be 
considered a systematic parameter survey of possible intermediate-scale 
conditions, motivated by the typical range of sub-kpc conditions seen 
in our galaxy-scale simulations. 
Dark matter is present, but is
dynamically irrelevant at these scales.  We have also varied the mass
of the central black hole, but at these scales it is still dynamically
unimportant.  Although the initial conditions for our calculations are
drawn from galaxy-scale simulations, the dynamics on small-scales
depends primarily on a few key properties of the simulation ($f_{\rm
  gas}$ and $B/T$), and is thus insensitive to many of the details of
the galaxy on larger scales. 

Our intermediate scale simulations
typically involve $\simeq 10^{6}\,$particles, with a force resolution
of a few pc and a particle mass of $\approx 10^{4}\,\msun$.  The
duration of the simulation is $\sim 10^{7}-10^{8}\,$yr -- this is many
dynamical times at small radii, but small compared to the dynamical
time at larger radii.\footnote{
To ensure there are no later-time phenomena of interest, and to study the 
relaxed structure of the stellar remnant produced by each re-simulation, we evolve 
most for $\gtrsim2\times10^{8}$ years, by which time all the gas is exhausted. 
We find that there is no qualitatively new behavior at these later times.}
  These re-simulations can thus be thought of as
a probe of the instantaneous behavior of the gas at small radii given
the inflow conditions set at larger radii.

We discuss a number of numerical tests and variations about this basic
methodology in Appendix~\ref{sec:numerical}.  Specifically, we show
that our results are not sensitive to including properties of the
larger-scale galaxy in which the simulation should be embedded, such
as the $\sim$ kpc-scale tidal potential.  They are also not sensitive
to whether we initialize an axisymmetric gas/potential distribution,
or whether the initial condition includes the non-axisymmetric modes
present in the larger-scale simulation.  The reason is that
instabilities due to self-gravity can grow exponentially from
shot-noise in the simulation even given an initially axisymmetric
structure. Thus the presence of initial asymmetries on $\sim 100$ pc
scales does not have a significant effect on the resulting transport
of gas to smaller radii; the transport is determined by the presence
(or absence) of internal instabilities in the gas on small scales.
Finally, because the ``initial'' densities on small scales are
intentionally initialized to be low relative to their values later in
the simulation, our results are not particularly sensitive to how we
initialize gas at small radii in the re-simulation; this is
important because there is no reliable information about these
small-scales in our larger-scale simulation.

To check that our re-simulation approach has not introduced any artificial behavior, 
we have run a small number of ultra-high resolution galaxy scale simulations, the inner properties of 
which can be compared to the intermediate scale re-simulations summarized here.
For these very high resolution calculations there is continuous inflow from 
large scales, so they can be self-consistently evolved for many dynamical times. The expense of these calculations, 
however, limits the survey of initial conditions possible. 
We have 6 such simulations: three mergers, identical in mass and 
geometry to our canonical case in Table~\ref{tbl:galaxy}, 
but with initial $f_{\rm gas}=0.2,\,0.4,\,0.8$, and $\sim10^{7}$ particles, 
which gives SPH smoothing lengths of $\sim10\,$pc. While not quite as high-resolution as our re-simulation runs, 
these provide an important check on the results of the latter and are run self-consistently for $4\times10^{9}\,$yr.   
We will show results from these ultra-high resolution simulations at several points; 
we find that they are quite similar to our re-simulation runs, 
thus supporting the methodology used for most of our calculations.     

The very high resolution merger calculations also allow us to follow the binary 
BH pair to much smaller separation (our assumptions lead to rapid merger 
below $\approx10\,$pc in these calculations).  We confirm, in these cases, 
that the BH-BH merger precedes most of the gas inflows at $\gtrsim 10$ pc, so 
that the assumption that the binary has merged is probably reasonable for our
re-simulation calculations.   Moreover, the gas mass at $\sim 10$ pc 
is large ($\sim M_{\rm BH}$) in the merger simulations.   Thus if gas-rich 
reservoirs indeed drive rapid BH-BH coalescence, the rapid merger of the 
two BHs should be a reasonable assumption on all of the  scales that we simulate
\citep[see e.g.][]{escala:bh.mgr.idealized,perets:binary.bh.loss.cone.filled.by.massive.perturbers,
mayer:bh.binary.sph.zoom.sim,perets:massive.perturber.bh.mgr,
dotti:bh.binary.inspiral,cuadra:binary.bh.mergers.w.gas.disks}. 

\vspace{-0.4cm}
\breaker
\subsection{Nuclear Scales: From 10\,pc to 0.1\,pc}
\label{sec:sims:10pc}

\begin{footnotesize}
\ctable[
  caption={{\normalsize Nuclear-Scale Resimulations ($\sim 0.1-100$ pc)}\label{tbl:nuclear}},center,star
  ]{lcccccccccccccc}{
\tnote[ ]{Parameters describing our nuclear-scale re-simulations of the 
sub-$100$\,pc regions from intermediate-scale simulations.
{\bf (1)} Simulation name/ID.
{\bf (2)} Minimum smoothing length (in pc).
{\bf (3)} Equation of state parameter (Figure~\ref{fig:qeos}). 
{\bf (4)} Initial gas fraction of the disky/cold component.
{\bf (5)} Initial scale height of the disky component.
{\bf (6)} Black hole mass ($M_{\sun}$). 
{\bf (7)} Initial bulge or nuclear stellar cluster mass, 
inside the given radius. 
{\bf (8)} Initial maximum surface density of the disky/cold component (gas plus stars). 
{\bf (9)} Initial mass of the disky component (gas plus stars) inside a given radius 
(does not include the BH mass or, if significant, nuclear star cluster/bulge mass). 
Dark matter is insignificant on these scales.  \\ 
}
\tnote[a]{
Central ``hole'' is extended in disk out to $5\,$pc.
}
\tnote[b]{
Simulations with no central deficit of matter; the initial density from the larger-scale simulation is extrapolated in to $r\rightarrow0$. Also expanded into series of initial conditions, as described in Appendix~\ref{sec:numerical}. 
}
}{
\hline\hline
\multicolumn{1}{c}{Simulation} &
\multicolumn{1}{c}{$\epsilon$} &
\multicolumn{1}{c}{$\qeos$} & 
\multicolumn{1}{c}{$f_{\rm gas}$} & 
\multicolumn{1}{c}{$h/R$} & 
\multicolumn{1}{c}{$M_{\rm BH}$} & 
\multicolumn{2}{c}{$M_{\rm b/cl}(<R)$} & 
\multicolumn{1}{c}{$\Sigma_{\rm d}(0)$} & 
\multicolumn{3}{c}{$M_{\rm d}(<R)$  $[M_{\sun}]$} \\
\multicolumn{1}{c}{Name} &
\multicolumn{1}{c}{[pc]} &
\multicolumn{1}{c}{ } & 
\multicolumn{1}{c}{($100$\,pc)} & 
\multicolumn{1}{c}{($100$\,pc)} & 
\multicolumn{1}{c}{$[M_{\sun}]$} & 
\multicolumn{1}{c}{$10$\,pc} & 
\multicolumn{1}{c}{$50$\,pc} & 
\multicolumn{1}{c}{$[M_{\sun}\,{\rm kpc^{-2}}]$} & 
\multicolumn{1}{c}{$1$\,pc} & 
\multicolumn{1}{c}{$10$\,pc} & 
\multicolumn{1}{c}{$50$\,pc} \\
\multicolumn{1}{c}{{\bf (1)}} &
\multicolumn{1}{c}{{\bf (2)}} &
\multicolumn{1}{c}{{\bf (3)}} &
\multicolumn{1}{c}{{\bf (4)}} &
\multicolumn{1}{c}{{\bf (5)}} &
\multicolumn{1}{c}{{\bf (6)}} &
\multicolumn{2}{c}{{\bf (7)}} &
\multicolumn{1}{c}{{\bf (8)}} &
\multicolumn{3}{c}{{\bf (9)}} \\
\hline
{\bf Nf8h1c0} & 0.3 & 0.0,0.25 & 0.75 & 0.16,0.26 & 2.9e7 & 1.8e5 & 3.0e5 & 7.5e10 & 1.0e5 & 1.5e6 & 1.0e8 \\ 
{\bf Nf8h1c0hol}\tmark[a] & 0.3 & 0.25 & 0.75 & 0.24 & 2.9e7 & 1.8e5 & 3.0e5 & 7.4e10 & 1.0e3 & 1.2e6 & 1.0e8 \\ 
{\bf Nf5h2c1} & 0.1 & 0.25 & 0.50 & 0.24 & 3.0e8 & 1.2e7 & 8.0e7 & 1.2e11 & 3.6e5 & 2.4e7 & 1.6e8 \\ 
{\bf Nf28h1c1} & 0.1 & 0.25 & 0.19,0.75 & 0.27 & 3.0e7 & 1.3e7 & 2.8e7 & 7.5e10 & 3.6e5 & 2.4e7 & 1.6e8 \\ 
{\bf Nf7h12c0dsk} & 0.1 & 0.25 & 0.70 & 0.23 & 2.9e7,3.0e8 & 0.0 & 0.0 & 4.5e9 & 1.0e4 & 0.9e6 & 6.0e6 \\ 
{\bf Nf5h1c0} & 0.1 & 0.25 & 0.48 & 0.25 & 2.9e7 & 0.0 & 0.0 & 1.2e11 & 3.0e5 & 2.4e7 & 1.6e8 \\ 
{\bf Nf5h1c2} & 0.1 & 0.25 & 0.48 & 0.25 & 2.9e7 & 3.0e7 & 9.0e7 & 1.2e11 & 3.0e5 & 2.4e7 & 1.6e8 \\ 
{\bf Nf8h1c0thin} & 0.1 & 0.125 & 0.75 & 0.16,0.27 & 3.0e7 & 1.8e5 & 0.3e7 & 7.6e10 & 2.0e5 & 1.5e7 & 1.0e8 \\ 
{\bf Nf5h1c1thin2} & 0.1 & 0.125,0.25 & 0.50 & 0.07,0.14 & 3.0e7 & 3.0e5 & 1.2e7 & 1.2e11 & 2.0e5 & 2.4e7 & 1.6e8 \\ 
{\bf Nf8h1c1qs} & 0.1 & 0,0.125,0.25,1 & 0.75 & 0.28 & 3.0e7 & 3.0e6 & 1.4e7 & 7.3e10 & 2.0e5 & 1.7e7 & 1.2e8 \\ 
{\bf Nf8h2c1} & 0.1 & 0.25 & 0.75 & 0.2,0.33 & 3.0e8 & 3.6e6 & 1.4e7 & 7.6e10 & 2.0e5 & 1.9e7 & 1.2e8 \\ 
{\bf Nf1h1c1low} & 0.1 & 0.25 & 0.08 & 0.25 & 3.0e7 & 3.5e6 & 1.4e7 & 1.7e11 & 4.7e5 & 3.7e7 & 2.5e8 \\ 
{\bf Nf3h1c1mid} & 0.1 & 0.20 & 0.26 & 0.23 & 3.0e7 & 3.5e6 & 1.4e7 & 2.1e11 & 6.0e5 & 4.6e7 & 3.0e8 \\ 
{\bf Nf6h12c2dsk} & 0.1 & 0.25 & 0.57 & 0.22 & 2.9e7,3.0e8 & 7.2e6 & 2.6e7 & 4.4e9 & 1.1e5 & 8.1e6 & 3.4e7 \\ 
{\bf Nf8h1c3dskM} & 0.1 & 0.25 & 0.75 & 0.30 & 2.9e7 & 1.6e7 & 7.1e7 & 7.4e10 & 3.5e5 & 3.0e7 & 1.7e8 \\ 
{\bf Nf8h1c1dens} & 0.1 & 0.25 & 0.75 & 0.25 & 3.0e7 & 3.6e6 & 1.4e7 & 3.8e11 & 1.1e6 & 8.1e7 & 5.3e8 \\ 
{\bf Nf8h1c1ICs}\tmark[b] & 0.1 & 0.25 & 0.75 & 0.28 & 3.0e7 & 3.1e6 & 1.4e7 & 7.3e10 & 2.0e5 & 1.7e7 & 1.2e8 \\ 
{\bf Nf8h1c1thin} & 0.1 & 0.125,0.18 & 0.75 & 0.08,0.17 & 3.0e7 & 3.1e6 & 1.4e7 & 7.3e10 & 2.0e5 & 1.7e7 & 1.2e8\\ 
{\bf Nf2h2b2} & 0.1 & 0.20 & 0.20 & 0.25 & 3.0e7 & 6.5e6 & 2.0e7 & 3.0e11 & 8.1e4 & 7.0e7 & 4.2e8 \\ 
{\bf Nf8h2b2} & 0.1 & 0.20 & 0.80 & 0.25 & 3.0e7 & 1.3e7 & 4.0e7 & 1.5e11 & 4.7e4 & 3.5e7 & 2.1e8 \\ 
{\bf Nf2h2b4} & 0.1 & 0.20 & 0.20 & 0.25 & 3.0e7 & 6.5e6 & 2.1e7 & 1.5e11 & 4.3e4 & 3.5e7 & 2.1e8 \\ 
{\bf Nf8h2b4} & 0.1 & 0.20 & 0.80 & 0.25 & 3.0e7 & 1.3e7 & 4.0e7 & 7.7e10 & 2.4e4 & 1.7e7 & 1.1e8 \\ 
{\bf Nf2h2b5} & 0.1 & 0.20 & 0.20 & 0.25 & 3.0e7 & 6.5e6 & 2.1e7 & 7.7e10 & 2.1e4 & 1.8e7 & 1.1e8 \\ 
{\bf Nf28h2b6} & 0.1 & 0.20 & 0.20,0.80 & 0.25 & 3.0e7 & 1.1e7 & 3.7e7 & 3.8e10 & 1.1e4 & 8.8e6 & 5.3e7 \\ 
{\bf Nf8h2b8} & 0.1 & 0.20 & 0.80 & 0.25 & 3.0e7 & 1.3e7 & 4.1e7 & 1.5e10 & 4.7e3 & 3.5e6 & 2.1e7 \\ 
{\bf Nf28h2b9} & 0.1 & 0.20 & 0.20,0.80 & 0.25 & 3.0e7 & 1.1e7 & 3.8e7 & 9.6e9 & 2.7e3 & 2.2e6 & 1.3e7 \\ 
{\bf Nf8h2b1h} & 0.015 & 0.20 & 0.80 & 0.25 & 3.0e7 & 6.4e6 & 2.0e7 & 1.5e11 & 5.2e4 & 3.5e7 & 2.1e8 \\ 
{\bf Nf8h2b3L} & 0.1 & 0.20 & 0.80 & 0.25 & 3.0e7 & 2.3e7 & 1.1e8 & 1.5e11 & 9.3e3 & 3.9e7 & 4.9e8  \\
%
{\bf Nf8h2b4q} & 0.1 & 0,0.02,0.06,0.12, & 0.80 & 0.25 & 3.0e7 & 1.3e7 & 4.0e7 & 7.7e10 & 2.4e4 & 1.7e7 & 1.1e8 \\ 
 &  & 0.25,0.35,0.5,0.7,1 &   &   &   &   &   &   &   &   &   \\ 
\hline\hline\\
}
\end{footnotesize}

The characteristic initial scale-lengths of the nuclear disks in our
intermediate scale calculations are $\sim 0.2-0.5\,$kpc.  As we
discuss in \S \ref{sec:results:transport}, if the gas fraction is
sufficiently large, instabilities quickly develop that transport
material down to $\sim 1-10\,$pc, near the resolution limit of our
intermediate scale calculations.  Material begins to pile up at these
radii because the BH mass dominates the potential and the efficiency
of large-scale modes decreases at small radii.  In order to understand
the dynamics on yet smaller scales, we therefore repeat our
``re-simulation'' methodology once more.  The approach is identical to that
described above, but this time using the intermediate-scale
simulations with resolution of $\lesssim 10\,$pc as our ``parent
simulation'' from which to motivate the initial conditions.

We again carried out $\sim 50\,$ such simulations, typically with
$\sim 10^{6}\,$particles, and a force/spatial resolution of $\sim
0.1$\,pc (particle mass $\approx100\,\msun$).  The properties of these
simulations are summarized in Table~\ref{tbl:nuclear}.  The
simulations are evolved for $\sim 10^{5}-10^{6}\,$yr; this is large
compared to the dynamical time at the smallest radii $\sim 0.1$ pc,
but very small relative to the dynamical time of the larger-scale
simulations from which the initial conditions are drawn.
The characteristic spatial scale of the re-simulated material is initially
$\sim 10-30\,$pc.  As described in Appendix ~\ref{sec:numerical}, we
carried out a number of numerical tests of the robustness of these
simulations.

At radii $\sim$ pc, the parameters that determine the dynamics are
largely the equation of state of the gas, the mass of the BH, the mass
of the nuclear disk formed by the inflow from larger scales, and the
gas fraction of that nuclear disk. Since the BH dominates the
spherical component of the potential at these radii, the ``bulge''
mass at these radii is only of secondary importance; we include it but
find that it makes little difference.

As in \S \ref{sec:sims:100pc}, we have checked the results of these ``re-simulations'' by 
carrying out a small subset of ultra-high resolution runs. These extend from $\sim 0.3-1000$ pc and follow inflow from larger scales deep into the potential of the BH; because they resolve larger spatial scales than our typical "nuclear scale" simulation, these can be run self-consistently for $2\times10^{8}$\,yr. 
Specifically, we have five such high-resolution 
intermediate scale simulations (see Table~\ref{tbl:intermediate}),
three with initial $f_{\rm gas}=0.8$ (a low, intermediate, and high $B/T$ case), 
and two with $f_{\rm gas}=0.2$ (low and high $B/T$). 
They have $\sim10^{7}$ particles and gravitational softening lengths 
of $\sim0.3\,$pc. 
We show the results from these runs explicitly at several points; we find that they are completely consistent with our survey of re-simulations, which cover a larger parameter space of galaxy/BH properties, but are more limited in dynamic range.

It is important to note up-front that our simplified treatment of the
ISM physics becomes particularly suspect on nuclear scales $\sim$ pc.
At these radii, our assumption that we can average over the dynamics
of stellar winds, supernovae, HII regions, etc.\ and define an
effective ISM equation of state may break down.  Nonetheless, we
believe that the efficient angular momentum transport found here is
likely generic, so long as some of the gas is prevented from forming
stars and the gas fraction is sufficiently high that instabilities
generated by self-gravity are initiated.  The fact that the main
sequence lifetime of a massive star is {\it longer} than the local
dynamical time on small scales probably increases the efficacy of
stellar feedback and decreases the fraction of the gas turned into
stars per dynamical time
\citep{murray:molcloud.disrupt.by.rad.pressure}.

At scales $\ll 0.1\,$pc, the potential is fully Keplerian and viscous
heating is sufficient to stabilize the disk against its own
self-gravity (i.e., $Q \gtrsim 1$) \citep{goodman:qso.disk.selfgrav}.  At these
radii, the system begins to approach a traditional accretion disk.
Given the cessation of star formation and the deep potential well of
the BH, we assume that the inflow rate at $\sim 0.1$ pc is a
reasonable proxy for the true accretion rate onto the BH.  Because our
simulations are not well-suited to describe the physics of the disk on
scales $\lesssim 0.1$ pc, we do not perform a further ``zoom in.''

\vspace{-0.8cm}
\section{Overview: From kpc to sub-pc Scales}
\label{sec:results:transport}

\begin{figure*}
    \centering
    \scaleup
    \plotsideup{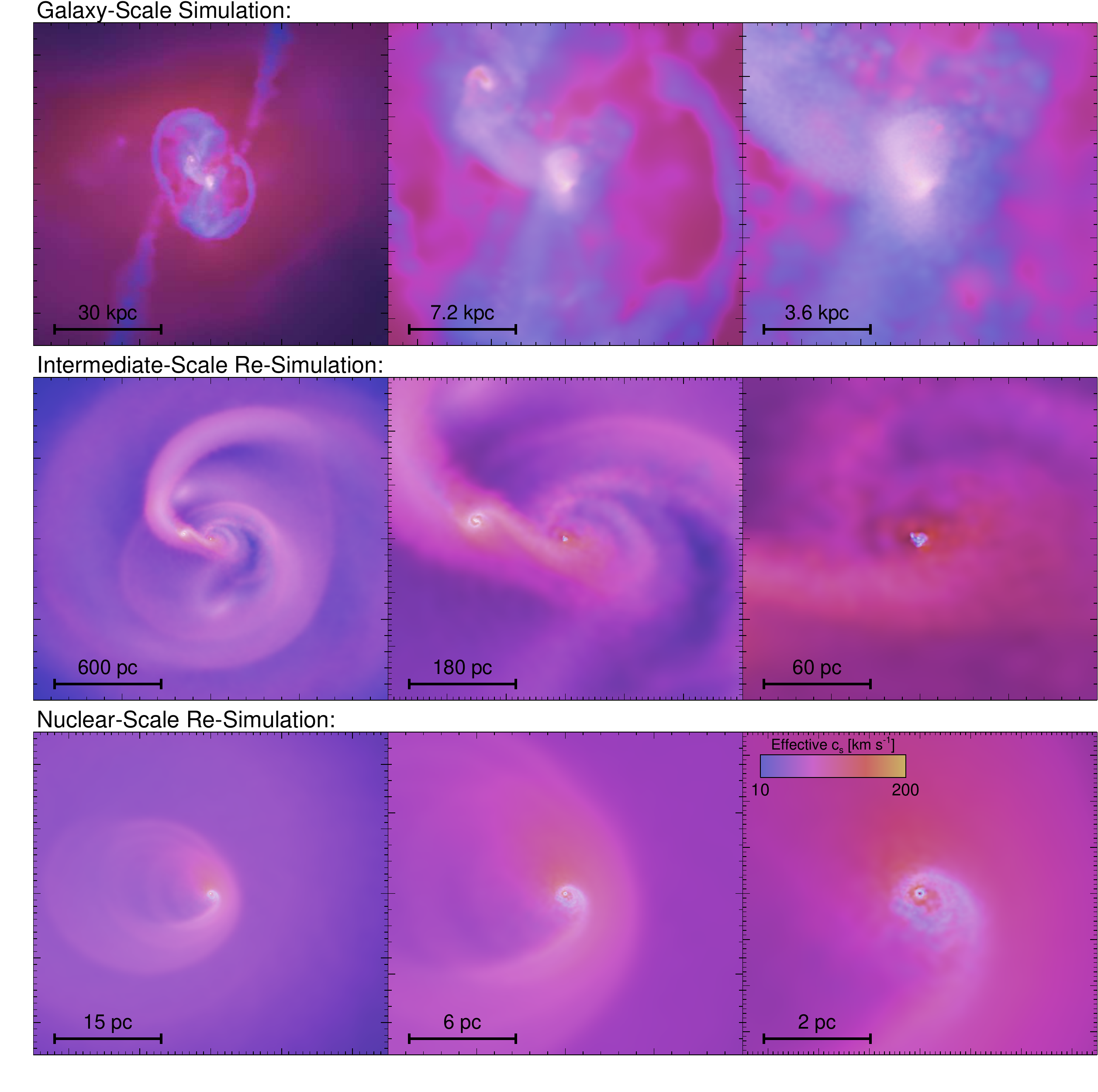}
    \caption{Example of our multi-scale simulations used to follow gas
      flows from $\sim100\,$kpc to $\sim 0.1\,$pc.  Each row is a separate simulation, with the initial conditions for the intermediate and nuclear-scale simulations taken from the output of the larger-scale runs in the row above it.           Each panel shows
      the projected gas density (intensity) and effective sound speed
      (color; blue is gas with an effective $c_{s}\sim10\,{\rm
        km\,s^{-1}}$, through yellow at $\sim 100-200\,{\rm
        km\,s^{-1}}$).  Each image is rotated to project the 
      gas density ``face on'' relative to its angular momentum vector.
      From top left to bottom right, panels zoom in to the nuclear
      region around the BH, with resolution spanning a factor $\sim
      10^{6}$ in radius.       {\em Top:} Large-scale gas-rich
      galaxy-galaxy major merger simulation, just after the
      coalescence of the two nuclei (run {b3ex(co)} in
      Table~\ref{tbl:galaxy}).  The apparent second nucleus is actually a
      clump formed from gravitational instability.  {\em Middle:} A
      higher-resolution re-simulation of the conditions in the central
      kpc (run {If3b3midRg} in
      Table~\ref{tbl:intermediate}). Despite the fact that the
      background potential is largely relaxed on these scales, the
      very large gas inflows lead to a strongly self-gravitating disk
      on $\sim0.5\,$kpc scales that develops a strong spiral
      instability, leading to efficient angular momentum transport to
      $\sim 10\,$pc.  Again, some clumping appears (there is only one
      nucleus).  {\em Bottom:} High resolution re-simulation of the
      central $\sim 30$ pc of the intermediate-scale simulation,
      with a resolution $\sim0.1\,$pc (run {Nf5h1c2} in
      Table~\ref{tbl:nuclear}).  The potential is quasi-Keplerian,
      suppressing traditional bar and spiral instabilities, but the
      large inflows lead to a self-gravitating system that
      develops a standing eccentric disk mode (single-armed $m=1$).
      The stellar and gaseous eccentric disks precess relative to one
      another on $\sim1-10\,$pc scales and drive efficient inflows of
      $\sim10\,\msun\,{\rm yr^{-1}}$ into the central $0.1\,$pc.
      \label{fig:zoom}}
\end{figure*}

Using the numerical simulations described in \S \ref{sec:sims}, we now
describe how gas is transported from $\gg $kpc scales to $\ll$pc
scales.  Initially, our discussion is somewhat qualitative; we focus
on emphasizing the key physics at play and our key results.  In
\S~\ref{sec:stability} we discuss the relevant stability criteria
more quantitatively, and outline some specific criteria necessary for
``interesting'' gas inflow.

Figure~\ref{fig:zoom} shows an illustrative example of the results of
our re-simulations on various scales.  We plot gas surface density maps, with color
encoding the gas effective sound speed, from scales of $\sim100\,$kpc
to $<1\,$pc.  The simulation in this case is a fairly gas-rich major
merger ($f_{\rm gas}\sim30-40\%$ at the time of the final coalescence)
of two $5\times10^{10}\,\msun$ baryonic mass galaxies. The smaller scale re-simulations 
were carried out just after the coalescence of the two
nuclei, which is near the peak of star formation activity, but when
the system is still quite gas rich. The initial systems had
pre-existing bulges of $\sim1/3$ the disk mass and BHs initialized on
the $M_{\rm BH}-\sigma$ relation ($\sim10^{7}\,\msun$).  Each panel is
rotated so that the view is close to ``face-on'' with respect to the
total angular momentum vector of the gas plotted in the image.  Viewed
edge-on, much of the gas forms a modestly thin ($H/R\lesssim 0.3$)
disky distribution at all radii.

\vspace{-0.5cm}
\subsection{Large Scales: Mergers and Bars}
\label{sec:results:transport:large}

From $\sim100\,$kpc to $\gtrsim 100\,$pc, the gas flows are
well-resolved by our galaxy-scale simulation. In mergers, the final
collision of the two galaxies yields strong torques that efficiently
cause most of the gas to flow to the center on a timescale
approximately equal to a few dynamical times. This process has been
described in detail in e.g.\ \citet{hopkins:disk.survival}, and
references therein, but for completeness we briefly summarize the
important physics. The secondary/merging galaxy does not {\em
  directly} torque the gas. Rather, the torques on the gas are
dominated by {\em local} torques from {\em stars originally in the
  same disk as the gas}.  The merger induces triaxial ``sloshes'' and
bar-like structures in the stars, i.e., non-axisymmetric modes,
supported by radial and/or random orbits.  These are also induced in
the gas. However, because the gas is dissipational, the gaseous modes
slightly lead those in the stars. The stellar disturbance, being
physically close to, trailing, and in near-resonance with the gas,
produces a strong torque that removes angular momentum from the gas
and easily dominates the total torque \citep[torques directly from the
secondary galaxy, from the primary halo, or hydrodynamic torques from
shocks or internal clump collisions, are all $\lesssim 10\%$ effects;
see e.g.][]{barneshernquist96,barnes:review,hopkins:disk.survival}.

\begin{figure*}
    \centering
    \scaleup
    \plotsidesmall{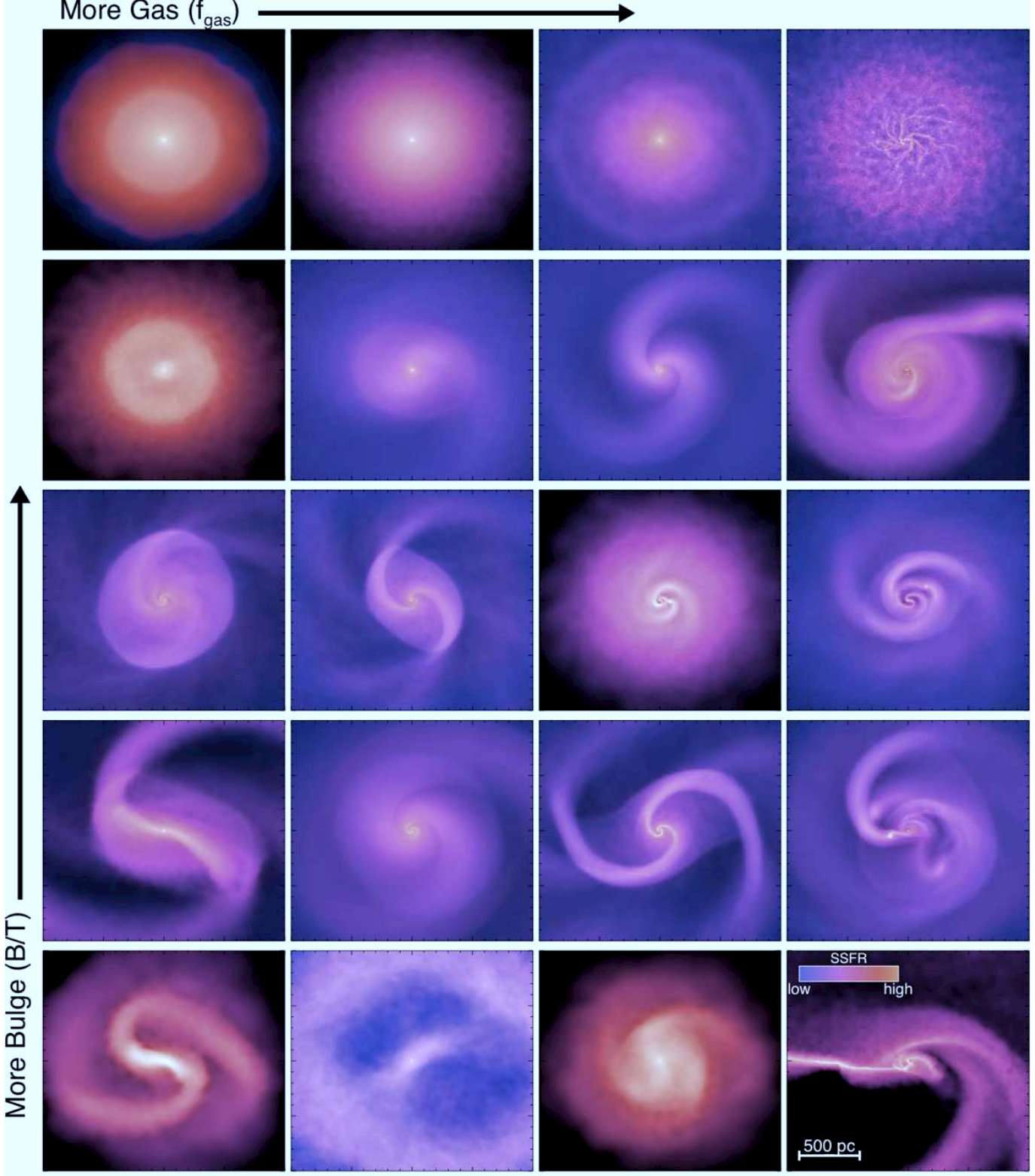}
    \caption{Images of the morphologies of the secondary instabilities
      that develop on intermediate scales ($\sim 100$ pc), as a
      consequence of galactic-scale torques bringing gas into the
      central kpc. Each panel shows projected gas density (intensity)
      and local star formation rate (color, from blue through yellow)
      in the central kpc. We sort the simulations by the parameters
      that have the largest impact on the dynamics: the bulge-to-disk
      ratio (disk being both stellar and gaseous) within $300\,$pc and
      gas fraction $M_{\rm gas}/(M_{\rm gas}+M_{\ast,\ \rm disk})$:
      from {\em top} -- bulge-dominated systems with $B/T\gtrsim0.8$
      to {\em bottom} -- disk-dominated systems with $B/T\lesssim0.1,$
      and from {\em left} -- systems with $f_{\rm gas}\lesssim0.1$ to
      {\em right} -- systems with $f_{\rm gas}\gtrsim0.8$. Global
      instabilities become more prominent with decreasing $B/T$; local
      instabilities (clumping/fragmentation and/or 
      more tightly wound spirals) become more prominent
      with increasing $f_{\rm gas}$. The simulations at fixed B/T
      and/or $f_{\rm gas}$ are similar, but include variation in e.g.\
      the initial mass profile shapes; this contributes to the
      dramatic diversity of morphologies on nuclear scales.  Almost
      all systems that are not entirely bulge-dominated develop strong
      large-scale instabilities that efficiently transport gas to
      $\sim 10\,$pc. Observationally, these would be categorized as a
      mix of bars, spirals, nuclear rings, crossed rings,
      fragmentation/clump instabilities, and single or triple-armed
      spirals. \label{fig:100pc.morphologies}}
\end{figure*}

After the two galactic nuclei coalesce, the disturbances in the
stellar component of the galaxy relax away in a number of crossing
times. Until they relax, gas inflows continue.  Moreover, the
coalescence of the nuclei completes much more rapidly, in a timescale
close to a single crossing time, at least in a {\em major} merger.
Thus, a significant fraction of the gas inflows can occur in the
background of a rapidly relaxing stellar potential, in the wake of the
nuclear coalescence. This is the stage illustrated in
Figure~\ref{fig:zoom}.  

The gas that loses angular momentum flows in to radii $\sim
0.5-1\,$kpc (for an $\sim L_{\ast}$ system), where it participates in
a nuclear starburst and builds a dense central stellar mass
concentration, critical for establishing the structural properties and
size of the remnant spheroid.  At these scales, the system is often
gas-dominated for a short period of time owing to these inflows
(provided the merger is sufficiently gas-rich).  However, as the gas
forms stars, the central region will quickly become more
stellar-dominated; because these stars form out of the gaseous disk,
in the relaxing potential --- they are {\em not} themselves violently
relaxed.  This is important for the subsequent evolution of the system
because of the presence of disk instabilities that would be suppressed
by a larger dispersion-supported (spherical) component in the very
central region.

The general scenario summarized here can be applied not just to major
mergers, but also to minor mergers, fly-by encounters, and even
sufficiently bar-unstable stellar disks.  The details will be
different, but the qualitative steps above, and the exchange of
angular momentum between gas and stars, is robust, ultimately leading
to inflow to sub-kpc scales. The subsequent evolution depends largely
how much material is efficiently channeled to small radii (relative to
the bulge and BH mass), not on how that material gets there.

\vspace{-0.5cm}
\breaker
\subsection{Morphology and Gas Transport From 1 kpc to 10 pc}
\label{sec:results:nuclear}

\subsubsection{General Behavior}
\label{sec:results:nuclear:general}

The gas infalling from large radii begins to ``pile up'' at radii
$\sim 0.1-1$ kpc, rather than continuously flowing in to yet smaller
radii. This is because the torques from the disturbances at large
radii become less efficient at small radii. This happens for three
reasons: (1) In the merger context, the stellar perturbation at small
radii relaxes after coalescence, decreasing the efficiency of gas
inflow.
(2) The rapid gas inflow implies that the system becomes increasingly
gas-dominated at radii $\sim 100$ pc, even if the initial disk gas
fraction is low, $\sim0.1$.  Because the primary angular momentum sink
of the gas is the local stars, when the system becomes locally
gas-dominated, angular momentum transfer is actually less efficient
\citep[see][]{hopkins:disk.survival,hopkins:disk.survival.cosmo}.  (3) The gas can
encounter the equivalent of an inner Linblad resonance.  This is
especially important for unstable gas bars, minor mergers, and
disturbances induced by early passages.
For the case of coalescence following major mergers, the disturbance
is {\em not} a single mode, but a series of modes at all scales. As
such, there is often no formal inner Linblad resonance or angular
momentum barrier (each mode may have such a barrier, but these are
spread over a wide range of scales; there is thus a means to overcome
the barrier associated with any single mode).

Figure~\ref{fig:zoom} shows the outcome of gas pile-up in the central
kpc using an intermediate-scale re-simulation ({\it middle row}). In this case, the
intermediate scale simulation is a high-resolution re-simulation of the larger-scale gas
distribution at a given epoch in a gas-rich major merger.  The gas
density reached from the larger-scale inflows is quite large -- $\sim
10^{10}\,\msun$ worth of gas has formed a disky component with a scale
length of $0.3\,$kpc and an average surface density of
$\sim10^{10}\,\msun\,{\rm kpc^{-2}}$. This is a large fraction of the
galaxy mass -- larger than the pre-existing bulge within these
radii. The small-scale gas disk is therefore strongly
self-gravitating.  Indeed, we see from Figure~\ref{fig:zoom} that it
quickly develops unstable, non-axisymmetric modes.\footnote{These
  modes develop almost identically even if we initialize the
  re-simulation to be perfectly smooth and remove all external tidal
  forces (see Appendix~\ref{sec:numerical}). It is thus not sensitive
  to the larger-scale environment; rather, the system is simply
  strongly globally unstable.}  This is essentially the ``bars within
bars'' scenario predicted by \citet{shlosman:bars.within.bars},
although the morphology of the system is clearly not a simple bar;
this is an important point to which we return below.  The strength of
the modes that develop in this re-simulation {\em depend} on the fact
that there is star formation in the gas -- as the gas turns into stars
in situ, those stars develop non-axisymmetric modes, and the two
precess relative to one another.  As in the galactic-scale torques
discussed above, this produces particularly efficient angular momentum
transport.  These processes ultimately lead to a significant amount of
gas flowing down to $\sim 10\,$pc.

\vspace{-0.3cm}
\subsubsection{Diversity in Morphologies and Inflow Strengths}
\label{sec:results:nuclear:diversity}

For our fiducial parameterization of the ISM equation of state, the
disk-to-bulge ratio and gas fraction on $\simeq 0.1-0.3$ kpc scales
have the largest influence on the dynamics and angular momentum
transport in our intermediate-scale simulations (for discussion of the
role of $q_{\rm EOS}$ and sub-resolution physics, see
Appendix~\ref{sec:ism}).  Figure~\ref{fig:100pc.morphologies}
illustrates this by showing the $\sim100$\,pc scale morphology of a
representative subset of our intermediate scale re-simulations, each
after a couple of local dynamical times of evolution
($\sim10^{7}-10^{8}\,$yr).  These are sorted by gas fraction $f_g$ and
$B/T$.\footnote{The other parameters of the simulations are not all
  identical, representing the properties of the simulations from which
  they are selected, but the key qualitative behavior depends
  primarily on these two parameters.}

As Figure \ref{fig:100pc.morphologies} shows, systems with very large
$B/T\gtrsim 0.9$ (top row) are globally stable, as expected
analytically.  In the extremely gas-rich, large $B/T$, case (top
right), local Toomre-scale instabilities develop -- if such small
clumps were infinitely long-lived, their orbits would decay via
dynamical friction and allow for some transport of gas to the center.
However, because the clumps are dense, they quickly turn into stars --
such a mechanism is largely a means to move {\em stars} to the
galactic nucleus, not gas (and in fact leads to nuclear stellar
clusters much larger than those observed; see Appendix \ref{sec:ism}
for a more detailed discussion).  Previous claims that such
clump-sinking could efficiently fuel BH growth \citep[see e.g.\ the
discussion in][and references therein]{wada:torus.mol.gas.hydro.sims,
  kawakatu:disk.bhar.model} have neglected star formation and have
thus dramatically over-estimated the inflow of gas via this process.
Local clumping instabilities like these do, of course, occur, forming
molecular clouds and star clusters.  However, there are both
theoretical and observational arguments that suggest that they quickly
dissolve on a few local dynamical times, probably via feedback from
some combination of stellar winds, HII regions, and radiation pressure
\citep{larson:gmc.scalings,blitz:gmc.properties,
  krumholz:gmc.lifetimes.model,allen:stellar.cluster.lifetime.short.vs.gal,
  bonnell:gmc.lifetimes.from.shock.turbulence,
  elmegreen:rapid.gmc.collapse.disruption}.
We also note that, under certain conditions, such local instabilities could instead produce small scale, tightly wound spiral waves instead of leading to fragmentation \citep[e.g.][]{lodato:2004.acc.disk.spiralwaves,rice:maximum.viscous.alpha,
boley:2006.protoplanetary.disk.w.cooling}; however, this generally occurs 
when the cooling time is comparable to or larger than the dynamical time, which is not the case in these simulations 
(although it will also depend on the turbulent decay time which, if sufficiently large, can 
allow turbulence to suppress runaway clumping and star formation).

\begin{figure*}
    \centering
    \scaleup
    \plotsidesmall{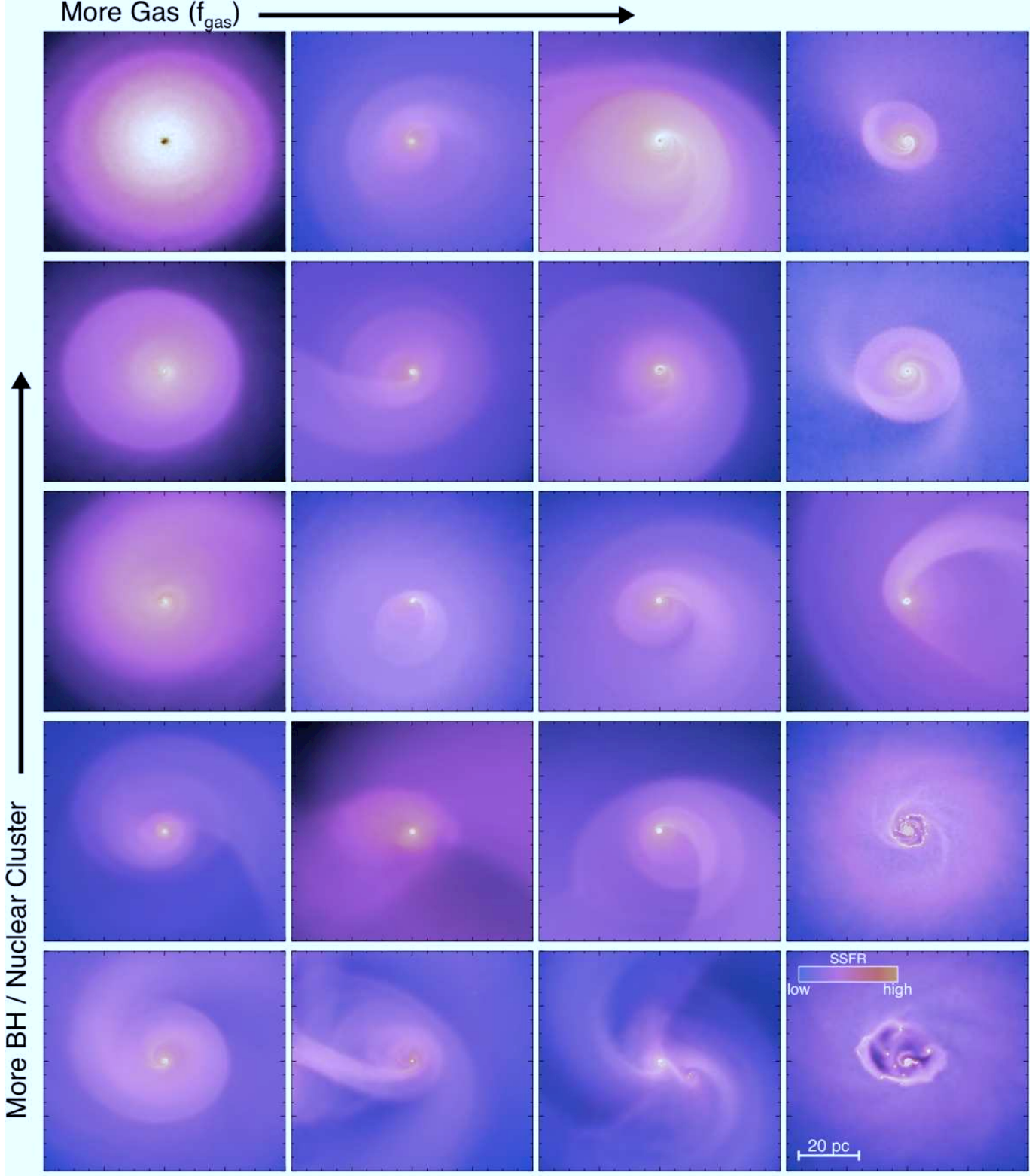}
    \caption{Images of the instabilities that develop in our small-scale ($\sim 10$ pc)
    nuclear re-simulations with $0.1\,$pc resolution.   
    Each panel shows projected gas density (intensity) and local 
    star formation rate (color, from blue through yellow) in the central kpc. 
    Each of these simulations can
    be thought of as a simulation of the 
    corresponding nuclear scales from Figure~\ref{fig:100pc.morphologies}. 
    The simulations extend into the  BH radius of influence. 
    The primary parameters of importance are the ratio of the 
    gas mass to the BH mass (or BH plus bulge/star cluster mass,
    when the latter is present) and the gas fraction in the 
    disky component: {\em top} to {\em bottom} is decreasing 
    BH/stellar mass while the disk gas fraction increases 
    from {\em left} to {\em right}.   A strong $m=1$ mode is 
    generic for reasonable BH/stellar mass and gas fractions -- 
    this corresponds to an eccentric,  globally precessing (non-winding) disk 
    (or single-armed spiral), a mode that is special to the quasi-Keplerian potential. 
    The resulting torques drive inflows of up to 
    $10\,\msun\,{\rm yr^{-1}}$ at $<0.1\,$pc scales 
    (Figure \ref{fig:time.dependence}), sufficient to fuel a luminous quasar.
\label{fig:10pc.morphologies}}
\end{figure*}

Figure~\ref{fig:100pc.morphologies} demonstrates that the strength of
global non-axisymmetric modes increases dramatically with decreasing
$B/T$. In fact, as soon as the bulge and disk are comparable (second
row from top), prominent modes appear.  In the disk-dominated cases
(bottom row), very large angular momentum transport occurs even in
cases without much total gas.  Of course, at each $B/T$, increasing
the gas content makes the system more vulnerable to local
instabilities as well -- usually making the overall inflow more clumpy and
time-variable.  In several cases, the large-scale mode (e.g.\ a spiral
arm) becomes self-gravitating and globally fragments (not necessarily
into smaller sub-units, but as a whole), leading to a major
coalescence -- what is almost (dynamically speaking) a scaled-down
merger in the central regions! 

Figure~\ref{fig:100pc.morphologies} also demonstrates an important
point seen in Figure~\ref{fig:zoom}. Although the instabilities seen
in our simulations qualitatively resemble the ``bars within bars''
idea of secondary instabilities once the gas density is sufficiently
high, the morphologies vary widely, and are {\em not} restricted to
traditional bars (although these certainly do appear).  At similar gas
fraction and $B/T$, we find that the strength of angular momentum
transfer is generally similar, but we also find that the visual
morphology (and the precise modes important for transport) can vary
widely, depending on time and on the details of the gas, stellar disk,
bulge, and halo profiles, and the precise equation of state.  Thus,
global quantities such as the mass profile and accretion rate are
comparatively robust, but the observational classification of these
systems would vary widely.\footnote{In detail, the angular momentum transport does depend on the structure of the unstable mode/perturbation. 
The precise dependence will be discussed in \papertwo.   At the dimensional level, the torques and inflow scale in the same manner independent of the detailed mode morphology, so long as the stellar torques are sufficiently strong to cause orbit crossing and shocks in the gas.   The details of the specific modes driving such shocks amounts to numerical factors of $\sim$a few in the torques.}
Indeed, Figure~\ref{fig:100pc.morphologies}
shows traditional bars and spirals, nuclear rings, crossed or barred
rings, single or three-armed systems, flocculent disks, and clumpy,
irregular morphologies. These all appear, with no obvious preference
for one or another as a whole, in our simulations.

This feature of our simulations may account for a number of
observational results in the literature.  For example, surveys of AGN
have often found that although nuclear {\em bars} on these scales only
appear in some fraction of sources (not necessarily much larger than
the fraction of non-AGN in which they appear), there are ubiquitous
asymmetric gas structures of {\em some} sort, similar to those modeled
here \citep[see e.g.][]{martini:seyfert.host.morph,
  garcia.burillo:torques.in.agn.nuclei.obs.maps.no.inflow,
  haan:nuga.gas.dynamics.maps,krips:nuclear.disk.torus.obs.seyferts,
  laine:nested.bars.in.seyferts,peletier:seyfert.morph.imaging,
  sakamoto:bar.driven.mol.gas.transport}.
We discuss this further in \S~\ref{sec:discussion}.

\vspace{-0.4cm}
\subsection{Towards the Accretion Disk: Eccentric Disks at $\sim$ Parsec}
\label{sec:results:nuclear:BH}

From $\sim 500$ to $\sim10\,$pc, our intermediate-scale simulations
successfully demonstrate efficient angular momentum transport via a
wide range of gravitationally unstable modes. Near the smallest radii
in these simulations, however, the systems encounter yet another
angular momentum barrier.  At that point, gas has reached the BH
radius of influence, i.e., the BH begins to contribute non-trivially
to the potential, which becomes quasi-Keplerian. This halts further
inflow because the disk is no longer strongly self-gravitating and so
is less susceptible to global modes.  In addition, the gas generally
encounters an inner Linblad resonance associated with the
intermediate-scale bar.

Indeed, it is widely appreciated that both ``bars within bars'' and
the direct or induced torques due to perturbations from mergers, close
passages, and large-scale bars do not produce efficient angular
momentum transfer at radii $\sim 0.1-10 \,$pc \citep[see e.g.\ the
discussion in][and references
therein]{athanassoula:bar.orbits,athanassoula:bar.vs.cmc,
  shlosman:bars.within.bars,heller:secondary.bar.instability,
  begelman:direct.bh.collapse.w.turbulence}.
As a consequence this is often considered the most
difficult-to-explain regime of gas inflow and angular momentum
transport.

\begin{figure*}
    \centering
    \scaleup
    \plotside{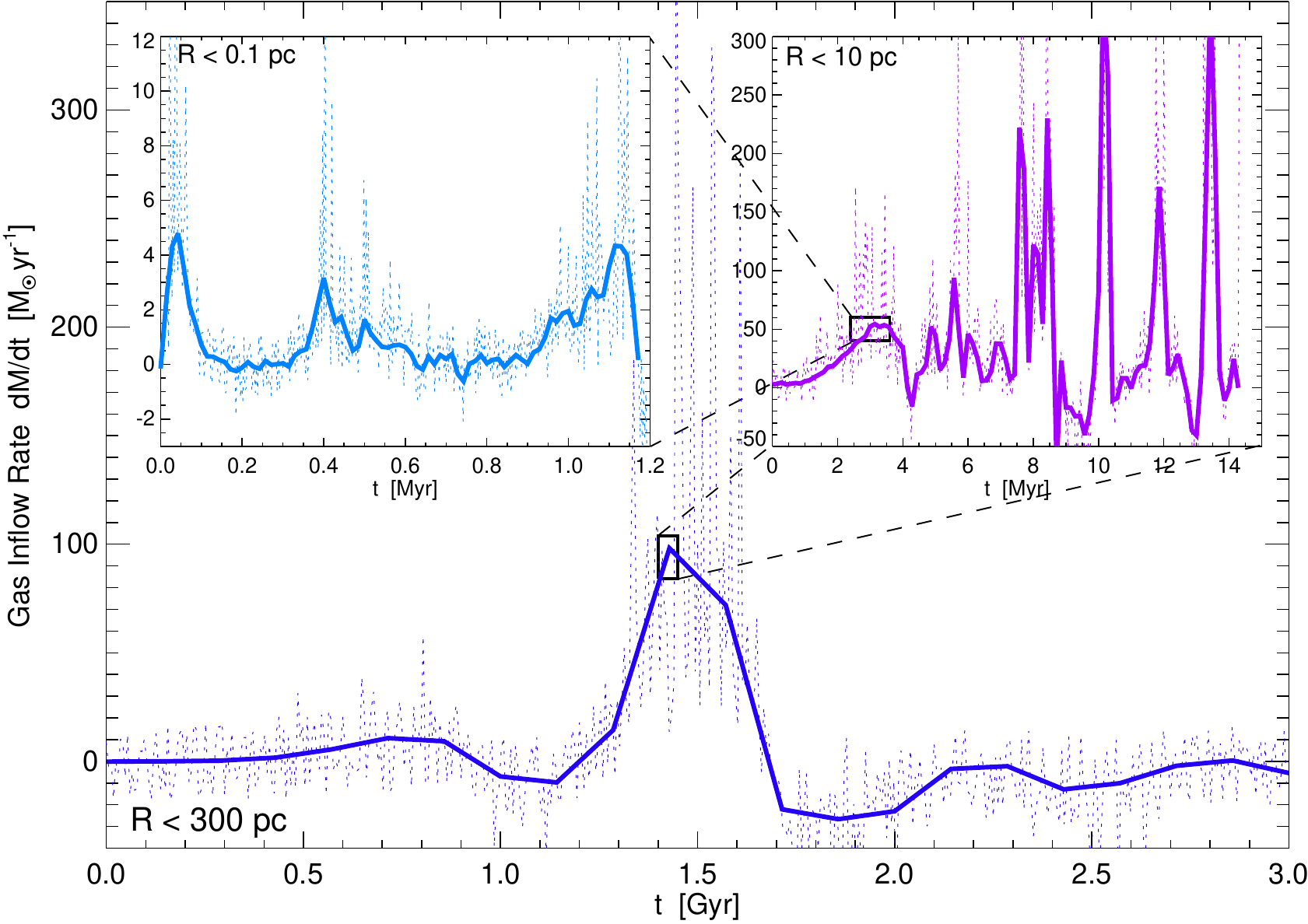}
    \caption{Time-dependent inflow rate through various radii, 
    for the example simulation shown in Figure~\ref{fig:zoom}. 
    {\em Main:} 
    Gas inflow rate into the central $300\,$pc in the 
    large-scale galaxy merger simulation, as a function of time since the beginning of the merger. 
    The large spike follows the coalescence of the two nuclei. Dotted line shows the instantaneous 
    inflow rates; solid line shows the 
    inflow rate averaged over five local dynamical times.
    {\em Top Right:} 
    Inflow rates into the central $\sim 10$ pc in the intermediate 
    scale re-simulation for a small time interval just after coalescence.
    {\em Top Left:} Inflow rates into $0.1$ pc in our smallest-scale
    resimulation, initialized with conditions from the previous
    intermediate-scale
    simulation after it has evolved for several local dynamical times.
    The accretion rate reaches $1-10 M_\odot \, {\rm yr^{-1}}$,
      sufficient to fuel a luminous quasar.
      Note the episodic, highly variable nature of accretion/inflow at
      each scale. 
      \label{fig:time.dependence}}
\end{figure*}

We find, however, that efficient angular momentum transfer continues
in our smaller-scale re-simulations for sufficiently gas-rich, disk-dominated
systems.  Figure \ref{fig:10pc.morphologies} shows this with images
spanning a range of $f_g$ and $B/T$; we will discuss this physics in
more detail in \S \ref{sec:results:duty} \& \ref{sec:stability}.
These results show that, so long as the gas density is sufficiently
large (relative to the BH mass), the system develops a precessing
eccentric disk (an $m = 1$ mode) that drives gas down to sub-pc scales
$\sim 0.1$ pc.  As before, stars rapidly form out of the disk, leading
to a similar mode in both the stars and gas; these modes precess about
the BH relative to one another with slightly different pattern speeds,
leading to crossing orbits, dissipation of energy and angular momentum
in the gas, and thus net inflow.  This is, once again, an instability
that depends primarily on the presence of sufficient gas at small
radii in the first place.  We find that this condition is met at some
point(s) in time in all of our simulations with significant gas mass
and instability at $\sim100\,$pc, i.e., in those simulations that meet
the ``bars within bars'' criteria in \S \ref{sec:results:nuclear} and
Figure \ref{fig:100pc.morphologies}.

\vspace{-0.5cm}
\breaker
\section{Inflow Rates and Gas Properties}
\label{sec:results:duty}

\begin{figure*}
    \centering
    \scaleup
    \plotside{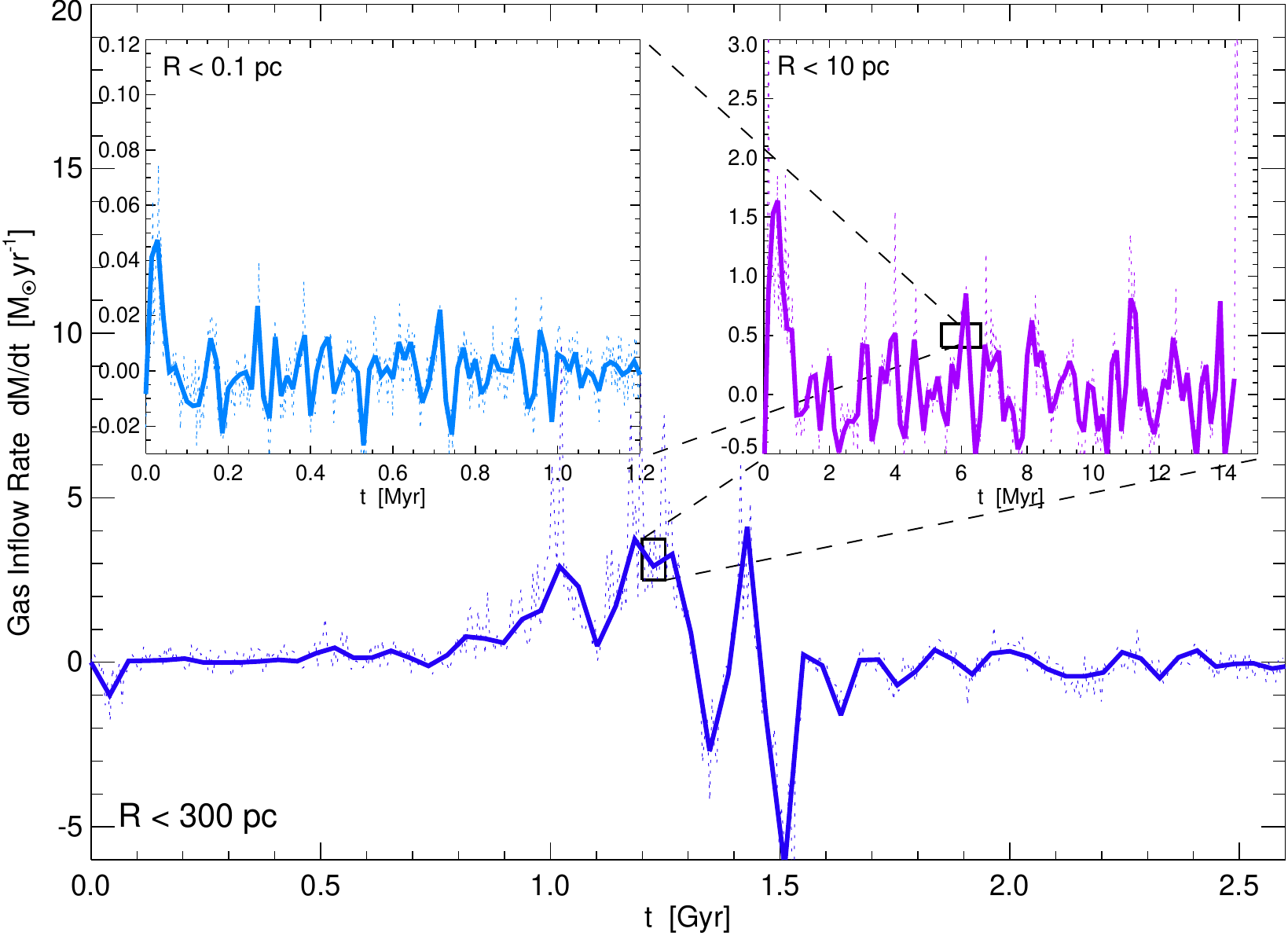}
    \caption{Time-dependent inflow rate through various radii, as in
      Figure~\ref{fig:time.dependence}, but for a simulation in which
      the large-scale inflow into the central kpc is weak (simulation
      {barex} in Table~\ref{tbl:galaxy}).  Specifically, this is a
      gas-rich disk with $B/T=0.3$, which is moderately
      bar-unstable. Compared to a major merger, the inflow to $< 300$
      pc is much weaker, and the presence of an inner Linblad
      resonance at small radii leads to both outflow and inflow from
      the bar. Because of the much lower gas supply to the inner part
      of the galaxy, secondary instabilities do not efficiently develop that can  transport gas further in.
      \label{fig:time.dependence.stable}}
\end{figure*}

Figure~\ref{fig:time.dependence} shows the time-dependent inflow rates
through several annuli, for the same set of multi-scale re-simulations
shown in Figure~\ref{fig:zoom}. These demonstrate and quantify the
general scenario summarized above: on large scales, coalescence and
the final stages of the merger drive a large quantity of gas into the
central few hundred pc, with inflow rates $\sim100-300\,\msun\,{\rm
  yr^{-1}}$. The total duration of this phase is a few times
$10^{8}$\,yr.  During this time, the gas accumulates at small radii
$\sim 100$ pc; simulating the dynamics on this scale for
$\sim10^{7}\,$yr, we find that secondary gravitational instabilities
develop that drive further inflows into the central $\sim10\,$pc, with
inflow rates $\sim10-200\,\msun\,{\rm yr^{-1}}$.  Zooming in yet again
during one epoch of significant inflow to $\lesssim 10$ pc, our
smallest-scale simulations resolve the rapid formation of an eccentric
nuclear disk around the central BH; the accretion rate into the
central $0.1$ pc, which is likely a reasonable proxy for the accretion
rate onto the BH, reaches $\sim1-10 \,\msun\,{\rm yr^{-1}}$.  This is
sufficient to fuel a luminous AGN at the Eddington rate.

Figure~\ref{fig:time.dependence} also demonstrates that the
small-scale accretion rate can be highly time-variable. This is in
part a consequence of the accretion of individual clumps/clouds (e.g.,
Fig. \ref{fig:10pc.morphologies}), but is also a consequence of the
fact that gravitationally unstable perturbations rapidly grow,
dissipate, and generate other structures; depending on, e.g., the
state of precession of the stellar versus gaseous disk, the system can
transition between inflow and outflow at a given radius.  Even on the
largest scales, the inflow is still highly variable, although is
coherent over a time much longer than the dynamical time because it is
driven by the global torques involved in the merger.
Because of the variability in $\dot M$ on different scales, we do
{not} expect every merger (or isolated galaxy with a large-scale bar),
at every time, to exhibit significant inflow from large scales all the
way down to the BH.  This is important -- after all, a large fraction
of observed mergers are not bright quasars.  In addition, the large
variation in the physical conditions at small radii for a relatively
fixed set of conditions at larger radii demonstrates that great care
must be taken when trying to correlate the galactic structure at $\sim
0.1-1$ kpc with the BH accretion rate in order to constrain the
physics of AGN fueling.

Figure~\ref{fig:time.dependence.stable} shows the same calculation of
the inflow rates, but for re-simulations of an isolated disk galaxy
with a moderate bar instability (the case in
Table~\ref{tbl:galaxy}). The system develops a fairly strong bar at
about a kpc (representing a perturbation to the potential of $\delta
\Phi/\Phi \sim 0.15$), but this is still much weaker than the major
merger case shown in Fig \ref{fig:time.dependence}; moreover, the
pre-existing bulge with $B/T=0.3$ means that there is an inner Linblad
resonance at a few hundred pc. The net result is that the overall
inflow is much weaker, and there can be outflow as well as inflow on
small scales (the system overcomes the Lindblad resonance for short
periods of time, leading to inflow followed by outflow as it
re-equilibrates). Re-simulating smaller scales, the weak inflows
produced by the bar lead to low gas mass ($\sim1-10\%$) relative to
the bulge mass on small scales, and global secondary instabilities do
not develop.  Inflow/outflow rates from the bar at $0.3-1\,$kpc are
characteristically $\sim0.5-10\,\msun\,{\rm yr^{-1}}$, similar to
observed local barred galaxies
\citep{quillen:inflow.along.bar.ngc7479,jogee:H2.masses,haan:nuga.gas.dynamics.maps};
at $\sim10\,$pc they do not exceed $1\,\msun\,{\rm yr^{-1}}$, and at
$<0.1\,$pc they are characteristically $\ll 0.1\,\msun\,{\rm
  yr^{-1}}$.  At these low accretion rates our calculations are not
reliable, but the key point is that in the low gas fraction limit,
gravitational torques drive very little accretion from $\sim$ kpc to
$\lesssim 0.1$ pc, even in the presence of a modest bar at $\sim$ kpc.

\begin{figure}
    \centering
    \scaleup
    \plotone{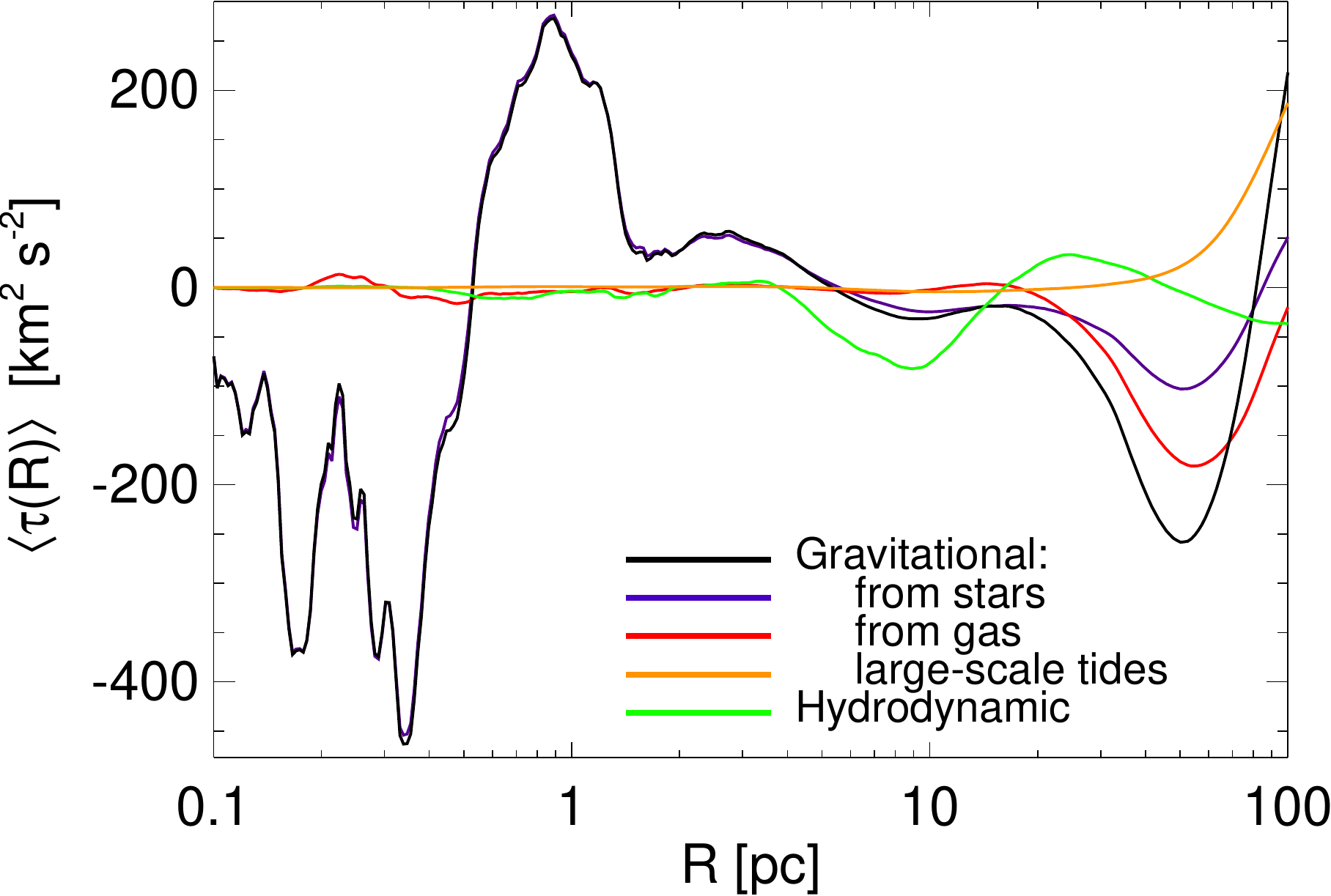}
    \caption{Radial profile of the specific torque in a representative
      nuclear scale simulation ({Nf8h1c1qs} in Table
      \ref{tbl:nuclear}) during a stage of rapid BH accretion.  The
      torque per unit mass acting on the gas is averaged over narrow
      radial annuli at one time; negative torques decrease the gas
      angular momentum.  The total gravitational torque (black line)
      is divided into the torque from stars in the same disk as the
      gas (purple), gas (red), and the contribution from the
      large-scale tidal field (orange).  The hydrodynamic
      torques from pressure forces and artificial viscosity (green) are comparatively 
      small. Inflow at all radii is dominated by gravitational
      torques, largely from nearby stars in the same $m=1$ pattern as
      the gas.
      \label{fig:torque.map}}
\end{figure}

Figure~\ref{fig:torque.map} shows an example of the torques driving
the inflows in our high accretion rate simulations. We show the radial
torque profile at one time in one of our nuclear-scale simulations,
broken down into gravitational torques from the stars and gas in the
nuclear disk (versus the large-scale tidal field) and hydrodynamic
torques (largely from pressure forces); note that negative torques
drive inflow.  The average gravitational torque in an annulus between
$R_{0}$ and $R_{0}+\Delta R$, from sources $j$, is defined by
\begin{align}
\nonumber {\bf \tau}_{\rm grav,\,j} & \equiv \frac{1}{M_{\rm gas}[\Delta R]}
\int_{R_{0}}^{R_{0}+\Delta R} \,
\rho_{\rm gas}({\bf r})\,{[}{\bf r}\times {\bf a_{\rm grav,\,j}}{]}_{z}\,{\rm d}r\,{\rm d}\cos{\theta}\,{\rm d}\phi \\ 
& = \frac{1}{M_{\rm gas}[\Delta R]}\,\sum_{i\ {\rm in}\ \Delta R} m_{i}\, 
{[}{\bf{r}_{i}} \times \nabla\Phi_{j}({\bf r_{i}}){]_{z}} 
\end{align}
where $M_{\rm gas}[\Delta R] = \sum m_{i}$ is the gas mass in the
annulus; the sum over $i$ in the second expression refers to the gas
particles inside the annulus, while $\Phi_{j}$ is the gravitational
potential generated by the torquing particles of interest (gas, stars,
etc.).\footnote{ For small separations between two particles i and j,
  $|{\bf{r}_{i}-{r}_{j}}|<\epsilon$, we include the GADGET force
  softening to be self-consistent (equivalent, inside a softening
  length, to the potential of a Plummer sphere), but this has little
  effect on our results.}  The radius ${\bf r}$ is defined with
respect to the BH, and we focus on the $z$ component of ${\bf \tau}$,
where the $z$ axis is defined to be the angular momentum vector of the
disk (torques in the other directions are negligible, so this is
almost identical to plotting $|{\bf \tau}|$).  The torques from
pressure forces are similarly defined by
\begin{align}
\nonumber {\bf \tau}_{\rm P,\,j} & \equiv \frac{1}{M_{\rm gas}[\Delta R]}
\int_{R_{0}}^{R_{0}+\Delta R} \,
\rho_{\rm gas}({\bf r})\,{[}{\bf r}\times {\bf a_{\rm P}}{]}_{z}\,{\rm d}r\,{\rm d}\cos{\theta}\,{\rm d}\phi \\ 
& = \frac{1}{M_{\rm gas}[\Delta R]}\,\sum_{i\ {\rm in}\ \Delta R} m_{i}\, 
{[}{\bf{r}_{i}} \times \frac{1}{\rho({\bf r_{i}})}\nabla P({\bf r_{i}}){]_{z}} 
\end{align}
where the pressures and densities are determined in standard fashion 
from the SPH quantities. Note that the BH can and does move 
(with typical amplitude $R\sim 0.1\,R[M_{d}(<R)\sim 0.1\,M_{\rm BH}]$) in response 
to the $m=1$ modes on nuclear scales; but since we are interested in the inflow 
{onto the BH itself}, we evaluate these torques about its instantaneous position 
(rather than, say, the center-of-mass of the system). The results are qualitatively 
similar in Figure~\ref{fig:torque.map} if we fix the central position 
(albeit quantitatively altered within $\ll 10\,$pc at a factor $\sim2$ level), but are less directly relevant to interpreting inflow onto the BH. 

The qualitative behavior shown in Figure \ref{fig:torque.map} is
representative of all of our simulations.  At essentially all radii,
gravitational torques dominate hydrodynamic torques.  Moreover, the
gravitational torques themselves are dominated by torques from stars,
not the torques of the gas on itself; specifically, the stars that are
important are in the same asymmetric perturbation as the gas.  This is
also the case on larger scales ($\gtrsim100\,$pc), for both mergers
and barred systems
\citep[see][]{barneshernquist96,hopkins:disk.survival}.  Torques from
the spherical component of the system (e.g.\ the halo and/or bulge
stars) are negligible at all radii.  Unsurprisingly, the torques from
the large-scale tidal field, defined here as the torques from the $\gg
100\,$pc scales of the parent simulation from which the initial
conditions of this re-simulation were drawn, become significant only
at the outer boundary of our re-simulation (see Appendix
~\ref{sec:numerical:outer}).  Figure \ref{fig:torque.map} shows that
there are several sign changes in the torque profile, reflecting the
specific state of the system at this time; the torque is time
variable, but we find there are sign changes as well in a
time-averaged sense.  Overall, the net rate of change of the angular
momentum of the gas very closely tracks the time-averaged
gravitational torque from the stars.  Hydrodynamic torques never
induce very strong torques (greater than those shown), whereas the
stellar gravitational torques can be a factor of $\sim
10-100$ larger than in Figure \ref{fig:torque.map} at some times.

\begin{figure}
    \centering
    \scaleup
    \plotone{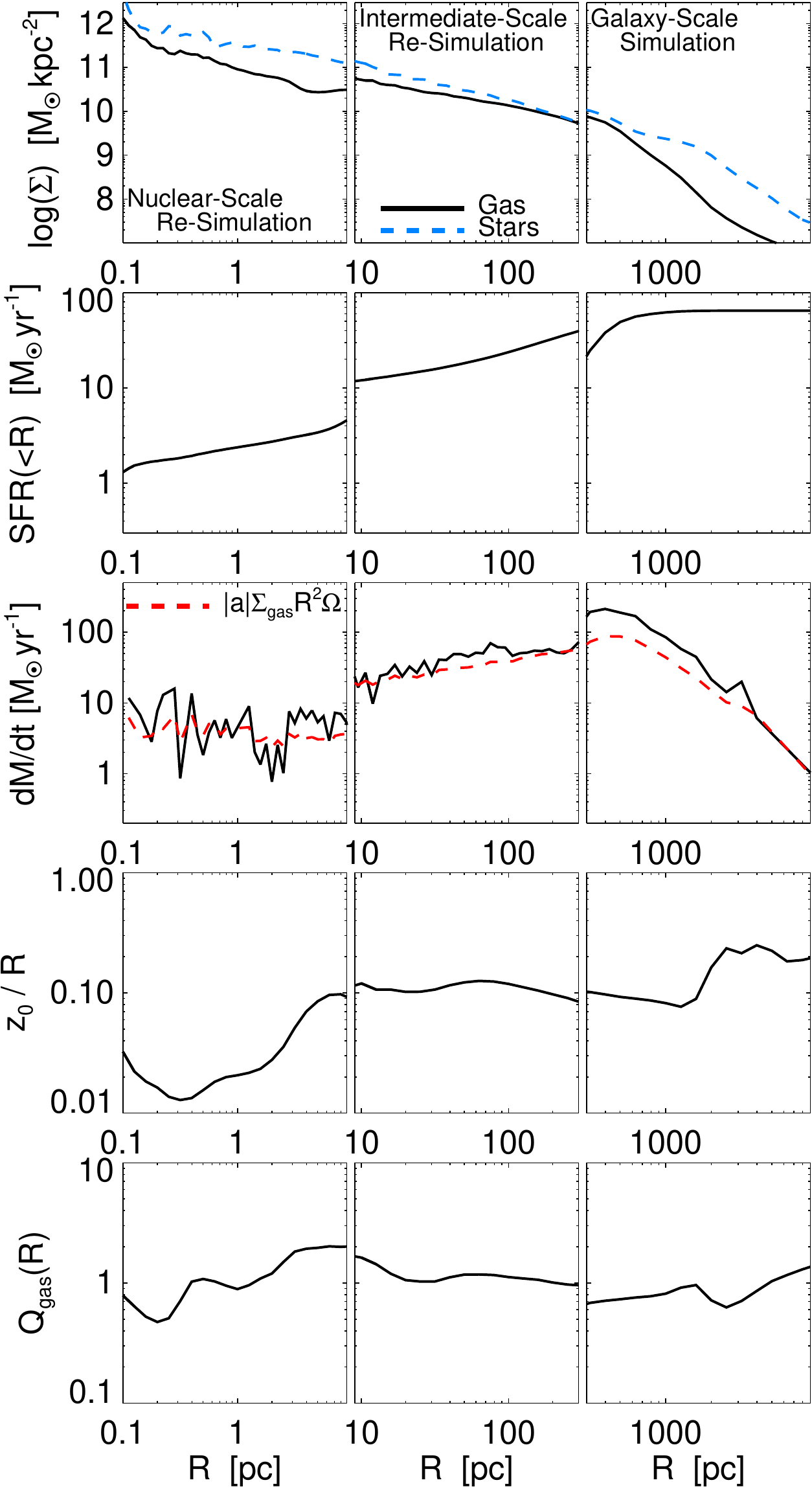}
    \caption{
   Properties of representative galactic (right panel), intermediate (middle), and nuclear-scale (left) simulations as a function of radius at a single time.  The simulations are the same as in Figures~\ref{fig:zoom} \& \ref{fig:time.dependence} and the time chosen corresponds to near the peak of activity, with significant gas inflow to small radii.  Note that the initial/boundary conditions for the smaller-scale simulations were taken from the larger-scale simulations shown, so the results are reasonably continuous  at the radial boundaries (despite the fact that these do not represent a self-consistent single simulation).    
    {\em Top:} Gas and stellar surface density. 
    {\em Second-from-Top:} Enclosed star formation rate in a given radius.
    {\em Middle:} Instantaneous inflow/outflow rate through the 
    annulus (black line); note that there are several sign changes as a function of radius. Also shown is
    the order of magnitude estimate of the accretion rate produced by gravitational torques from a 
    non-axisymmetric mode of amplitude $|a|$, $\dot M \sim |a|\,\Sigma_{\rm gas}\,R^{2}\,\Omega$ 
    (dashed red line;  \S~\ref{sec:results:duty}).   
    {\em Second-from-Bottom:} Scale height $z_{0}/R$ of the gas.
    {\em Bottom:} Toomre $Q$ parameter of the gas. 
 \label{fig:representative}}
\end{figure}

The fact that the torques on the gas are dominated by stars is robust, and occurs for two reasons. First, the stars contribute significantly to the mass on the scales $\gtrsim 0.1$ pc that we focus on here (where star formation can occur).   For typical star formation efficiencies, it is difficult to have the gas mass $\gg 50\%$ of the total mass for a reasonable fraction of the lifetime of the system. 
On large scales, galaxies are known to not be so gas rich (although they certainly can reach $\sim 50 \%$ gas fraction). On small scales, 
if the star formation efficiency is a few percent per dynamical time, then even a pure gas inflow 
from larger scales  is likely to only remain gas-dominated 
for $\sim10$ local dynamical times (on the smallest scales, $\sim10^{6}\,$yr)
-- thus for the majority of the time during which inflow is continued, the system will contain a significant stellar mass.   This is consistent with direct observations of sub-kpc regions of starburst 
galaxies \citep{downes.solomon:ulirgs,bryant.scoville:ulirgs.co} 
and the $\sim 1-10$ pc nuclear scales around AGN \citep{hicks:obs.torus.properties}.    
On sufficiently small scales, $\lesssim 0.01-0.1$ pc, star formation will become inefficient, 
but at precisely those radii, we have 
(by definition) essentially reached the $\alpha$-disk.

\begin{figure}
    \centering
    \scaleup
    \plotone{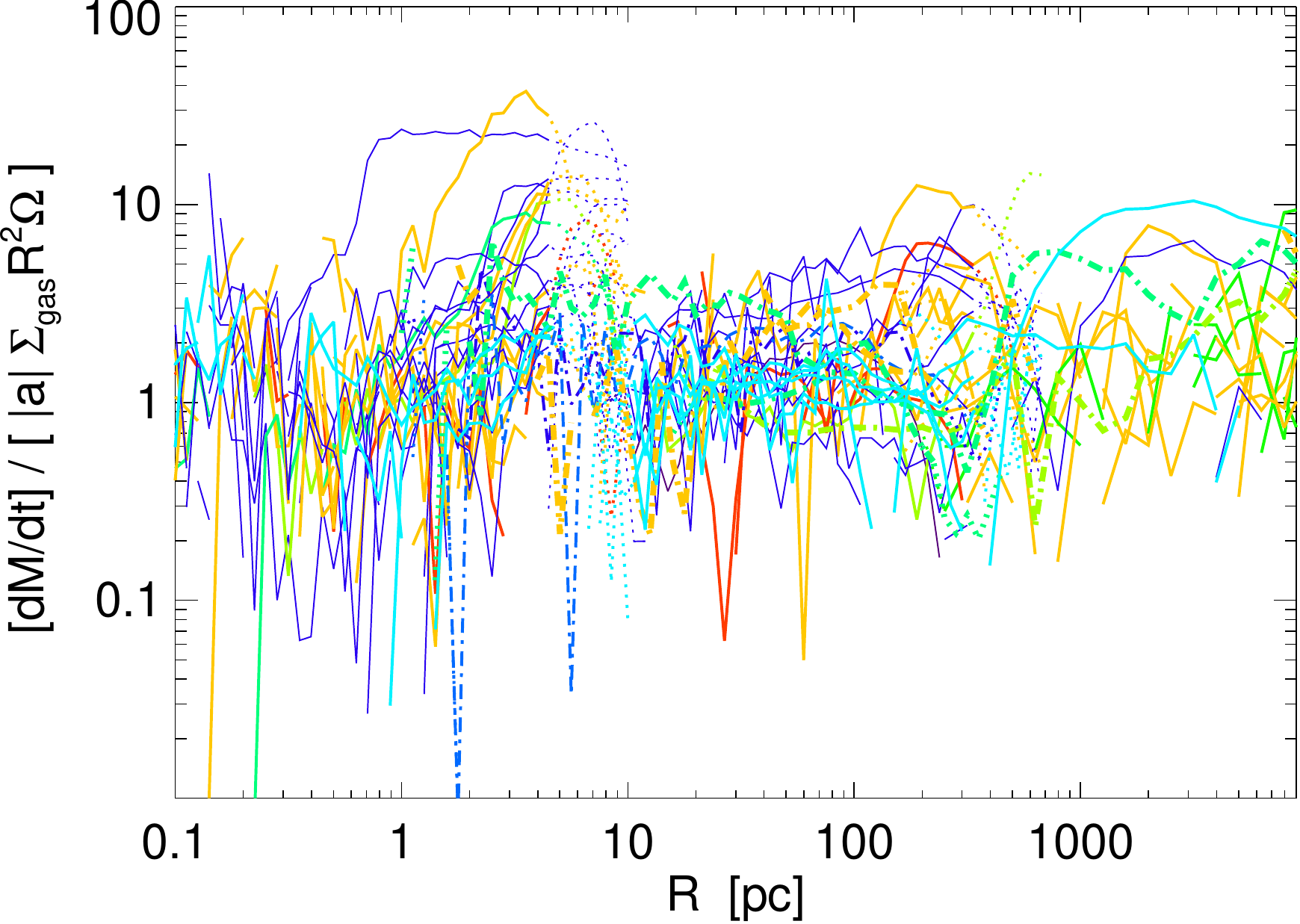}
    \caption{Comparison of the true instantaneous inflow rate in the
      simulations (${\rm d}M/{\rm d}t$) to the simple dimensional
      scaling $|a|\,\Sigma_{\rm gas}\,R^{2}\,\Omega(R)$ expected for
      pure gravitational torques, where $|a|$ is the magnitude of the
      strongest non-axisymmetric mode at a given radius $R$ (directly
      measured in the simulations; 
      with typical values $\sim 0.01-0.3$).  Each solid line denotes a
      different simulation, with dotted lines showing the radii near the
      boundaries of our re-simulations, where the exact profile shape
      is suspect; colors correspond to initial gas fractions as in
      Figure~\ref{fig:stability.criteria}.  
      Thick dot-dashed lines correspond to ultra-high resolution simulations. 
      The simulations are shown
      during the active/strong inflow phases, as in
      Figure~\ref{fig:time.dependence}.  The inflow rate in the
      simulations is consistent with being produced by gravitational
      torques over $5$ orders of magnitude in radius, with relatively
      little scatter ($0.3-0.5\,$dex at all radii).      \label{fig:inflow.vs.viscous}}
\end{figure}


Even if the gas mass is comparable to or greater than the stellar mass, 
the self-torque of the gas on itself is much weaker than the torques from the 
stars on the gas. This has been demonstrated in detail
in the case of large-scale perturbations from 
galaxy mergers, bars, and spiral waves \citep{noguchi:merger.induced.bars.gas.forcing,
barneshernquist96,barnes:review,berentzen:gas.bar.interaction,hopkins:disk.survival}. 
We show that it is also true for smaller radii in Figure  \ref{fig:torque.map}.
We discuss this in detail in a subsequent paper (\papertwo; in preparation), but briefly outline the key physics here.  

It is well-known that the magnitude of the self-torque in a pure gaseous or pure stellar disk 
is second-order in the mode amplitude \citep[$\propto |a|^{2}$, where 
$|a|$ is the fractional magnitude of the asymmetry; see e.g.][]{lynden-bell:1972.spiral.amplification}. 
Moreover, \citet{kalnajs:1971} showed (with the exact solution for a logarithmic spiral) that the torques are further suppressed by powers of $\sim kR$ and $m$, where $k$ ($m$) is the radial (azimuthal) wavenumber of the mode. 
This is despite the fact that the {instantaneous} torque on gas/stars moving through 
a perturbation has no such suppression and is linear in $|a|$. For small mode amplitudes, the gas orbits periodically (in the test-particle limit) in response to the perturbation and the positive and negative linear terms exactly cancel because of the symmetry on either ``side'' of the mode.   In a mixed stellar+gas system, however, the two perturbations are not exactly in phase because of dissipation in the gas.   Perhaps most importantly, the gas can be driven into shocks by a sufficiently strong perturbation in the stars, leading to strong dissipation.   As a result, the symmetric response of the gas is broken, and the net torque from stars on the gas is linear in $|a|$ and largest in the global limit.  At the order of magnitude level, the inflow rate is thus given by \be \dot{M}_{\rm grav} \sim |a|\,\Sigma_{\rm
  gas}\,R^{2}\,\Omega.
\label{eq:mdotgrav}
\ee  Note that expressed in terms of a viscosity $\nu$ such
that $\dot M \simeq 3 \pi \nu \Sigma_{\rm gas}$, equation
(\ref{eq:mdotgrav}) corresponds to $\nu \simeq (|a|/3 \pi) V_c R$,
where $V_c = R \Omega$ is the local circular velocity.  Equivalently,
the inflow time of the gas is $\sim |a|^{-1} \Omega^{-1}$. 
Typical values of $|a|$ are $\sim0.01-0.3$ in our simulations 
when significant inflow is present (see \S \ref{sec:stability}).
We stress that, as described above, equation~\ref{eq:mdotgrav} differs fundamentally from the dimensional expectation for self-torques, which is second-order in $|a|$ and 
suppressed by terms in $|kR|$ and $m$.   

Figure~\ref{fig:representative} shows a number of properties of our fiducial simulations from Figures~\ref{fig:zoom} \& \ref{fig:time.dependence}, for each of the three spatial scales we consider: galactic ($\sim$ kpc), intermediate ($\sim 10-100$ pc), and nuclear ($\sim 0.1-10$ pc).   We show the stellar and gaseous surface densities, the local star formation rate, the gas inflow rate in comparison to equation~\ref{eq:mdotgrav}, and the vertical scale height and Toomre Q of the gas disk.   We stress that Figure~\ref{fig:representative} shows three {\it independent} simulations, but that the smaller-scale simulations were initialized with properties drawn from the larger-scale simulations; this explains the approximate, but not exact, continuity between the different scales.  Although Figure~\ref{fig:representative} is quite instructive, it can also be misleading to focus on the results of an individual simulation.  The reason is that there is large variation in time and potentially large scatter introduced by modest differences in galaxy properties (due to, e.g., large-scale fragmentation of a spiral arm biasing the results of a given simulation).   For this reason we believe that it is extremely important to consider the statistical properties of a large number of simulations that vary the precise initial conditions, gas fraction, bulge to disk ratio, etc.   In what follows we now discuss the statistical properties of the simulations summarized in Tables~\ref{tbl:galaxy}-\ref{tbl:nuclear}.  In doing so, we will show a number of Figures that contain the results of many of our simulations.  In these Figures, the critical point to focus on is less the results of any given simulation (which can be hard to pick out), but rather the ensemble properties of all of the simulations (e.g., median, scatter, ...).  Recall that the parameter space spanned by our simulations includes systematic surveys in the gas fraction, bulge (or BH)-to-disk ratio, and sub-grid equation of state of the gas. 

Figure \ref{fig:inflow.vs.viscous} compares the order of magnitude estimate in equation~\ref{eq:mdotgrav} with the true inflow rate in the simulations, as a function of radius; as in Figure~\ref{fig:representative}, we show the results of simulations on all three of the spatial scales we have simulated, with dotted lines showing results near the radial boundary of a given calculation.  In calculating $\dot M_{\rm grav}$ in equation~\ref{eq:mdotgrav}, we approximate $|a|$ using the magnitude of the larger of the $m=1$ and $m=2$ Fourier component of the surface density distribution within $R$ (see
\S~\ref{sec:stability}).  The results in Figure
\ref{fig:inflow.vs.viscous} are at the peak of activity for each
simulation with significant inflow, i.e., when $\dot{M}$ is
near-maximum.  Modulo a constant numerical prefactor, the simple
scaling in equation (\ref{eq:mdotgrav}) works well over the entire
dynamic range of our simulations, with a scatter of $\sim0.3-0.5\,$dex
at any radius. The same equation also works for our ultra-high resolution simulations, 
which provide a smooth interpolation over the boundaries in our typical 
``re-simulation'' calculations.  The agreement in Figure \ref{fig:inflow.vs.viscous}  demonstrates clearly that the inflow rates do not follow the second-order expectations from self-torques in the absence of dissipation. 
In \papertwo, we show that more detailed modeling of
the gas response (accounting for e.g.\ the details of the perturbation structure) 
can both predict the overall amplitude of $\dot M$ in Figure \ref{fig:inflow.vs.viscous}  
and reduce the scatter to $<0.3\,$dex.

It is also important to note that the inflow rate in equation~\ref{eq:mdotgrav} is {\em independent of the sound speed $c_s$,  i.e., of the thermodynamics of the gas.}  Rather, the inflow rate is
determined by the magnitude of the non-axisymmetric gravitational
potential.  This is one of the reasons why we believe that our results
are likely to be relatively robust, even given the significant
uncertainties in the ISM physics.  By contrast, if the transport were
dominated by local viscous stresses, the accretion rate would be given
by $\dot M_{visc} \sim 3 \pi \nu \Sigma_{\rm gas}$, where $\nu \simeq
\alpha c_s^2/\Omega$.  For large-scale spiral waves, the angular
momentum flux associated with the wave corresponds to an effective
viscosity that is at most $\nu \sim c_s r$
\citep{goodman:qso.disk.selfgrav}.  Comparing this with the inflow
rate induced by orbit crossing (eq. [\ref{eq:mdotgrav}]) highlights
that the dominant effect of the non-axisymmetric potential is that the
stellar potential induces crossing orbits in the gas; the direct
transport by spiral waves in the gas is smaller by at least a factor
of $\sim c_s/V_c$.  We find that in simulations that are initially $100\%$ gas (even those initialized with already-strong mode amplitudes), the torques tend to remain fairly weak 
for a few dynamical times until the stellar component is non-negligible, at which point strong shocks appear, the stellar and gaseous modes develop a phase-shift, and the inflow rates rise sharply.   The pure gas limit is, however, more sensitive to the thermodynamics of the gas and should be studied in more detail in future work.  In our calculations, we find that it is critical to include stars,
gas, and star formation to properly capture inflow in galactic nuclei: it is the interplay between the stars and gas, and the competition between star formation and inflow, that sets the net accretion rate at small radii onto the BH.

\begin{figure}
    \centering
    \scaleup
    \plotone{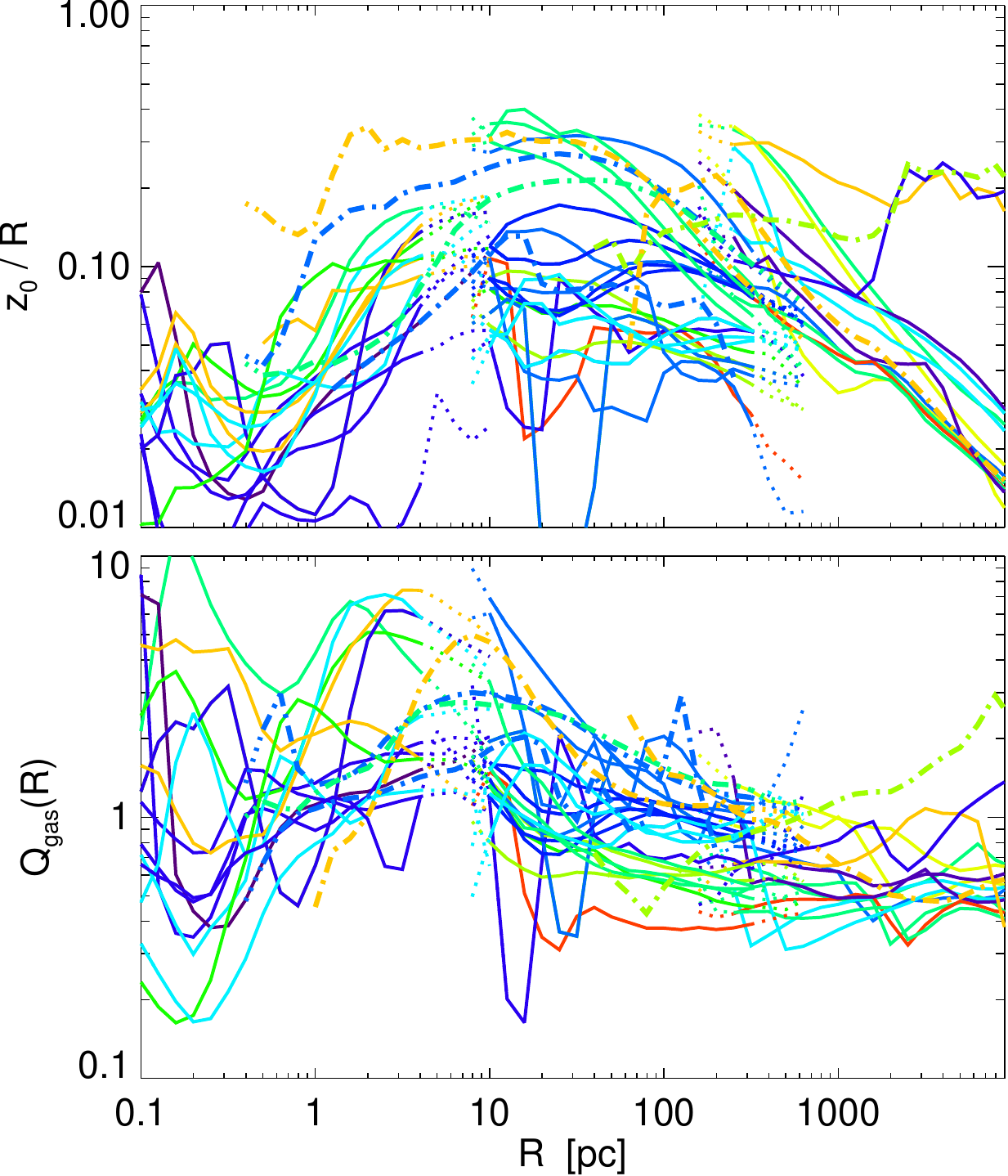}
    \caption{Properties of the gas disks in our simulations.     
    {\em Top:} Scale height $z_{0}/R$ of the gas as a function of  radius, for simulations with significant inflows.
    {\em Bottom:} Toomre $Q$ parameter of the gas. 
    The inflows are robustly self-gravitating ($Q \sim 1$) and disky, 
    with $z_0/R \lesssim 0.2$.    Each solid line denotes a
      different simulation, with dotted lines showing the radii near the
      boundaries of our re-simulations, where the exact profile shape
      is suspect; colors correspond to initial gas fractions as in
      Figure~\ref{fig:stability.criteria}.  
      Thick lines correspond to ultra-high resolution simulations. 
      The range in behavior emphasizes the importance of a large, 
      statistically representative suite of simulations. 
      \label{fig:gas.Q}}
\end{figure}

Our interpretation and analysis of the inflow rates assumes that the gas 
resides in a self-gravitating, at least modestly thin, disk.   
To demonstrate this explicitly, Figure \ref{fig:gas.Q} shows the 
vertical scale height and Toomre Q parameter of the gas as a 
function of radius.  For each simulation, at a given radius $R$  we determine the 
scale height $z_{0}$ of the gas by fitting the gas density distribution 
to a Gaussian with dispersion $z_{0}$ (after projecting to the 
plane perpendicular to the net angular  momentum vector). 
At large radii, there is scatter owing to the difference 
between mergers (with sizeable $z_{0}/R$) and unstable disks. 
From  $\sim1-100\,$pc, $z_{0}/R$ ranges from 
$\sim0.05-0.25$; this is determined by our assumed
ISM equation of state, as well as kinematic features such as twists and misalignments. 
At the smallest radii $z_{0}/R$ begins to decrease rapidly because the 
potential is increasingly dominated by the BH (the circular 
velocity increases but the maximum sound speeds are finite). 
Note that, at all radii, the minimum SPH smoothing length 
$h_{\rm sml}\lesssim z_{0}$, so that the disk thickness is at least 
modestly resolved.   Our simulated disks are always reasonably thin because the cooling time is always short compared to the dynamical time, so that the sound speed in the simulation reliably tracks that of our subgrid model.  

Figure \ref{fig:gas.Q} also shows the Toomre $Q$ parameter of the gas, defined as 
$Q\equiv c_{s}\,\kappa/\pi\,G\,\Sigma_{d}$, where $\kappa$ is the local epicyclic frequency 
and $\Sigma_{d}$ is the total surface density of the disk (gas plus stars). 
With some scatter, systems largely lie near $Q\sim1$. 
This is broadly expected for self-regulating disks;
in the simulations, $Q \sim 1$ is a consequence of star formation, 
the assumed stellar feedback, and gravitational dynamics. 

\vspace{-0.5cm}
\breaker
\subsection{Gas Density Profiles}
\label{sec:results:profiles}

\begin{figure}
    \centering
    \scaleup
    \plotone{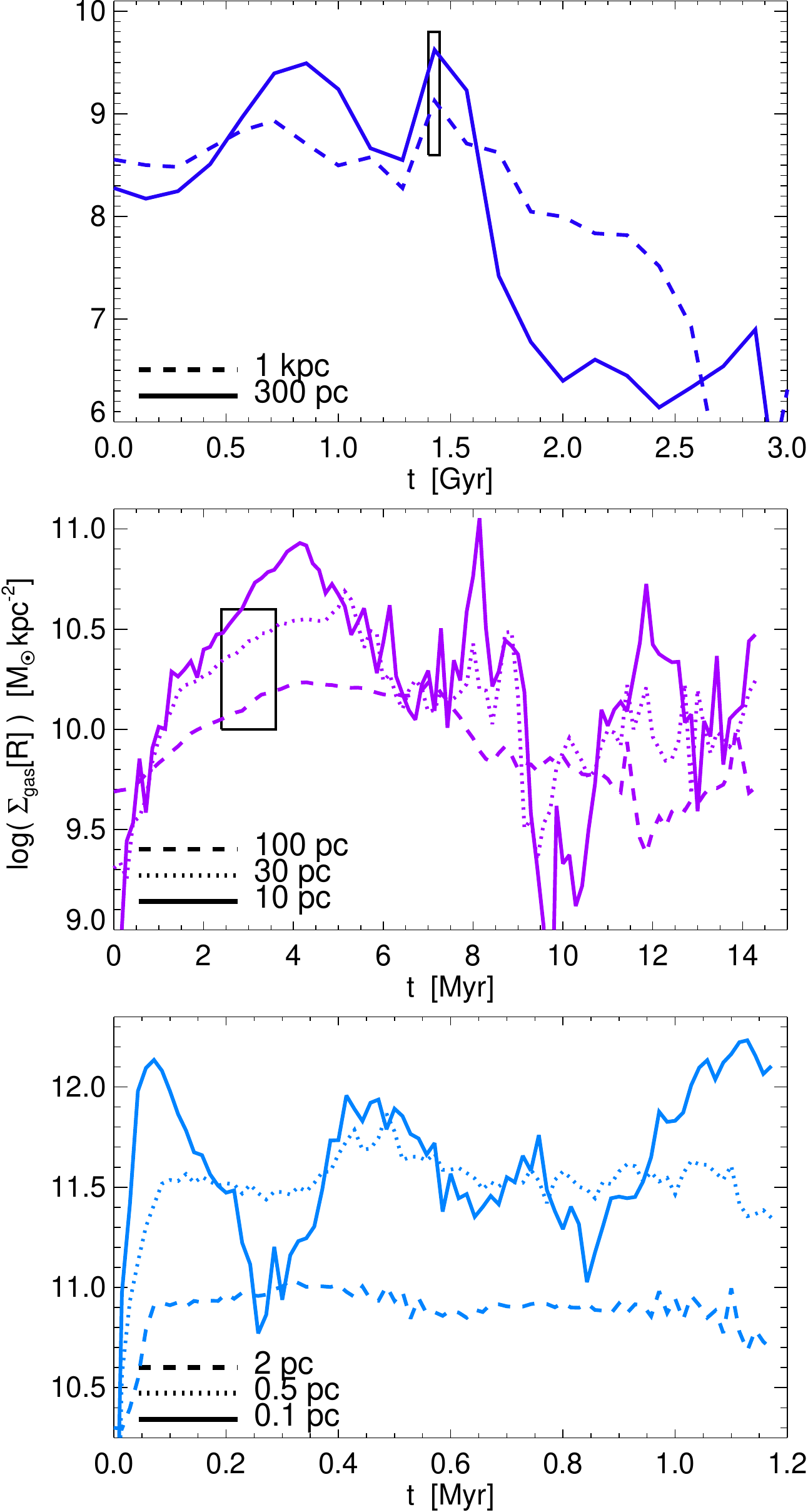}
    \caption{Gas surface density $\Sigma_{\rm gas}$ at different 
    radii as a function of time, for the  simulations shown in Figure~\ref{fig:time.dependence}. 
    {\em Top:} Large scale galaxy merger simulation. The box 
    shows the time interval re-simulated in the panel below. 
    {\em Middle:} Intermediate-scale 
    re-simulation. The box again shows the time interval for the smaller scale re-simulation. 
    {\em Bottom:} 
    Nuclear-scale re-simulation. Comparing with 
    Figure~\ref{fig:time.dependence},  high inflow rates 
    are generally correlated  with higher local surface densities, but the relation is not one-to-one. 
    \label{fig:sigma.r.t}}
\end{figure}

Figure~\ref{fig:sigma.r.t} shows the gas surface density as a function
of time, at different radii, in the high-inflow simulation from
Figures~\ref{fig:zoom} \& \ref{fig:time.dependence}.\footnote{Note
  that for the re-simulations, the density at small $r$ is
  intentionally initialized to be extremely small, but rapidly rises
  at very early times.}  Comparison of Figures
\ref{fig:time.dependence} \& \ref{fig:sigma.r.t} clearly indicates
that there is a reasonable correlation between high inflow rates and
high gas surface densities at the same radii.  However, the
correlation is not necessarily one-to-one because of the complex
time-dependent dynamics on small scales. In some cases, the inflow
leads the high surface density (e.g., $0.4\,$Myr at $0.1\,$pc or
$\sim2-4\,$Myr at $10\,$pc), in which case the large $\dot M$ causes
the high $\Sigma$, rather than the other way around.  Occasionally,
$\dot M$ and $\Sigma$ can even be anti-correlated (e.g., at $10\,$Myr
on $10\,$pc scales).

\begin{figure*}
    \centering
    \scaleup
    \plotside{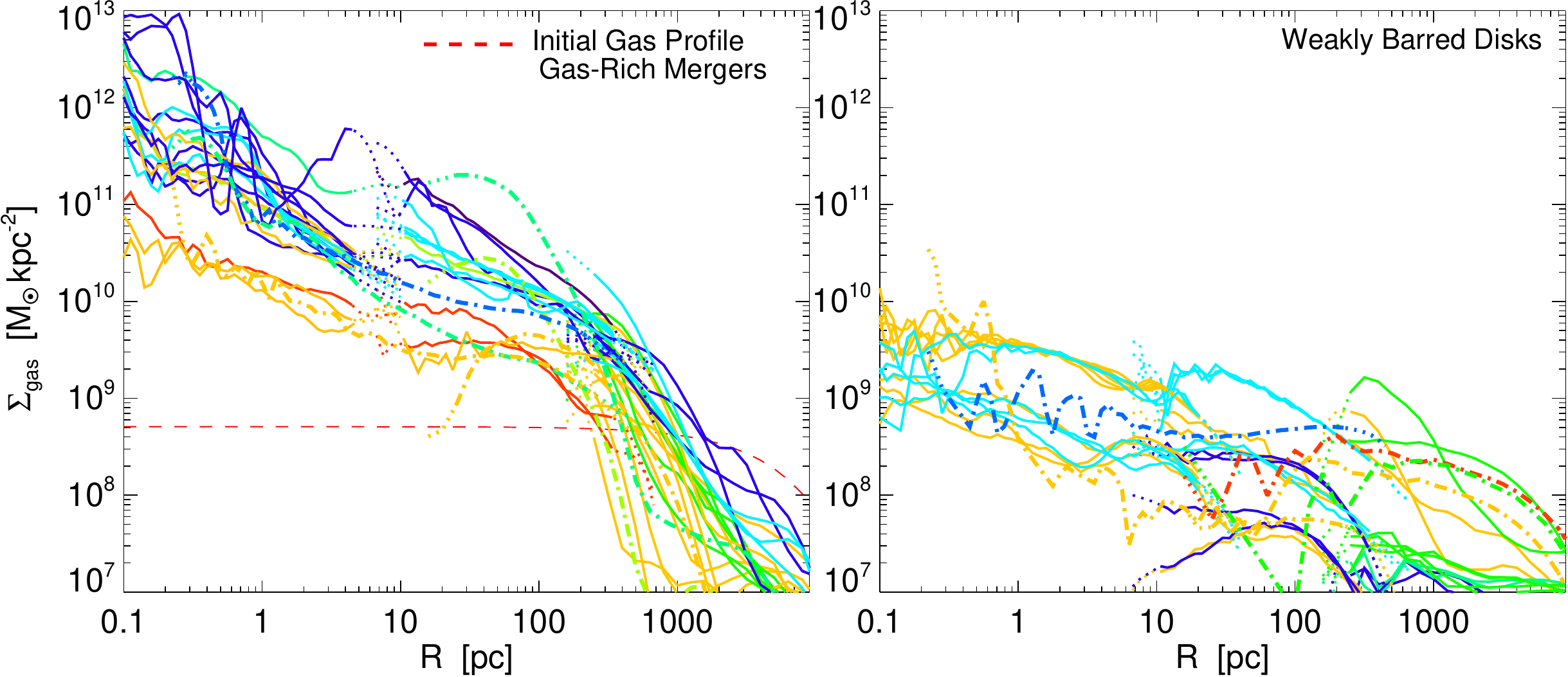}
    \caption{{\em Left:} Gas surface density profiles during the
      strong inflow phases, i.e., when the accretion rate at small
      radii peaks.  Each solid line denotes a different simulation,
      with dotted lines showing the radii near the boundaries of our
      re-simulations, where the exact profile shape is suspect; colors
      correspond to initial gas fractions as in
      Figure~\ref{fig:stability.criteria}.   Thick lines are the ultra-high resolution simulations.
       Each simulation consists of re-simulations of the nuclear dynamics during gas-rich
      galaxy-galaxy mergers, as in Figures~\ref{fig:time.dependence}
      \&\ \ref{fig:sigma.r.t}.  Dashed red line shows the initial gas
      density profile of the large-scale simulations (run b3ex(ic) in
      Table~\ref{tbl:galaxy}).  The gas density profile is
      quasi-steady over the entire active phase, with star formation
      offsetting inflow; the time variation in $\Sigma_{\rm gas}$ for
      one example is shown in Figure~\ref{fig:sigma.r.t}. {\em
        Right:} Gas surface density profiles for simulations of
      isolated barred-disk galaxies and the corresponding
      re-simulations.  In cases with a very small pre-existing bulge,
      the central gas density can increase by an order of magnitude,
      but not the $2-4$ orders of magnitude seen in the left panel.
      \label{fig:profiles.active}}
\end{figure*}

Figure~\ref{fig:profiles.active} shows the gas density profiles as a
function of radius for our ensemble of simulations (including 
the individual simulations shown in Figures
\ref{fig:time.dependence} \& \ref{fig:time.dependence.stable}); for the
gas-rich systems in the left panel, these surface density profiles are
at the peak of activity, i.e., when $\dot M$ is relatively high.
During these active phases, the gas reaches a quasi-equilibrium
density distribution.  For a given annulus, the net accretion rate to
smaller radii is typically a small fraction of the total rate at which
gas is supplied to that annulus.  Thus to first approximation, the
surface density can be estimated by considering the competition
between star formation and inflow in a given annulus.  If the star
formation rate surface density scales $\propto \Sigma_{\rm
  gas}^{3/2}$, as in \citet{kennicutt98}, then setting the star
formation rate inside some radius equal to the time-averaged inflow
rate $\dot{M}_{\rm in}$ gives an expected $\langle \Sigma_{\rm
  gas}\rangle \propto \dot{M}_{\rm in}^{2/3}\,R^{-4/3}$.  Given $\dot
M_{in} \propto R^{1/2}$, which is reasonably consistent with
Figure~\ref{fig:time.dependence}, this implies $\Sigma_{\rm gas}
\propto R^{-1}$, similar to the results for the gas rich systems in
Figures~\ref{fig:sigma.r.t}-\ref{fig:profiles.active} (although
slightly steeper than the numerical results).
There is of course significant variation: we show the results from 
our large suite of simulations to emphasize the variation introduced by initial and boundary 
conditions, treatment of star formation and gas physics, etc., but 
also to highlight the robust average behavior and the dependence on 
global quantities such as gas fraction and inflow rate at larger radii (see Fig.~\ref{fig:stability.criteria} for the color scheme). 

Note that the gas surface densities achieved on small scales can be 
extremely large, $\sim 10^{11}-10^{12}\,\msun\,{\rm kpc^{-2}}$ in the central $\sim0.1-10\,$pc; this is comparable to, 
or even somewhat larger than, the highest-surface density stellar systems 
known \citep[][]{hopkins:maximum.surface.densities,hopkins:sb.ir.lfs}.  
And it is a factor of $\sim 10^4$ larger than the initial gas 
surface density at small radii (shown in Fig.~\ref{fig:profiles.active} for comparison)!    
Assuming that a significant fraction of the gas eventually turns into
stars, the relic stellar density profiles expected from these gas-rich
simulations are similar to the observed profiles of nearby ``cusp''
ellipticals \citep[e.g.,][]{jk:profiles}, which are indeed
believed to be direct descendants of gas-rich mergers. However,
estimating the ``final'' stellar profile at small radii, where our
simulations are run for only a modest number of local dynamical times,
requires a careful correction for the effects of duty cycle, so we
defer a more detailed comparison to future work.

Figure~\ref{fig:profiles.active} (right panel) also shows the gas
surface density profiles for cases in which the initial large-scale
gas inflows are not strong (e.g.,
Fig. \ref{fig:time.dependence.stable}).  Most of these simulations
represent either extremely gas-poor mergers or (more commonly) systems
that are weakly bar unstable or bar unstable but with significant
bulge components (such that bars form efficiently, but inflow stalls
at a large inner Linblad resonance).  In such cases, the large-scale
dynamics cannot increase the gas density at sub-kpc scales very far
above the initial value of $\sim 10^{8}-10^{9}\,\msun\,{\rm
  kpc^{-2}}$.  Without this enhancement, the system is stable against
secondary instabilities, and so there is little gas inflow at small
radii (e.g., Fig. \ref{fig:time.dependence.stable}).

In both the strong and weak inflow cases in Figure \ref{fig:profiles.active}, our small subset of 
ultra-high resolution simulations (dot-dashed thick lines; 6 galaxy-scale runs that resolve down to 
$\sim10\,$pc, and 5 intermediate-scale runs that extend to $<1$\,pc) 
are fully consistent with the larger suite of lower resolution re-simulation runs.
These high resolution simulations have continuous inflow from large to small radii, 
and can therefore be run self-consistently for a much larger number of 
dynamical times than our typical resimulation. 
In spite of this advantage, we find that our resimulation technique yields very similar results. 

\vspace{-.65cm}
\breaker
\section{Conditions for (In)Stability}
\label{sec:stability}

In the preceding sections, we have presented results for cases in
which gas inflow from large scales both is, and is not, sufficient to
trigger secondary instabilities at small radii, leading to efficient
gas inflow down to $\sim 0.1\,$pc. Here we provide  a more quantitative assessment
of the conditions under which there is significant inflow of gas to sub-pc scales.

\begin{figure*}
    \centering
    \scaleup
    \plotsidesmallest{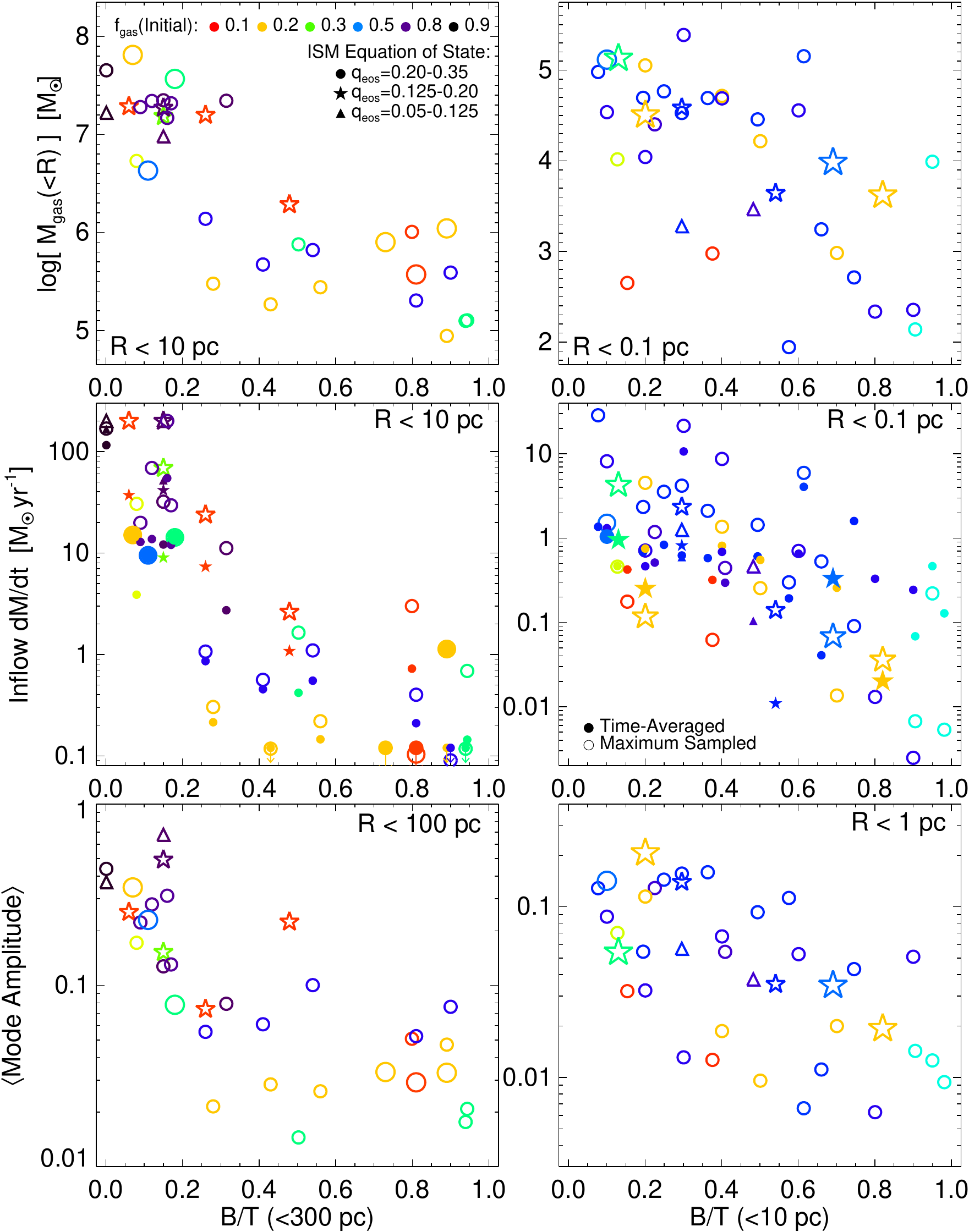}
    \caption{Inflow strengths as a function of bulge to total ratio
      (B/T) for several initial gas fractions (colors) and ISM
      equations of state $\qeos$ (symbol type), for a subset of our
      simulations. Large symbols in each refer to the subset of 
      ultra-high resolution simulations. 
      {\em Top:} Total gas mass inside an inner radius of
      $10\,$pc ({\em left}; from our intermediate-scale
      re-simulations) or $0.1\,$pc ({\em right}; from our
      nuclear-scale re-simulations), near the peak of accretion in
      each simulation.  ``Bulge'' (B) refers to any spherical term in
      the potential (bulge+halo+black hole).  {\em Middle:} Gas inflow
      rate through the relevant inner radius.  Solid points denote the
      time-averaged inflow rate over the full re-simulation while open
      points are the maximum seen in any individual snapshot (to give
      some sense of the duty cycle).  {\em Bottom:} Amplitude of the
      dominant non-axisymmetric mode in the stars (generally $m=2$ at
      intermediate scales, $m=1$ at nuclear scales), averaged over the
      peak/strong inflow phase.  Inflow properties scale broadly with
      $B/T$ as expected, but the large scatter and
      sensitivity to other parameters makes a large parameter survey
      critical.
      \label{fig:stability.criteria}}
\end{figure*}

Figure~\ref{fig:stability.criteria} shows how several different
measures of the efficiency of angular momentum transport and gas
inflow depend on the initial gas fraction $f_{\rm gas}$ and the
``bulge'' to total ratio $B/T$ -- the same parameters that strongly
influence the morphology of the gas (Fig.~\ref{fig:100pc.morphologies}
\&\ \ref{fig:10pc.morphologies}). We show results for a few different
equations of state $\qeos$ (see Fig.\ref{fig:qeos}). The quantities we
use to define the efficiency of gas inflow are the gas mass within
$10$ and $0.1\,$pc $M_{\rm gas}(<R)$ (top panel), the inflow rates
within the same radii (middle panel) and the formal amplitude of the
non-axisymmetric gravitational perturbations at $\sim 1$ and $\sim
100$ pc (bottom panel).  For the inflow rate (middle panel), we show
both the time average and ``peak'' values to convey the variability as
a function of time in the simulation.

The mode amplitude is measured in the standard fashion from the
surface density distribution at a given radius, using
\begin{equation}
|a_{m}(R,t)| = \frac{||\int_{0}^{2\pi} \Sigma(R,\phi)\,
\exp{(i\,m\,\phi)\,{\rm d}\phi} ||}
{\int_{0}^{2\pi}\Sigma(R,\phi)\,{\rm d}\phi}\ .
\end{equation}
We measure the amplitude from the stellar disk since it is slightly
more robust to local clumping, but using the gas surface density gives
similar results.  For simplicity, we show the results for the
largest-amplitude mode in each case, but this is almost always $m=1$
in the nuclear-scale simulations (right panels) and an $m=2$ bar or
$m=1$ spiral in the intermediate-scale simulations (left panels).  We
measure the relevant amplitude at radii slightly larger than the radii
where the inflow is measured, since it is this non-axisymmetry that
drives the inflow (it is also somewhat less noisy at larger radii).

The ``bulge-to-total'' ratio in Figure~\ref{fig:stability.criteria} is
defined as the mass fraction in {\em all} spherical components within the specified radius, since these all serve to
suppress instability in the same manner (e.g. any black hole, bulge, halo, and/or nuclear star cluster would
qualify). Gas fraction is defined as
the gas fraction of the {\em disky} component within the same radius.

Figure~\ref{fig:stability.criteria} shows clearly that 
there is a very strong trend of inflow strength with $B/T$, in the expected 
sense. Strongly disk-dominated systems have characteristic inflow 
rates 3 orders of magnitude higher than 
strongly ``bulge-dominated'' systems; likewise mode amplitudes 
and gas masses inside small radii are higher by $1-2$ 
orders of magnitude. 

There are several interesting additional results in
Figure~\ref{fig:stability.criteria}.  First, within the range of
$q_{\rm eos}$ that we consider (motivated by the observations in
Figure~\ref{fig:qeos}), the inflow properties do not depend
significantly on the ISM physics. This is consistent with our previous
arguments; so long as the sub-resolution model is sufficient to
prevent catastrophic fragmentation and runaway star formation,
gravitational torques that are reasonably independent of the ISM
microphysics dominate the angular momentum redistribution.    Note that there is also no distinguishable difference between the results of our ultra-high resolution runs (shown with large symbols) and our standard resimulations. 
Second, the inflow properties depend much more strongly on B/T than on
the initial gas fraction; this is because the stellar disk generates
the same instabilities as the gas disk. In fact, the collisional
nature of gas means it cannot support certain strong modes
\citep{toomre:spiral.structure.review}, so the transport can be less
efficient in the fully gas-dominated limit (the gas-rich systems
modeled here often do not develop strong modes until a non-trivial
fraction of the gas has turned into stars). 

Third, there is considerable scatter at all $B/T$. This is in no small
part due to the time-variable nature of the small scale dynamics (\S
\ref{sec:results:transport}).  In addition, however, a quantity such as $B/T$ is {\em not} a global invariant; it depends on both position and time in these simulations. As such, some simulations
with a given $B/T$ at the specified radius can have a larger or
smaller $B/T$ at some larger radius, leading to more or less efficient
inflow from those scales that in turn leads to more or less efficient
inflow to smaller scales.  Identifying $B/T$ at a given radius and
time is thus a crude measure of stability.  Moreover, the torques and inflow rates depend on the precise structure of the modes involved; a significant portion of the scatter in Figure~\ref{fig:stability.criteria} 
reflects the difference between e.g.\ bars, spiral arms, and loosely or tightly-wrapped waves.  This large scatter highlights the importance of using a large suite of simulations to conduct parameter surveys and
build statistical samples -- any individual system, studied for a
small amount of time, can be highly non-representative.

\vspace{-0.4cm}
\subsection{Intermediate (sub-kpc) Scales}
\label{sec:stability:intermediate}

For the intermediate scale results shown in the left panel of Figure
\ref{fig:stability.criteria}, we can gain some analytic understanding
of the behavior by following \citet{shlosman:bars.within.bars}, who
discuss the criteria for secondary instabilities (``bars within
bars'').  It is well established that a self-gravitating disk becomes
unstable to large-scale gravitational modes roughly when the
Ostriker-Peebles criterion is met, i.e., when $T_{\rm rot}/|W| >
t_{\rm crit}$, where $T_{\rm rot}\sim M_{d}\,V_{c}^{2}$ is the kinetic
energy of rotation (in the rotationally supported -- i.e.\
kinematically cold -- component), $|W|\sim G\,M^{2}/R$ is the
potential energy, and $t_{\rm crit}$ is a threshold value
$\simeq0.15-0.26$, depending on the nature of the instability (e.g.,
\citealt{bardeen:1975.disk.instability.crit,
  aoki:1979.gas.disk.instab.crit}).  If a mass fraction $f_{d}=1-B/T$
is in the disk, the condition for instability generically becomes
\begin{equation}
f_{d} > f_{\rm crit}(t_{\rm crit},\,a/R)
\label{eq:bars.in.bars} 
\end{equation}
where $a$ is the bulge scale length, $R$ is the disk scale length, and
the exact functional form of $f_{\rm crit}$ depends on the profiles of
the disk and bulge. For Kuzmin and Plummer profiles for the disk and
bulge, respectively, we find that this instability criterion becomes
\begin{align}
\frac{M_{d}[<R]}{M_{b} + M_{d}[<R]} > & 
\left(
\frac{3\pi\, t_{\rm crit}}{4[1-2\,t_{\rm crit}]}\,\frac{R}{a}
\right)^{1/2} & (R\ll  a)\\ 
 & 
\left(
\frac{3\pi\, t_{\rm crit}}{16}\,[1+(a/R)^{2}]\,\frac{R}{a}
\right) & (R\sim a)
\label{eq:bars.in.bars2} 
\end{align}
where $M_{d}[<R]$ is the total (stars + gas) disk mass within $R$ and
$M_{b}$ is the total bulge mass.

If the disky material of interest has a scale-length comparable to
that of the bulge (generally true for the intermediate-scale
simulations here), equation (\ref{eq:bars.in.bars}) simply becomes
$B/T\lesssim 1-t_{\rm crit} \simeq 0.7-0.8$.  This is reasonably
consistent with the fact that the inflow rate and interior gas mass in
the intermediate scale simulations in Figure
\ref{fig:stability.criteria} (left column) asymptote for $B/T \gtrsim
0.6-0.7$.

\citet{shlosman:bars.within.bars} show that an initial gas disk must
be compressed to at least $R_{\rm new}/R_{i} \sim (1.0-2.5)\,f_{\rm
  gas}^{2}/[1+0.2\,(B/T)^{2}]$ to satisfy equation
(\ref{eq:bars.in.bars}), again for systems in which the disk and bulge
initially have similar scale-lengths.  For the case of major mergers,
we can compare this to \citet{covington:diss.size.expectation} and
\citet{hopkins:cusps.ell,hopkins:cusps.evol}'s estimates of how much
angular momentum the gas loses during the merger.  These authors find
that the gas often inflows until it becomes self-gravitating -- as a
result, secondary bar instabilities should be ubiquitous.  More
quantitatively, they find $R_{\rm new}/R_{i}\approx f_{\rm
  gas}/(1+f_{\rm gas}/f_{0})$ with $f_{0}\approx0.2-0.3$. Using this,
we estimate that mergers with reasonable gas fractions $\gtrsim
0.3-0.5$ will lead to secondary instabilities and rapid inflow. For
isolated disk galaxies with bars, the critical criterion is whether the
inner Linblad resonance lies inside or outside of $R_{\rm new}$; this
will generally {\em not} be true in low-$f_{\rm gas}$ or high-$B/T$
systems.

\vspace{-0.5cm}
\subsection{Nuclear (pc) Scales}
\label{sec:stability:nuclear}

Inside the radius where the BH begins to dominate the gravitational
potential, it becomes increasingly difficult for system to be globally
gravitationally unstable in the sense of $T_{\rm rot}/|W| > t_{\rm
  crit}$.  Specifically, in the potential of the BH this criterion
implies that the disk would no longer be unstable
inside a radius $R_{\rm min} \sim 10\,{\rm pc}\,(M_{\rm
  BH}/10^{8}\,\msun)^{1/2}\, (\Sigma_{d}/10^{11}\,\msun\,{\rm
  kpc^{-2}})^{-1/2}$. 
This confirms our earlier claims (e.g., \S
\ref{sec:results:nuclear:BH}) that the character of the instabilities
responsible for angular momentum transport must change near the BHs
radius of influence.  Indeed, we see numerically that the transport on
small scales is dominated by $m = 1$ modes (e.g.,
Fig.~\ref{fig:10pc.morphologies}).

A full discussion of the origin of the $m = 1$ modes is beyond the scope
of this paper; we will present a detailed analytic analysis (in 
simplified BH+disk models) of their 
structure, growth rates, and pattern speeds in future work 
(in preparation). 
We can, however, present a simple analysis indicating
how we believe the modes arise.  For $m=1$ modes, $\Omega-\kappa/m
\rightarrow 0$ in a Keplerian potential.  
It is well-known that this
allows very low frequency (low pattern speed $\Omega_p$) modes to be
present in a Keplerian potential (e.g.,
\citealt{tremaine:slow.keplerian.modes}).  
Moreover, \citet{tremaine:slow.keplerian.modes} showed that the {\em only} modes that can be global and exert coherent strong torques over a large dynamic range in a quasi-Keplerian 
potential are these $m=1$ modes. 
We find that the modes in the simulations indeed have small $\Omega_p$, i.e., $\Omega_p \ll
\Omega$ and $m=1$, deep in the potential well of the BH.  Note that for a
collisionless particle, rather than the stellar+fluid disks of interest here,
this simply corresponds to a standing elliptical Keplerian orbit.  
Small $\Omega_{p}/\Omega$ (and wavenumber $m=1$) is important because it
implies near-resonance conditions at small radii; this allows the
stars to produce strong torques on the gas, which drives orbit
crossings, shocks, and angular momentum loss, even when the mass of
the disk at small radii is much less than the BH mass \citep[see
e.g.][]{chang:m31.eccentric.disk.model}. 
Following \citet{tremaine:slow.keplerian.modes} and expanding the equations of motion for a linear mode in a nearly-Keplerian potential 
(in the WKB limit) it is straightforward to show that the magnitude of the induced 
velocities ($\delta v/V_{c}$) from a given mode 
scale as $\delta v/V_{c} \sim (\delta \Sigma/\Sigma)\,M_{d}(<R)/M_{\rm BH} = 
|a|\,M_{d}(<R)/M_{\rm BH}$ for all $m\ne1$; for $m = 1$, however, $\delta v/V_{c}\sim |a|$.  In other words, the conventional wisdom that torques from global modes are strongly suppressed 
in the spherical potential of a BH is generally true, but the resonance condition that $\kappa \simeq \Omega$ in Keplerian potentials allows for large coherent eccentricities, 
self-gravitating collective effects, and torques from $m=1$ modes at arbitrarily small disk-to-BH masses. 
As a result, although we certainly see higher-$m$ modes in our small-scale simulations, the 
global structure is almost always dominated by a coherent global $m=1$ eccentric disk. 

However, \citet{tremaine:slow.keplerian.modes} also showed that slow modes with pattern speeds $\Omega_{p}\ll \Omega$, are
linearly stable in quasi-Keplerian potentials. How, then, do these
modes arise in the simulations?  To start, we consider the dispersion
relation for linear waves in the WKB limit \citep[see
e.g.][]{lau:spiral.wave.dispersion.relations}, which is given by
\begin{equation}
{\Bigl(}\frac{\omega-m\Omega}{\Omega}{\Bigr)}^{2} = 1+\nu
- 2\tilde f_{d} A \left(|kr|^2 + m^2\right)^{1/2}  + A \left(|kh|^2 + {\frac{m^2h^2}{r^2}}\right)
\label{eqn:dispersion}
\end{equation} 
where
\begin{equation}
\tilde{f}_{d} \equiv 
\frac{\pi\,\Sigma_{d}\,R^{2}}{M_{\rm enc}(<R)} \approx \frac{M_{d}(<R)}{M_{\rm tot}(<R)},
\end{equation}
\begin{equation}
A = 1 + {\frac{2 m^2 (3 - \nu)}{(\nu + 1)(|kr|^2 + m^2)}},
\end{equation}

\begin{equation}
\nu \equiv 1+2\,\frac{\partial \ln V_{c}}{\partial \ln R} \approx
\frac{{\rm d}\ln M_{\rm enc}(<R)}{{\rm d}\ln R} \ ,
\end{equation}
and $h = c_s/\Omega$ is the disk thickness.  We now consider the
limit of equation (\ref{eqn:dispersion}) such that $k r$
is modest.  Of course, our analytic treatment of such quasi-global
gravitational modes should be considered with due caution, as the WKB
approximation necessary to derive the dispersion relation is not
justified.  But this nevertheless points towards the conditions under
which global gravitational instability may be present.\footnote{ Note
  also that equation~\ref{eqn:dispersion} is derived with the WKB
  approximation, but including terms up to second order in
  $|k\,R|^{-1}$ (dropping out-of-phase terms), which makes it somewhat
  more accurate for the large-scale modes of interest here.
  Specifically, this leads to the additional enhancement of the growth
  rate $A\sim 1+\Gamma\,\sin{i}$, where $\Gamma\equiv \partial
  \ln{\Omega}/\partial \ln{R}$ and $i$ is the arm pitch angle; this
  approximates (at this order) the effects of the swing amplifier, an
  important contribution to the instability criterion.}  In this
limit, equation (\ref{eqn:dispersion}) will have growing modes if the
right-hand side is negative, which for a cold disk ($h \ll r$)
requires
\begin{equation}
  \tilde{f}_{d} > \frac{1}{2\,m}\, \frac{(1+\nu)^{2}}{7-\nu}\ .
  \label{instability}
\end{equation}
As $\nu\rightarrow 0$, i.e., as the potential becomes Keplerian,
equation~\ref{instability} suggests that $m = 1$ modes can become unstable when
$\tilde{f}_{d} \gtrsim 0.1$.  This is consistent with the fact that we
find the $m = 1$ modes at nearly all $B/T$
(Fig. \ref{fig:stability.criteria}).  Equation~\ref{instability} 
also allows for the instability of higher-$m$ modes, 
but these cannot exert coherent strong torques on the gas, 
nor can they propagate efficiently to small radii, so they do not dominate the structure or inflows 
we see in our simulations. 

Because the disk in general will have more mass at large radii, any
instability will grow outside-in; for massive disks, the most-unstable
point is where $M_{d}(<R) \sim 0.5-1\,M_{\rm BH}$ ($\tilde{f}_{d}\sim
1/3-1/2$).  Here modes can grow on a dynamical time.  Equation \ref{eqn:dispersion} shows that this growth is for fast modes: overstabilities with Re($\omega$)\,$\sim m\,\Omega$. In the
simulations, the slow modes typically dominate the torques at sub-pc
scales.  However, we find that the fast mode itself drives the slow
modes. Specifically, the $m=1$ mode first appears around the most
unstable point $R_{\rm crit}$ predicted above, where $M_{d}(<R_{\rm
  crit}) \sim 0.5-1\,M_{\rm BH}$.  This rapidly leads to an eccentric
disk at these radii.  As the instability causes gas to lose angular
momentum, the gas falls inwards, and begins turning into stars.  The
eccentric mass distribution at $R_{\rm crit}$ then in turn efficiently
{\em induces} a similar eccentricity at smaller radii. As noted in
\citet{tremaine:slow.keplerian.modes}, the $m=1$ slow mode may be
linearly stable, but it can be excited in the first place -- and will
be long-lived -- in response to an external asymmetry in the
potential.  Our interpretation of the simulations is that the unstable
self-gravitating disk near $\sim R_{\rm crit}$ provides the asymmetry
that generates the slow, low pattern speed, $m = 1$ mode at smaller
radii.  The pattern speed $\Omega_{p}=\Omega(R_{\rm crit})$ is
conserved as the mode moves in, but the circular speed $\Omega(R)$ is
rapidly increasing at smaller radii near the BH, so the mode at small
radii is indeed a slow mode -- for the typical parameters here, the
pattern speed is $\Omega_{p}\sim1-5\,{\rm km\,s^{-1}\,pc^{-1}}$, set
by the circular speeds at $\sim10-100\,$pc, where $M_{d}(<R)\sim
M_{\rm BH}$ (for comparison, $\Omega \gtrsim1000\,{\rm
  km\,s^{-1}\,pc^{-1}}$ at $R<1\,$pc).  The slow mode at small radii
responsible for the angular momentum transport is therefore not a
local small-scale instability, but part of a global instability
beginning at larger radii.  In the future, we plan to explore the
origin of this global $m = 1$ mode in more detail.  Calculating the
dynamics of the gas and stars, and including self-gravity and star
formation, are all important for capturing the inward propagation of
these modes.

Similar $m = 1$ modes have been studied previously as a mechanism for accretion in fluid disks 
\citep{adams89:eccentric.instab.in.keplerian.disks,shu:gas.disk.bar.tscale,
  ostriker:eccentric.waves.via.forcing,bournaud:2005.lopsided.disk.inflow}, both in the galactic context and in proto-planetary or proto-stellar disks \citep[e.g.,][]{banerjee:2004.protostellar.sphere.collapse,
  boley:2006.protoplanetary.disk.w.cooling,
  vorobyov:2006.protostellar.rapid.accretion,vorobyov:2009.secular.stellar.disk.evol,
  krumholz:2007.rhd.protostar.modes,
  cai:2008.protoplanet.disk.w.rad}. 
In a purely gaseous disk, wave packets can propagate through the OLR to $r\rightarrow\infty$, eventually becoming simple sound waves \citep{adams89:eccentric.instab.in.keplerian.disks}. Because the waves can freely escape to the outer boundary, infinite pure gaseous disks in nearly-Keplerian potentials do not support strong growing $m=1$ modes, and the mode growth is quite sensitive to the description of the disk boundary.   However, for a stellar disk, the mode cannot propagate beyond the 
OLR where $\Delta\equiv\kappa^{2}-m\,(\Omega-\Omega_{p})^{2}=0$. 
Refraction of the stellar waves at this boundary is important, and implies that global 
$m = 1$ modes can grow even when the disk extends to $R\gg R_{\rm OLR}$. 
As a result, we shall show in subsequent work that the $m=1$ 
modes in quasi-Keplerian stellar disks are relatively insensitive to the outer boundary conditions.   This is also implicit in \citet{tremaine:slow.keplerian.modes}, where for many of the modes the outer radius of the disk could be arbitrarily large.    The ability of stellar $m = 1$ modes to grow more readily again highlights the importance of considering the two component stellar + gas system when considering gas inflow in galactic nuclei.

\vspace{-0.6cm}
\breaker
\section{Accretion versus Star Formation}
\label{sec:results:sf}

\begin{figure*}
    \centering
    \scaleup
    \plotsidesmall{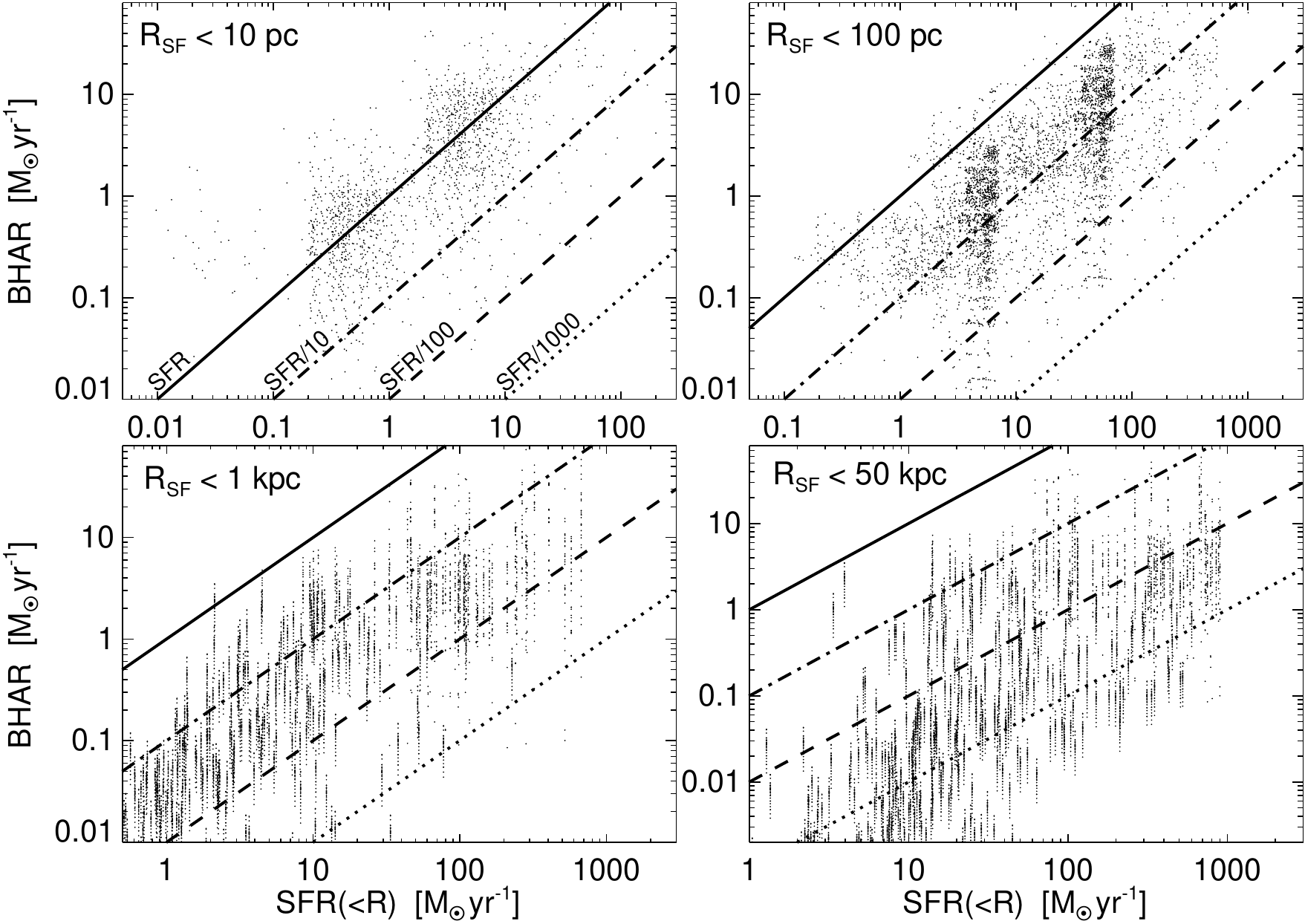}
    \caption{The predicted relationship between 
    the BH accretion rate (BHAR) and the star formation rate (SFR) inside
    a given radius.
    For each simulation, we plot the SFR 
    integrated within the indicated projected radius, at 
    different times during the simulation, compared with 
    the $<0.1\,$pc BH accretion rate from appropriate re-simulation.  
    The characteristic timescale for variability at small radii -- 
    which sets the variability in the BHAR --   
    is much shorter than the corresponding timescale at large radii; 
    at fixed SFR, each system can thus have a variety of BHARs.  
    Nevertheless, the BHAR correlates with SFR, with the least 
    scatter for the nuclear star formation (upper left).
    The correlations seen here are a natural consequence 
    of the fact that high gas densities, and thus high star formation 
    rates, are necessary to trigger secondary 
    instabilities and drive accretion on small scales. 
 \label{fig:mdot.vs.sfr}}
\end{figure*}

Given the gas inflows modeled here, our results imply a correlation
between the accretion rate onto the central BH -- and hence the AGN
luminosity -- and the star formation rates on larger scales.
Figure~\ref{fig:mdot.vs.sfr} shows this correlation for a number of
our simulations. We take the inflow rate at $0.1$\,pc as the AGN
accretion rate, which corresponds to a bolometric luminosity of
\begin{equation}
L_{\rm bol,\,AGN} = \epsilon_{r}\,\dot{M}_{\rm BH}\,c^{2} \approx 
5.7\times10^{45}\,
{\Bigl (}\frac{\epsilon_{r}}{0.1}{\Bigr )}\,
{\Bigl (}\frac{\dot{M}_{\rm BH}}{\msun\,{\rm yr^{-1}}}{\Bigr )}\,
{\rm erg\,s^{-1}}\ 
\end{equation}
where the radiative efficiency is $\epsilon_{r}=0.1$.
Figure~\ref{fig:mdot.vs.sfr} compares the BH accretion rate to the
total star formation rate inside a (projected) radius $R$, at a few representative
radii: true nuclear scales ($R<10\,$pc), more readily observable
nuclear scales ($R<100\,$pc), the smallest scales observable in most
moderate to high-redshift systems ($R<1\,$kpc), and the total star
formation rate, integrated over all radii ($R\rightarrow \infty$). 
For each point, in 
Figure~\ref{fig:mdot.vs.sfr}, the projected SFR$(<R)$ is technically the median over 
$\sim100$ sightlines, but projection effects are negligible. 
For the larger radii, we measure the SFR in an appropriate 
intermediate or galaxy-scale simulation. The BH accretion rate is 
then determined from that in the appropriate re-simulation or set of 
re-simulations matched to the galaxy scale simulation central properties 
at the given time. Because a given dynamical time in a galaxy-scale simulation 
corresponds to many dynamical times (and simulation outputs) on very small 
scales, the number of points is larger in the plots going out to larger scale. 

Not surprisingly, Figure~\ref{fig:mdot.vs.sfr} shows that there is a
correlation between accretion rate and star formation rate, on all
scales.  There is significant scatter in this correlation, although
the scatter is less when the SFR is evaluated inside a small nuclear
annulus.  If we assume a linear proportionality between BH accretion
rate and star formation rate in some annulus, we find
\begin{align}
\dot{M}_{\rm BH} & \sim \dot{M}_{\ast}(R<10\,{\rm pc}) \\ 
& \sim 0.1\,\dot{M}_{\ast}(R<100\,{\rm pc}) \\ 
& \sim 0.03\,\dot{M}_{\ast}(R<1\,{\rm kpc}) \\ 
& \sim 0.003\,\dot{M}_{\ast}({\rm total}) 
\end{align}
for the median relation in the simulations, which corresponds to a
bolometric luminosity ratio of $L_{\rm bol,\,SF}/L_{\rm bol,\,AGN}
\sim (0.01,\,0.1,\,0.3,\,2.5)$ at $(R<10\,{\rm pc},\,R<100\,{\rm
  pc},\,R<1\,{\rm kpc},\,R<\infty)$, respectively, or a ratio of the
star formation-powered infrared luminosity to the hard X-ray
luminosity from the AGN of $L_{\rm IR,\,SF}/L_{\rm HX,\,AGN} \sim
(0.2,\,2,\,6,\,60)$ in the same annuli.

In greater detail, the correlations are not precisely linear.
Parameterizing the correlation between star formation and BH accretion
as a power-law, $\dot{M}_{\rm BH} \propto \dot{M}_{\ast}^{\alpha}$, we
find that the correlations above are approximately valid for
$\dot{M}_{\rm BH}\sim 0.1-10\,\msun\,{\rm yr^{-1}}$, but $\alpha$
changes from near-linear at small scales ($R<10\,$pc) to somewhat
super-linear ($\alpha \sim 1.5$) on large scales.  For the total star
formation rate, we find
\begin{equation}
\dot{M}_{\rm BH}\sim 0.5\,{\msun\,{\rm yr^{-1}}}\,{\Bigl (} \frac{\dot{M}_{\ast}}{50\,{\msun\,{\rm yr^{-1}}}} {\Bigr )}^{1.5}\ .
\end{equation}
However, we caution that our simulations do not include the
appropriate physics to resolve the variety of processes that can
produce low accretion rates $\ll 0.1\,\msun\,{\rm yr^{-1}}$, nor do
they sample a large regime of galaxies with total star formation rates
$\lesssim1\,\msun\,{\rm yr^{-1}}$, so the fits above should not be
used in the limits of weak accretion and star formation. The
qualitative point is nonetheless important: non-axisymmetric
instabilities due to self-gravity become inefficient at low gas
densities (low star formation rates), and so other mechanisms must be
important for fueling at least some of the low-luminosity AGN
population.

It is notable that even during the active phases modeled here, the
{\em average} relation between accretion rate and total SFR is such
that the systems are most often not AGN dominated in a bolometric
sense (although the two are certainly comparable, in these phases).
In other words, even the most active systems will be consistent with
SFR-dominated properties except for a small fraction of the time; this
is consistent with observations such as the FIR-radio correlation
\citep{walter:2009.hyperlirg.in.highz.qso.host,
  riechers:2009.qso.edd.lim.starbursts, wang:highz.qso.ir}. However,
the scatter in accretion rates at fixed SFR is large, and most of the
growth in BH mass comes from the rare, short-lived times when the
systems scatter to higher $L_{\rm AGN}/L_{\rm SF}$.  Roughly, we can
estimate these duty cycles with the scatter in
Figure~\ref{fig:mdot.vs.sfr} -- about $\sim30\%$ of the systems in
``peak'' AGN phases have sufficient accretion rates ($\dot{M}_{\rm
  BH}\gtrsim0.01\,\dot{M}_{\ast}$) so as to be AGN-dominated.

The predictions in Figure \ref{fig:mdot.vs.sfr} can be compared to a
number of observations.  \citet{wu:liner.sfr.vs.bhar} compile {\em
  total} star formation rate estimators in $\sim100$ nearby AGN
ranging in luminosity from LINERS ($\dot{M}_{\rm BH}<0.01\,\msun\,{\rm
  yr^{-1}}$) to PG quasars ($\dot{M}_{\rm BH}\sim1-5\,\msun\,{\rm
  yr^{-1}}$).  The sample details are discussed in
\citet{satyapal:liner.sfr.vs.bhar}, but the authors find they are
reasonably robust to the choice of SFR or BH accretion rate indicator.
\citet{silverman:sf.in.agn.hosts} consider a similar compilation in
more luminous systems in the COSMOS survey, spanning a redshift range
$z\sim0.3-1.1$.  At total SFR values $>10\,\msun\,{\rm yr^{-1}}$,
these samples agree fairly well.  Specifically, for
$\dot{M}_{\ast}\sim10-1000\,\msun\,{\rm yr^{-1}}$, these different
compilations are consistent with a median $\dot{M}_{\rm
  BH}\sim0.001-0.01\,\dot{M}_{\ast}$, with factor of $\sim3$
dispersion; this is quite similar to our results shown in Figure
\ref{fig:mdot.vs.sfr}. On the other hand,
\citet{ho:agn.fb.supresses.sf} used OII line emission to argue that
the star formation rate associated with luminous type 1 AGN is at best
comparable to the BH accretion rate, inconsistent with the median
integrated galaxy predictions in Figure \ref{fig:mdot.vs.sfr}.  This
may be because obscuration makes OII an imperfect proxy for star
formation in dense, obscured, starbursts.  In addition, luminous Type
1 AGN are probably, by selection, on the tail end of the BHAR/SFR
distribution (and could in principle be in an AGN feedback-dominated
stage of evolution not modeled here).  Indeed, \citet{kim:agn.sfrs}
followed up Ho's work and found enhanced star formation in type 2
quasars, similar to our predictions here.

\citet{wang:seyfert.nuclear.sf} consider spatially resolved estimates
of nuclear luminosity and emission-line (SFR) profiles in a range of
Seyferts and quasars (mostly at low redshift) at spatial scales from
$\sim0.05-3\,$kpc \citep[see][]{imanishi:2002.nuclear.sfr.vs.agn,
  imanishi:2003.nuclear.sfr.vs.agn,imanishi:2004.nuclear.sfr.vs.agn}.
Their results are similar to those in a variety of other studies
\citep[see e.g.][]{kauffmann:qso.hosts,kewley:agn.host.sf,
  evans:agn.host.sfr,schawinski:blue.spheroids}.  At both $R<100\,$pc
(where the observations span a dynamic range in SFR$(<100\,{\rm
  pc})\sim0.1-10\,\msun\,{\rm yr^{-1}}$) and $R<1\,$kpc (SFR$(<1\,{\rm
  kpc})\sim1-100\,\msun\,{\rm yr^{-1}}$), our results are reasonably
consistent with these observations, in both normalization and scatter.
A small sample is studied at extremely high resolution in
\citet{davies:sfr.properties.in.torus}, where star formation and AGN
luminosity are resolved at $\sim10-30\,$pc scales. They find that,
within these radii, the luminosity from star formation is $\sim1-5\%$
of that from the BH, again consistent with our prediction here.

\vspace{-.6cm}
\section{Discussion}
\label{sec:discussion}

We have presented high resolution smoothed particle hydrodynamic
simulations of the dynamics of gas in the central regions of galaxies,
in order to study gas inflow from galactic scales $\sim10\,$kpc down
to the scales of a canonical AGN accretion disk ($\sim0.1\,$pc). We
use the properties of galactic-scale simulations -- both mergers and
isolated galaxy disks -- to initialize new higher resolution
``re-simulations'' that focus on the $\lesssim 100$\,pc dynamics. This
technique is then applied a second time to follow the gas to $\sim
0.1$\,pc. By intention, our calculations do not employ particle
splitting and are not all exact realizations of the small-scale
dynamics associated with a single, specific larger-scale
simulation.  Our ``re-simulations'' should not be taken as such an exact realization. 
Instead, since it is not practical to directly solve the ``full'' problem (from $\sim10\,$kpc to $10^{-5}\,$pc), we have focused on understanding and isolating the key physics involved. 
To provide a more physical model, however, we use initial conditions drawn from a variety of larger galaxy-scale simulations, rather than an ad hoc set-up.  This approach allows us to
systematically survey, for the first time, how a large variety of initial conditions and galaxy properties affect gas dynamics in galactic nuclei.

It is still not feasible, with any numerical scheme, to simultaneously model the small scales of an AGN accretion disk and those of a galaxy for cosmological timescales.  For this reason, direct ``zoom-in'' approaches have thus far been fairly limited, usually evolving each region for just a few 
local dynamical times. 
Our numerical approach attempts to overcome this severe limitation. 
First, the smallest regions of each simulation are simulated for many local dynamical times. Even more importantly, we quantity the dynamics on small scales
using $\sim 100$ simulations of a large range of initial conditions and galaxy properties 
This approach does not require an exact mapping between one large-scale simulation 
and another on smaller scales (although we often interpret our results in this manner). For example, even if our nuclear-scale simulations of the central 
$\sim100\,$pc of a galaxy are not self-consistently matched to those of a
specific larger-scale simulation, they
still represent valid simulations of a given set of initial
conditions of a bulge+stellar disk+gas disk+BH system, evolved for
many dynamical times. As such, we can still use them to understand 
the physics of angular momentum transport and BH fueling at these radii (given these initial conditions). 
This methodology  allows us to survey a wide range of initial conditions and galaxy properties and isolate the physical parameters that are most important for the 
physics of gas inflows on small scales in galactic nuclei.
We have also checked our "re-simulation" approach by carrying out a small number of ultra-high resolution simulations that bridge the gap between different resimulated regions; we find in all cases that these ultra-high resolution simulations validate our resimulation methodology.

Our calculations demonstrate that for sufficiently gas-rich,
disk-dominated systems, which have a large inflow of gas into
$\sim$kpc scales, there is a cascade of non-axisymmetric gravitational
instabilities that ultimately produces BH accretion rates as high as
$\sim 10\,\msun\,{\rm yr^{-1}}$ at $\lesssim 0.1$ pc
(Fig.~\ref{fig:stability.criteria}).  Moreover, we have explicitly
shown that these conditions are satisfied during major mergers of
gas-rich galaxies, thus providing theoretical support for the
connection between mergers and BH growth. It is, however, also
important to stress that our work only implies that galaxy mergers are
a sufficient condition for fueling a central quasar, not that they are
necessary: similar inflows can be obtained in isolated gas-rich,
bar or spiral wave-unstable galactic disks.


In broad terms, our results support the ``bars within bars'' scenario
proposed in \citet{shlosman:bars.within.bars}. However, we show that
the secondary instabilities on intermediate scales ($\sim 10-100$\,pc)
exhibit a diverse range of morphologies, including standard nuclear
spirals, bars, rings, barred rings, one or three-armed spirals, and
irregular/clumpy disks and streams (Fig. \ref{fig:zoom} \&
\ref{fig:100pc.morphologies}). This is very important for comparing to
real galaxies: observations have generally found that nuclear bars are
not ubiquitous in AGN (perhaps not substantially more common than in
non-active systems). However, these same observations have found that
{\em some} form of asymmetric nuclear structure is ubiquitous, with
the most common features being nuclear spirals and rings very similar
to those seen in Figure~\ref{fig:100pc.morphologies} \citep[see][and
references therein]{
  martini:seyfert.host.morph,martini:nuclear.agn.morphologies,
  peletier:seyfert.morph.imaging,knapen:seyfert.morphology,
  laine:nested.bars.in.seyferts,kanpen:nuclear.region.in.bars.vs.host.prop,
  greene:2009.sigma.gas.in.agn}. Our work shows that these
observations are {not} necessarily in conflict with the hypothesis
that gravitational torques dominate inflow on sub-kpc scales in AGN.
Rather there is a large diversity in the observational manifestation
of these gravitational torques. ``Bars within bars'' should not be
taken too literally. Instead a more accurate characterization would
be: ``it's non-axisymmetric features all the way down'' (or ``stuff
within stuff''). 

This diversity in observational appearance owes both to the effects of
global parameters such as gas fraction and bulge-to-total ratio, and
more subtle variations induced by the relative scale lengths of disks
and bulges, the precise profile shapes, and the thickness/dispersion
in the sub-kpc bulge, stellar disk, and gaseous disk components.
Observationally, these parameters are all seen to vary considerably
even in isolated galaxies, let alone in chaotic merging systems; as a
result, we expect the morphological diversity exhibited in Figures
\ref{fig:zoom} \& \ref{fig:100pc.morphologies} to be the norm.

Once the gas reaches the BH radius of influence ($\sim10\,$pc for a
typical $\sim L_{\ast}$ system), the potential becomes quasi-Keplerian
and spherical (dominated by the BH itself); the system is thus no
longer bar unstable. However, the gas is still locally
self-gravitating and prone to forming stars if it does not
inflow rapidly. Indeed, this is traditionally the range of radii at
which it has been the most difficult to produce the large accretion
rates needed for luminous AGN and quasars
\citep{shlosman:midscale.accretion,thompson:rad.pressure}. We find,
however, that new large-scale non-axisymmetric modes arise at $\sim$
pc-scales and robustly allow for continued gas transport to smaller
radii.  Specifically, for sufficiently gas-rich, disky systems, we
find an eccentric/lopsided disk or a one-armed spiral instability (an
$m=1$ mode); see Figure~\ref{fig:10pc.morphologies}.  This leads to
large inflow rates of $\sim1-10\,\msun\,{\rm yr^{-1}}$ into the
central $0.1$ pc, comparable to what is needed to explain the
brightest quasars observed. At radii $\lesssim 0.01-0.1$ pc, the
accretion flow is no longer self-gravitating and {\it local} viscous
heating is sufficient to maintain $Q \gtrsim 1$ and transport gas to
the BH \citep{goodman:qso.disk.selfgrav,thompson:rad.pressure}; the
disk is approaching a canonical thin Keplerian accretion disk.

The $m = 1$ modes seen here have been studied previously as a
mechanism for accretion onto BHs and protostars
\citep{adams89:eccentric.instab.in.keplerian.disks,shu:gas.disk.bar.tscale,
  ostriker:eccentric.waves.via.forcing}, but have largely been
neglected in studies of fueling luminous AGN.  This is in part because
most previous simulations neglected either star formation or the
self-gravity of the gas and stars (owing to computational limits or,
of course, the fact that these complications are not present in the
case of a protostar).  Moreover, previous calculations have shown that
although near-Keplerian potentials can support the ``slow'' pattern
speed  $m = 1$ modes we find here, such modes are
linearly stable \citep{tremaine:slow.keplerian.modes}. Why, then, are
these modes ubiquitous in our simulations?  We find that the inclusion
of stars and gas, self-gravity, live star formation, and some model
for stellar feedback are all important for the behavior of these
instabilities.  In particular, we typically find a complex
interplay between the gas and stars (\S \ref{sec:results:duty} \&
\ref{sec:stability:nuclear}).  The system often develops the $m = 1$
modes first near co-rotation, where the disk and BH contribute
comparably to the potential; the mode growth is faster 
in the stellar distribution, which can support self-crossing orbits and which is much less sensitive to the outer properties of the disk. As gas streams move through the lopsided stellar distribution, they are
torqued, experience orbit crossing, and lose angular momentum and
energy \citep[as in, e.g.,][]{chang:m31.eccentric.disk.model}.  As the
gas moves to smaller radii, the stars at larger radii provide a less
efficient angular momentum sink.  However, star formation ensures that
new stars are formed in situ (themselves in a lopsided distribution),
allowing the gas to continue to flow in.  This process leads to the $m
= 1$ mode propagating inwards into the gravitational potential well of
the BH, from the larger radii where it originated. The torques and inflow rates are thus quite different in stellar+gas systems as compared to pure 
gas disks.  In particular, orbit crossing induced by the stars can lead to much more 
efficient gas inflow; we will study this analytically in \papertwo. 

Our calculations show that the inflow of gas on sub-kpc scales,
including the BH accretion rate at $\lesssim 0.1$ pc, is highly time
variable (Fig. \ref{fig:time.dependence}). This variability is a
result of gravitationally unstable modes forming, damping, and,
re-forming, and the fact that individual clumps or dense regions can
dominate the inflow rate at a given time. The ``duty cycle'' for large
inflow rates is modest, even in systems with a large time averaged
accretion rate.  In one of our most violent simulations, a massive
nuclear ring forms at $\sim$kpc; owing to an inner Linblad-like
resonance, the system has an outflow from sub-kpc scales, and inflow
from the merger at larger radii. Since the outflow/inflow rates are
large, the ring soon becomes strongly self-gravitating and globally
collapses, leading to one of the largest net accretion rates into
$\sim10\,$pc in any of our simulations. However, more than $\sim95\%$
of the time during the ``active'' phase, the system has strong outflow
from the sub-kpc region. Moreover, because the gravitational modes
driving accretion at large scales (e.g.\ the merger), at $\sim 100$ pc
(secondary instabilities), and at $\sim$pc (the lopsided nuclear disk)
are physically separate, their variability on short timescales is not
fully coupled. This is clearly important when relating observed
torques on large scales to AGN accretion rates on small scales. In
many observed AGN or barred disks (with a proper inner Linblad
resonance), there are no strong inflows (and may even be outflows)
observable at $\sim$ kpc
\citep{block:obs.bar.torque.maps,garcia.burillo:torques.in.agn.nuclei.obs.maps.no.inflow,
  haan:nuga.gas.dynamics.maps,stoklasova:torque.maps.1kpc.seyferts,
  durbala:obs.isolated.gal.torque.maps}. Our simulations demonstrate
that this can in fact be the case even in systems with large
time-averaged accretion rates. 

 The key parameter determining the qualitative behavior at each scale
  in our simulations is the ratio of the total mass in a rotationally
  supported disk (gas and
  stars) to the pre-existing mass in any spherical component (bulge,
  halo, or black
  hole); see Figure~\ref{fig:stability.criteria} and \S~\ref{sec:stability}. There
  is, of course, a considerable literature studying the instability
  of self-gravitating disks in disk/bulge systems, which is
  applicable to the
  intermediate scales ($\sim 10-100$ pc) in our simulations 
  \citep[see e.g.][]{bardeen:1975.disk.instability.crit,athanassoula:disc.instab.criteria,
  narayan:87.bar.tscale.shearing.sheet,raha:bar.instabilities,
  christodoulou:bar.crit.1,earn.sellwood:95.nbody.bar.stab,
  bournaud:gas.bar.renewal.destruction,dubinski:bar.evol.sim.tests}. As outlined in
  \citet{shlosman:bars.within.bars}, most of these criteria amount to
  variations on the Ostriker-Peebles criterion
  that the kinetic energy of rotation inside some radius be $\gtrsim
  20\%$ of the potential energy. This
  requires increasing the surface density of disky
  material {\em within} the pre-existing bulge/halo effective radius
  to a value comparable to, or equal to, that of the bulge/halo.
  For major, gas-rich mergers ($f_{\rm gas}\gtrsim0.3$), this
  condition is almost always met. However, for  gas-poor
  mergers, or for isolated barred galaxies, it is not clear how often
  systems will in fact trigger
  secondary instabilities. We find cases where weak large-scale
  bars trigger no significant subsequent activity on smaller scales
  (Figure~\ref{fig:time.dependence.stable}). 

  In the gravitational potential of the central BH, the criterion for
  further inflow is somewhat altered, although we believe that it is
  qualitatively similar (\S \ref{sec:stability:nuclear}): lopsided $m
  = 1$ modes are dynamically important provided that the total mass in
  the disky component is $\sim0.1-1\,M_{\rm BH}$ in the central $\sim
  10$ parsecs (Figure \ref{fig:stability.criteria}; note the large scatter
  even at fixed disk to BH mass).  We find
  that this criterion is generally satisfied, so long as the larger
  scales ($\sim 10-100$ pc) are themselves unstable.  The critical
  barrier, then, to triggering a cascade of instabilities leading to
  significant BH accretion is the presence of a large self-gravitating
  gas inflow at $\sim$ a kpc.

  There is significant evidence that galaxies in fact meet the
  criteria outlined here for AGN fueling. Studies of ongoing gas-rich
  mergers find that the gas densities reach and exceed the
  self-gravity threshold, with gas densities similar to those
  predicted here
  \citep[Figures~\ref{fig:sigma.r.t}-\ref{fig:profiles.active}; see
  e.g.][]{tacconi:ngc6240.gasdynamics,
    hibbard.yun:excess.light,laine:toomre.sequence,
    iono:ngc6240.nuclear.gas.huge.turbulence, schweizer:ngc34.disk}.
  Similarly, studies of early-type galaxies find that the stellar
  surface densities can be separated into a distinct ``dissipational''
  component -- the stars at small radii which must have formed in a
  dissipational starburst -- and a ``dissipationless'' envelope of
  violently relaxed stars \citep[see][]{hopkins:cusps.fp}. The
  starburst relic component is inferred to have formed from
  self-gravitating gas; indeed, this appears to define its size-mass
  relation \citep{hopkins:cusps.ell}.  In addition, the criterion for
  significant gas inflow given a modest pre-existing bulge corresponds
  to a surface density in the rotationally supported component of
  $\sim10^{11}\,\msun\,{\rm kpc^{-2}}$ at $\sim10\,$pc
  (Figure~\ref{fig:profiles.active}), for an $\sim L_{\ast}$
  system. This is comparable to the characteristic stellar surface
  density observed at these radii in massive spheroids \citep[see][and
  references therein]{hopkins:maximum.surface.densities}.

  One of the strong predictions of this work is that the link between
  high gas surface densities and inflow leads to a correlation between
  BH accretion rate and star formation (\S~\ref{sec:results:sf}). We
  have assumed a star formation law in which $\dot \rho_* \propto
  \rho^{1.5}$, which
  corresponds to a fixed star formation efficiency per local dynamical
  time. This is supported by models of turbulence-regulated star
  formation, in which the efficiency per dynamical time is a weak
  function of ambient conditions such as the Mach number of the
  turbulence \citep{krumholz.schmidt}. With this assumption, we
  predict a correlation between BH accretion and the star formation
  at various scales in galactic nuclei, albeit one
  with significant scatter (Figure~\ref{fig:mdot.vs.sfr}). These
  predictions agree reasonably well with current
  observations
  \citep{wang:seyfert.nuclear.sf,wu:liner.sfr.vs.bhar,silverman:sf.in.agn.hosts}.
  In the future, it will be interesting to explore the possibility
  that  the star formation efficiency in galactic
  nuclei is significantly lower than the mean ISM value, as has been
  suggested by several authors \citep{thompson:rad.pressure,
    begelman:direct.bh.collapse.w.turbulence,larson:column.density.bh.vs.sf}.

  If the asymmetric stellar structures found here are long-lived
  (decay times of $\sim$ Gyr), they should in principle be observable
  around nearby massive BHs. One tantalizing possibility is that the
  ubiquitous eccentric stellar and gaseous disks we find at $\sim
  1-10$ pc may explain the nuclear stellar disk seen in M31, which has
  been the subject of considerable study \citep[see e.g.][]{lauer93,
    tremaine:m31.nuclear.disk.model,salow:nuclear.disk.models,
    sambhus:m31.nuclear.disk.model,peiris:m31.nuclear.disk.models,
    bender:m31.nuclear.disk.obs}.    \citet{hopkins:m31.disk} 
    examine this possibility in detail and show that the spatial scales $\sim1-10\,$pc, moderate
  eccentricities, and stellar masses $\sim0.1-1\,M_{\rm BH}$ found in our simulations are all
  comparable to those observed.    Eccentric nuclear disks have also been
  observed (albeit in less detail) around a number of other massive
  but inactive BHs, including NGC4486b \citep{lauer:ngc4486b}, M83
  \citep{thatte:m83.double.nucleus}, and VCC128
  \citep{debattista:vcc128.binary.nucleus}.  This strongly suggests that we can probe the the instabilities that produced the growth of massive BHs in the ``fossil'' relics of that
  accretion.

  The dominant uncertainties in the models presented here are our
  treatment of the ISM physics, star formation, and feedback. Lacking
  a full micro-physical understanding of these processes -- or the
  resolution to directly incorporate such a model -- it is necessary
  to adopt a sub-resolution prescription for their impact.     Our model is that stars form at all radii, in local, self-gravitating clumps,  with a fixed efficiency relative to the local dynamical time. In  addition, feedback from stars (supernovae, radiation, stellar
  winds/outflows, etc.) contributes, potentially along with cosmic
  rays and magnetic fields, to generating a turbulent velocity that is
  much larger than the thermal sound speed of the ISM. It is this
  turbulent ``sound speed'' that is included in our ISM equation of
  state (Figure~\ref{fig:qeos}).     Observations suggest that these
  assumptions are at least plausible even at the small scales we model
  here: a local Kennicutt-Schmidt relation holds in nuclear
  starbursts and on small scales in galaxies \citep{kennicutt98,
   kennicutt:m51.resolved.sfr,bigiel:2008.mol.kennicutt.on.sub.kpc.scales}
 and star formation efficiencies per dynamical time do not appear to
 evolve even in very dense regions \citep[comparable to the
 highest gas densities modeled here; see,  e.g.,][]{tan:mol.cloud.formation.times,
    krumholz:sf.eff.in.clouds}. Moreover, AGN are observed to have
  associated nuclear star formation on scales as small as a pc; the
  densities of stars formed {\em in situ} at these scales \citep{davies:sfr.properties.in.torus, hicks:obs.torus.properties} are reasonably consistent with our predictions.   Large turbulent and/or non-thermal gas velocity dispersions are also ubiquitous, from starburst nuclei on $\sim$kpc
  scales to the sub-pc scales around AGN probed by water masers  \citep{downes.solomon:ulirgs, westmoquette:m82.sb.core.gasdynamics,tacconi:ngc6240.gasdynamics,iono:ngc6240.nuclear.gas.huge.turbulence,kondratko:3079.acc.disk.maser}; these observations motivate our choice of sub-resolution feedback models (Figure~\ref{fig:qeos}).  To study the impact of ISM uncertainties on our results, we explicitly used different star formation efficiencies in some of our simulations;  although this does have a quantitative affect on the resulting gas densities and inflow rates, the physical processes that we have identified as driving accretion remain the same.

    As noted in \S \ref{sec:sims}, the primary consequence of the large  effective sound speed in our models is to increase the Jeans and  Toomre masses, thus suppressing the formation of small-scale
  structure. That is, we effectively smooth over the smaller scales on
  which our treatment of the ISM physics is not appropriate.  This implies that in our model,  most of the gas, most of the time, resides in relatively diffuse structures, rather than being bound to very compact clusters, as would be the case if we did not include stellar feedback and/or a sub-grid sound speed (see Appendix \ref{sec:ism}).  We  suspect that this approach is reasonable for the global  gravitational dynamics highlighted in this paper.  In particular,
  our calculations indicate that over a remarkably large dynamic range, from $0.1\,$pc to $10\,$kpc, the torques and angular momentum  transport are gravitational, and scale simply with the magnitude of
  the asymmetry in the potential, not with the sound speed of the gas  (Fig. \ref{fig:inflow.vs.viscous} and \S \ref{sec:results:duty}).  In \papertwo, we will present a more detailed comparison between
  analytic models and our numerical results that supports this conclusion.

  In our calculations, we have neglected AGN feedback in order to
  study how gas gets down to a BH in the first place.  To the extent
  that feedback is important on these scales, our simulations may
  better approximate the ``early'' stages of BH growth, when the BH is
  still relatively small (compared to, e.g., the still-forming bulge),
  and before it reaches a critical mass or luminosity at which
  feedback becomes dynamically important. When BH growth becomes
  self-regulated, inflow and outflow are coupled, and the problem of
  AGN fueling must be addressed with a model that includes AGN
  feedback as a function of gas properties and spatial scale.

  \breaker \acknowledgments We thank Phil Chang, Lars Hernquist, Norm
  Murray, and  Volker Springel for helpful discussions during the development of
  this work, and the referee for a careful reading of the text and a number of very useful suggestions.   PFH and EQ were supported by the Miller  Institute for Basic Research in Science, University of California  Berkeley. EQ was also supported in part by NASA grant NNG06GI68G and
  the David and Lucile Packard Foundation.
  \\

\bibliography{/Users/phopkins/Documents/lars_galaxies/papers/ms}

\begin{appendix}

\section{Numerical Tests}
\label{sec:numerical}

We have performed a number of tests of our methodology, in an effort
to ensure both that we have captured the most important physics in our
simulations, and to ensure that there are no artificial effects
occurring due to numerical or resolution artifacts. We describe some
of the key tests here.  For general tests of {\small GADGET-3}, we
refer the reader to
\citet{springel:gadget,springel:multiphase,springel:models,springel:entropy}.

\vspace{-0.5cm}
\subsection{Resolution Tests}
\label{sec:numerical:general}

For the initial galaxy-scale simulations (Table~\ref{tbl:galaxy}),
resolution tests have been extensively described previously
\citep{dimatteo:msigma,robertson:msigma.evolution,cox:kinematics,
  hopkins:cusps.ell}; these range from $\sim10^{5}$ particles to
$\sim10^{7}\,$particles, and spatial resolutions/softenings from
$\sim20\,$pc to $\sim100\,$pc. For our purposes, the quantities of
interest (e.g.\ the gas inflows to $\sim100\,$pc scales) are
well-converged \citep[see e.g.\ Appendix~B of][]{hopkins:cusps.ell}.

\begin{figure}
    \centering
    \plotone{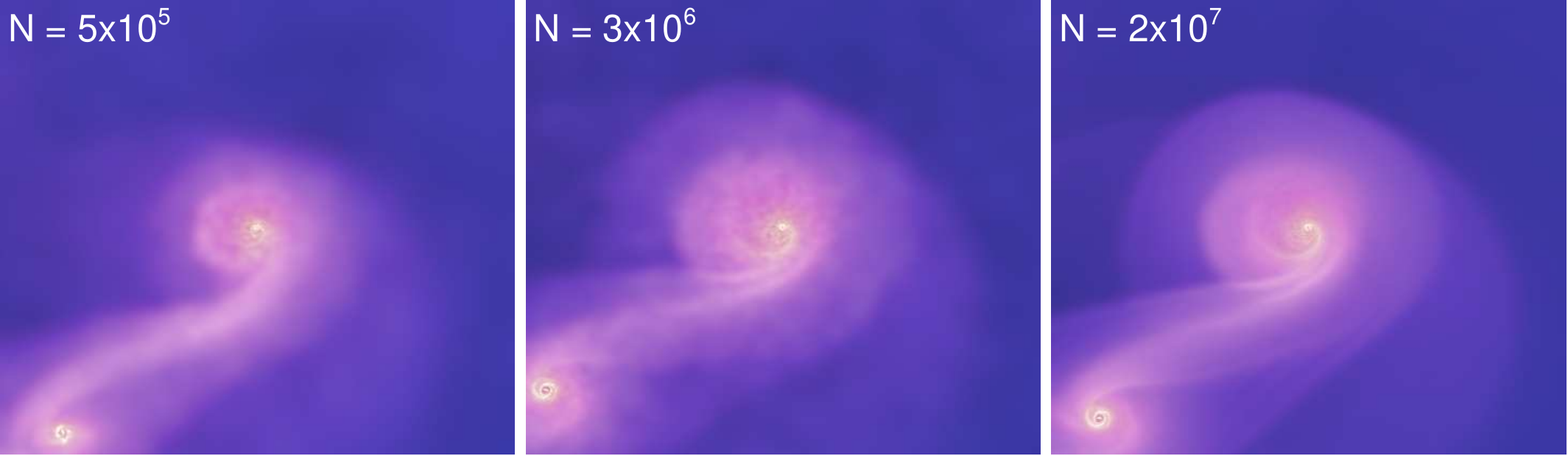}
    \caption{Example resolution test for a nuclear scale simulation.
    The same initial conditions  
    re-simulated with the particle number shown (lower-resolution 
    and nominal cases of Nf8h2b1h; see Table 3). For 
    $\gtrsim 0.5\times10^{6}$ particles, our results are converged.
    \label{fig:rtest}}
\end{figure}

We have performed a series of analogous resolution tests on our
intermediate (Table~\ref{tbl:intermediate}) and small-scale
simulations (Table~\ref{tbl:nuclear}), covering a similar span in
particle number.  For the intermediate-scale simulations, we have
varied the SPH smoothing length and gravitational softening length
from $<1\,$pc to $\sim10\,$pc.  We find that the results of
particular importance for this paper, e.g., the total gas mass that
loses angular momentum and flows to smaller scales, are well-converged
for all the resolutions $\lesssim10\,$pc used here. This is not
surprising, because for physical reasons (discussed in the main text),
the material reaches an angular momentum barrier at $\sim10\,$pc.  For
our small-scale simulations, we find similar behavior for all force
resolutions $\ll 1\,$pc (experimenting with a range 
of softenings from $0.02-1\,$pc), and with good convergence at higher particle
numbers $\gtrsim$ a few $10^{5}$. 
Given the range of resolution explored, comparison with 
Figure~\ref{fig:gas.Q} demonstrates that, especially in our high-resolution studies, 
even the verticle structure of the gaseous disks is well-resolved 
(in the highest-resolution cases, the typical $h_{\rm sml}/R<0.01$). 
Figure~\ref{fig:rtest} shows an example resolution test 
varying from $5\times10^{5} - 2\times10^{7}$ particles.

\begin{figure*}
    \centering
    \scaleup
    \plotsidesmall{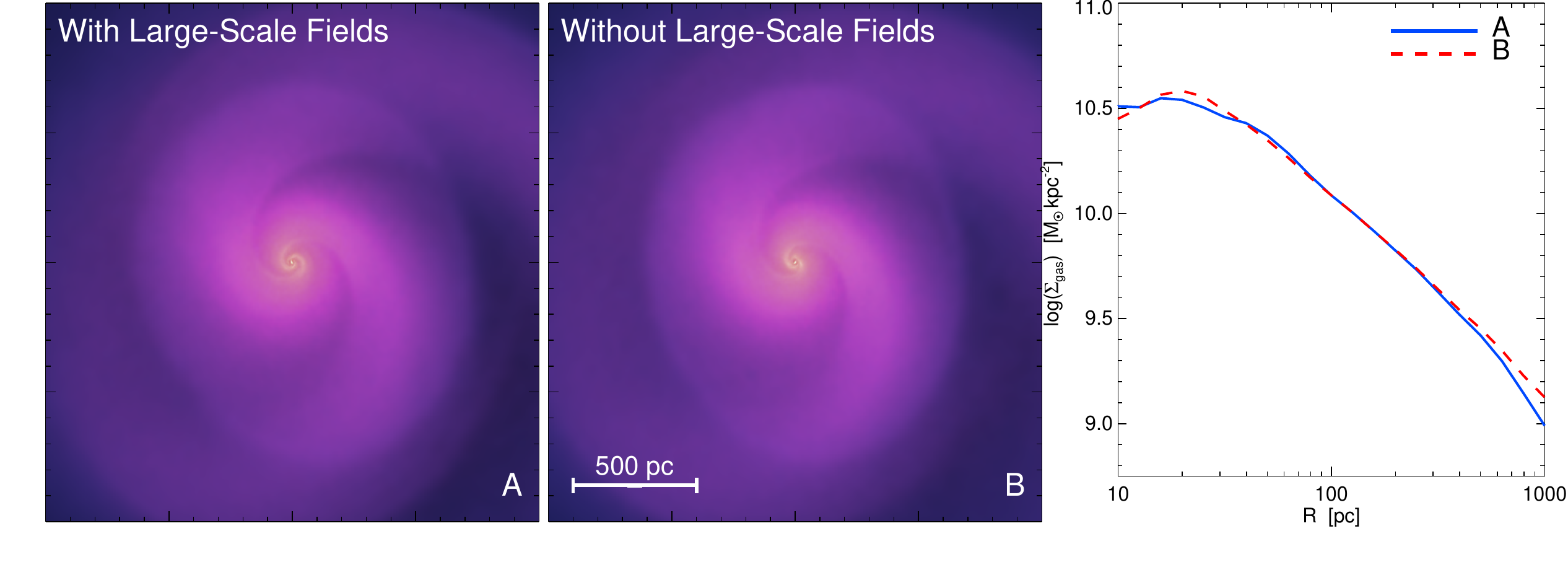}
    \caption{Simulations testing the effects of large-scale tidal forces in our 
    re-simulations. {\em Left:} Re-simulation of the central 
    $\sim 10\,$pc-kpc of a larger-scale galaxy merger simulation, 
    in which the full time-dependent 
    potential from the large-scale simulation is analytically 
    included in the re-simulation.  We show the gas density 
    and star formation rate density as in Figure~\ref{fig:100pc.morphologies}, 
    three large-scale dynamical times after the beginning 
    of the simulation. {\em Center:} Same re-simulation, but 
    without the additional potential from the large-scale 
    material; i.e., the gravitational potential is only from 
    the ``live'' material being re-simulated in the central kpc. 
    The image is at the same point in time. {\em Right:} Gas 
    density profiles of both re-simulations at the peak of the 
    resulting inflows. Because the density  declines rapidly 
    with radius, the large-scale density/potential fields make 
    no significant difference to the dynamics at the these scales.
    \label{fig:tidal.test}}
\end{figure*}

\vspace{-0.4cm}
\subsection{Do the Results Depend on Large-scale Tidal Fields?}
\label{sec:numerical:outer}

Because we are simulating the properties of large-scale
non-axisymmetric systems, one might worry that the tidal forces from
large radii could be important.  This is certainly the case, e.g., in
cosmological simulations.  In the present context, the question is
whether we need to include the matter distribution at $\sim10\,$kpc in
real-time to understand the dynamics of material at $\lesssim
100\,$pc.

We assess this directly as follows. For a given re-simulation of the
nuclear regions, we determine the matter distribution of the
larger-scale simulation from which the re-simulation is drawn and
analytically fit the non-axisymmetric contributions to the potential
at all radii.\footnote{To do this, we adopt the radial basis expansion set proposed 
in \citet{hernquist:potential.expansion}, with standard spherical harmonics. 
This is similar to the density modeling described in \S~\ref{sec:sims}, and can be performed 
to arbitrary order, but only the first few terms are not noise-dominated. 
}
Figure~\ref{fig:tidal.test} compares the results from a re-simulation
in which we include the large-scale non-axisymmetric potential at
$\gtrsim$ kpc with our standard approach, in which we do not include
this potential and in which the matter at large radii is not included
in the re-simulation (so the information is lost). This is simulation
{If9b5}  in Table \ref{tbl:intermediate}, a
$\sim0.01-1$ kpc re-simulation of the galactic nucleus during a
galaxy-galaxy merger simulation at the peak of nuclear activity, just
after the coalescence of the two galactic nuclei.  At large scales,
tidal tails are still present (these will take up to several Gyr to
fully relax), and the system is largely unrelaxed outside of a few
hundred pc, so this is a time when the large-scale effects would a
priori be the most important.  In the simulations shown in
Figure~\ref{fig:tidal.test} ({\em left} panel), we analytically fit
the large-scale tidal field in the galaxy-galaxy merger simulation at
multiple times, and interpolate it in time in our re-simulation.

Figure~\ref{fig:tidal.test} shows that there is effectively no
difference between the simulations with and without the large-scale
tidal field, at any of the radii of interest.  Including large-scale
tidal forces does somewhat truncate the large spiral wave pattern that
appears as a secondary instability in the intermediate-scale
simulation, but this truncation occurs only on the largest radii
$\gtrsim500\,$pc, radii that we are {\em not} trying to model in
detail in the re-simulation.  Most importantly, the scales which the
re-simulation is intended to accurately represent, $\lesssim100\,$pc,
are indistinguishable in the two simulations.
The reason for this is ultimately that the mass profile is a
relatively steep function of radius.  If a mass $\delta M$ is enclosed
in an annulus at a radius $r$, then the tidal force it produces is
$\propto |a| \delta M /r^{3}$, where $|a|$ is the magnitude of the
non-axisymmetric component of the mass distribution at that annulus.
The contribution of a given large radius to the tidal force at small
radii is thus $\propto \rho(r)$, the local mass density.  This,
however, is always a decreasing function of (increasing) radius, so
that material at larger radii makes little contribution to the tidal
force.
This can be compared with the role of tidal forces in determining the
angular momentum of large-scale dark matter structures and ultimately
(via these structures) dark matter halos; since the large-scale matter
field has $\rho\sim $constant on average, there are significant
contributions to tidal forces from all radii. This is why including
the information about large-scale tidal fields is critical for
simulations of cooling and galaxy formation, but it is not important
when following the further inflows of material in individual galaxies.

We have performed similar experiments to those shown here for other
intermediate-scale simulations (both in mergers and isolated
galaxies), and for our smallest-scale simulations ($\sim0.1-10\,$pc),
and reach the same conclusions in each case.

\vspace{-0.3cm}
\subsection{Do the Results Depend on  Initial Conditions?}
\label{sec:numerical:seeding}

A reasonable question about our re-simulation technique is whether it
might be sensitive to the precise initial conditions in the nuclear
disk that is initialized at high resolution.  In particular, there is
little detailed information about the structure or dynamics of the
nuclear disk in the original larger-scale simulation, precisely
because it lacked the necessary resolution to study the small-scale
nuclear dynamics.  Here we show that our results do not depend on the
details of how we initialize the re-simulations in the nuclear region.
The fundamental reason for this is that the nuclear dynamics is
governed by instabilities that are self-consistently generated on
small scales.  These instabilities depend primarily on the gross
structural properties of the nucleus, and rapidly lose memory of the
initial conditions.

\begin{figure*}
    \centering
    \scaleup
    \plotsidesmallrotate{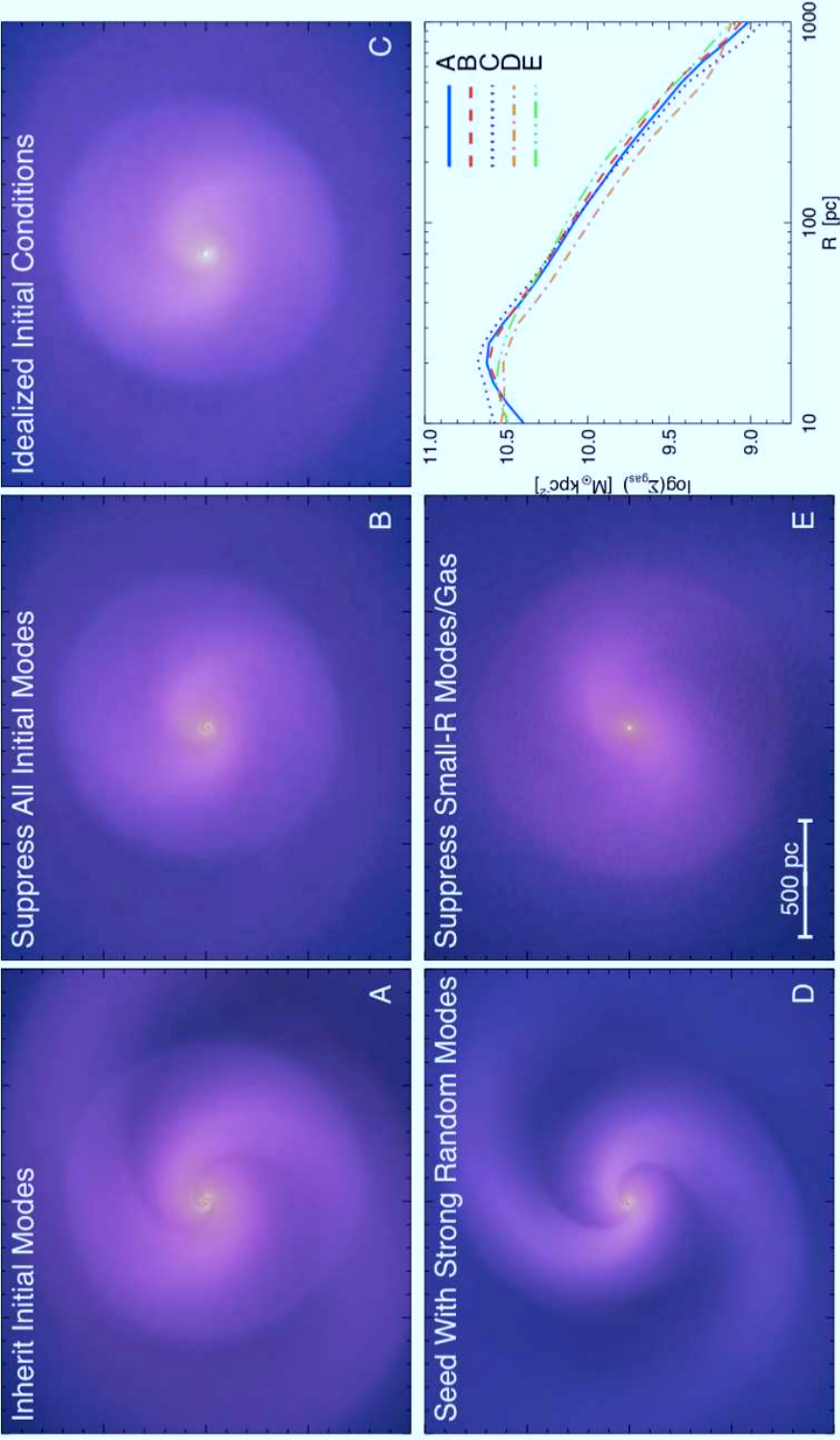}
    \caption{Effects of initial conditions on our re-simulations of
      galactic nuclei. Panels A-E show the gas density and star
      formation rate density $10^{8}$\,yr after the beginning
      of the simulation.  {\em Bottom right:} Surface density profiles
      of all five simulations at the same time.  {\em A:}
      Re-simulation of the central $\sim$kpc from a larger-scale
      merger simulation (as in Figure~\ref{fig:tidal.test}). The
      initial conditions are taken exactly from the larger-scale
      simulation, i.e., with inherited non-axisymmetric modes.  {\em
        B:} Same as A but in this case we initialize azimuthally 
        averaged (i.e., axisymmetric) profiles.  {\em C:} Even more idealized: we initialize an
      axisymmetric bulge+disk+halo+BH system with analytic profile
      shapes; the mass and radius of each component is the same as
      that of the material in the original large-scale simulation at
      these radii, but the profile shapes are not.  {\em D:} Same as 
      A, but with random seed non-axisymmetric modes having order 
      unity random amplitudes and phases {\em E:} Same as A, but 
      the gas at $\lesssim100\,$pc is removed so that there is initially a 
      central ``hole'' in the gas distribution.  
      Comparison of A-E demonstrates that the precise initial conditions 
      have little effect on the  mode dynamics and gas transport. 
      Morphologically, systems with stronger (weaker) initial seeds
       develop modes slightly more quickly (slowly), but this time 
       shift has little significant or lasting effect on the dynamics.
       \label{fig:seed.test}}
\end{figure*}

Figure~\ref{fig:seed.test} shows the results of five different
re-simulations, each with different initial conditions.  These are all
$\sim100\,$pc re-simulations of the central regions from a
galaxy-merger simulation (intermediate-scale simulation {If9b5} 
in Table \ref{tbl:intermediate}), just after the coalescence of the
two galaxy nuclei.  Panel A is a rigorous re-simulation of the matter
distribution in the central kpc of the large-scale simulation, with
the matter initialized ``as-is'' from the large-scale simulation.
Panel B is the same except we do not include any of the
non-axisymmetric modes in the density, pressure, potential, etc. in
the initial conditions; i.e., we azimuthally average the initial
conditions used in Panel A.  Panel C is even more idealized: the
system is axisymmetric and with exponential disk and Hernquist
profiles for the ``disky'' and ``spherical''
components in the re-simulation, respectively.  The masses and effective radii of
these components match the original simulation, but the profiles are
these simple analytic approximations rather than the true profile from
the larger-scale simulation.  To test whether the presence of
particular non-axisymmetries is important, Panel D includes
non-axisymmetric structure in the initial conditions, but the precise
modes are randomly chosen in phase and amplitude and so do not match
the actual non-axisymmetries from Panel A.  Finally, all of the above simulations
include gas at small radii in the initial conditions; this is because
the finite smoothing length in the larger-scale simulation spreads gas
out over a region $\sim 100$ pc even though the dynamics below this
scale is completely incorrect.  In order to be as conservative as
possible, Panel E considers an initial ``hole'' in the gas
distribution at radii that are not resolved in the large scale
simulation; for numerical reasons, the ``hole'' is a sharp power-law
cutoff inside $100\,$pc.  This could represent a very hard angular
momentum barrier, such as a strong inner Linblad resonance.

The images and surface density plots in Figure~\ref{fig:seed.test} are
$10^{8}$\,yr after the beginning of the simulation. 
There is remarkably little difference
in the mode structure visible in the images.  More quantitatively, the
surface density profiles are very similar, with significant gas inflow
to $\sim 10$ pc. This demonstrates that the precise initial conditions
used in the re-simulations are not important for most of our results.
The fundamental reason for this is that the instabilities that
dominate the dynamics arise from the {\em internal} structure of the
material; they depend on the mass and scale-lengths of the
rotationally and dispersion supported components, but not on the
initial seed amplitudes of various modes.  In the case of axisymmetric
initial conditions (Panel B), the initial perturbations take slightly
longer to appear because the initial power present to be amplified is
smaller (it is due to particle noise); and in the case of strong
initial modes (Panel D), the instability is slightly more developed at
the fixed time shown here. However, these time shifts have little
long-term effect on the evolution of the system.

\begin{figure*}
    \centering
    \scaleup
    \plotsidesmall{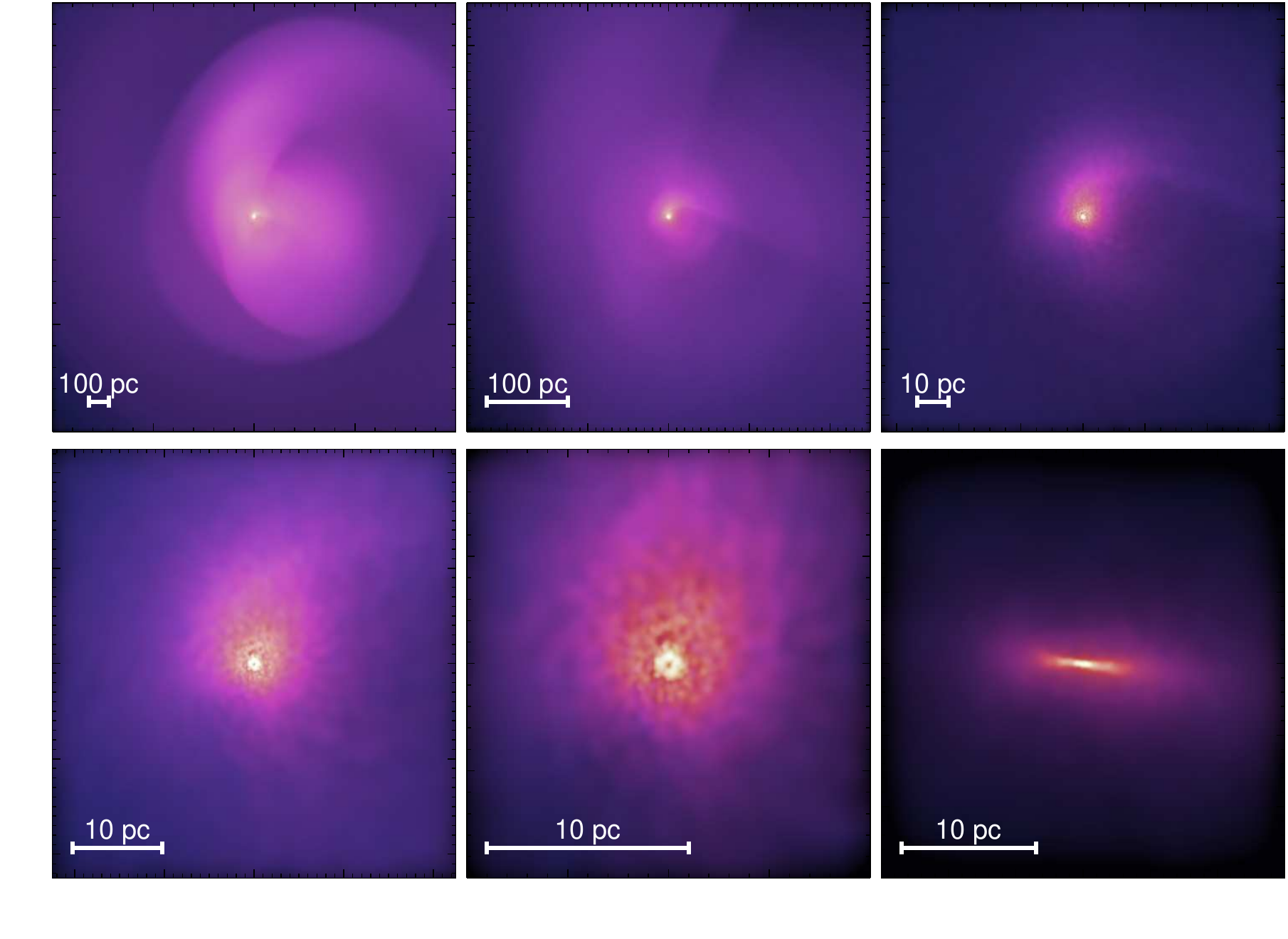}
    \caption{Illustration of one of our ultra-high resolution simulations 
    (Inf28b2h). The gas density distribution is shown face-on, as in Figure~\ref{fig:seed.test}, 
    in most panels, and edge-on at bottom right. 
    The initial conditions are those of a typical intermediate-scale re-simulation, 
    but with $\sim10^{7}$ particles and sub-pc resolution. 
    This obviates the need for a secondary ``re-simulation'' of the nuclear scales. 
    The large-scale modes fuel gas inwards and here lead to a large nuclear gas mass, 
    which then forms a lopsided $m=1$ mode around the BH and drives further 
    inflow. The eccentric nuclear gas disk is clearly visible, and similar to 
    those in our nuclear-scale re-simulations; in fact there is no statistical difference, 
    given similar boundary conditions. 
       \label{fig:ultrahighres.illustration}}
\end{figure*}

The robustness of our conclusions to resolution, tidal fields, and 
initial conditions is also demonstrated by the fact that our 
ultra-high resolution simulations, discussed at length in the text, 
give identical results to our standard re-simulated runs. These ultra-high resolution simulations 
allow us to bridge the boundaries of our standard re-simulations, 
and so do not suffer the potential negative effects considered here.
Figure~\ref{fig:ultrahighres.illustration} shows an illustrative example 
of one such simulation about mid-way through the period of peak 
activity. 
The simulation is initialized as one of our typical intermediate-scale 
re-simulations, but with $\sim10^{7}$ particles, giving a final 
resolution of $\lesssim$ pc.
On large scales, the dynamics is similar to our other 
intermediate-scale runs, with an $m=2$ mode forming rapidly 
(there are non-trivial mode amplitudes from $m=1-3$). 
The inflow leads to large gas masses at small radii; by the active phase shown in Figure~\ref{fig:ultrahighres.illustration},  the lopsided, eccentric 
nuclear disk is evident. The properties of this disk are 
statistically indistinguishable from those in our 
nuclear-scale re-simulations, given similar boundary conditions 
(compare e.g.\ Figure~\ref{fig:zoom}).

\section{ISM Physics}
\label{sec:ism}

\begin{figure*}
    \centering
    \scaleup
    \plotsidesmall{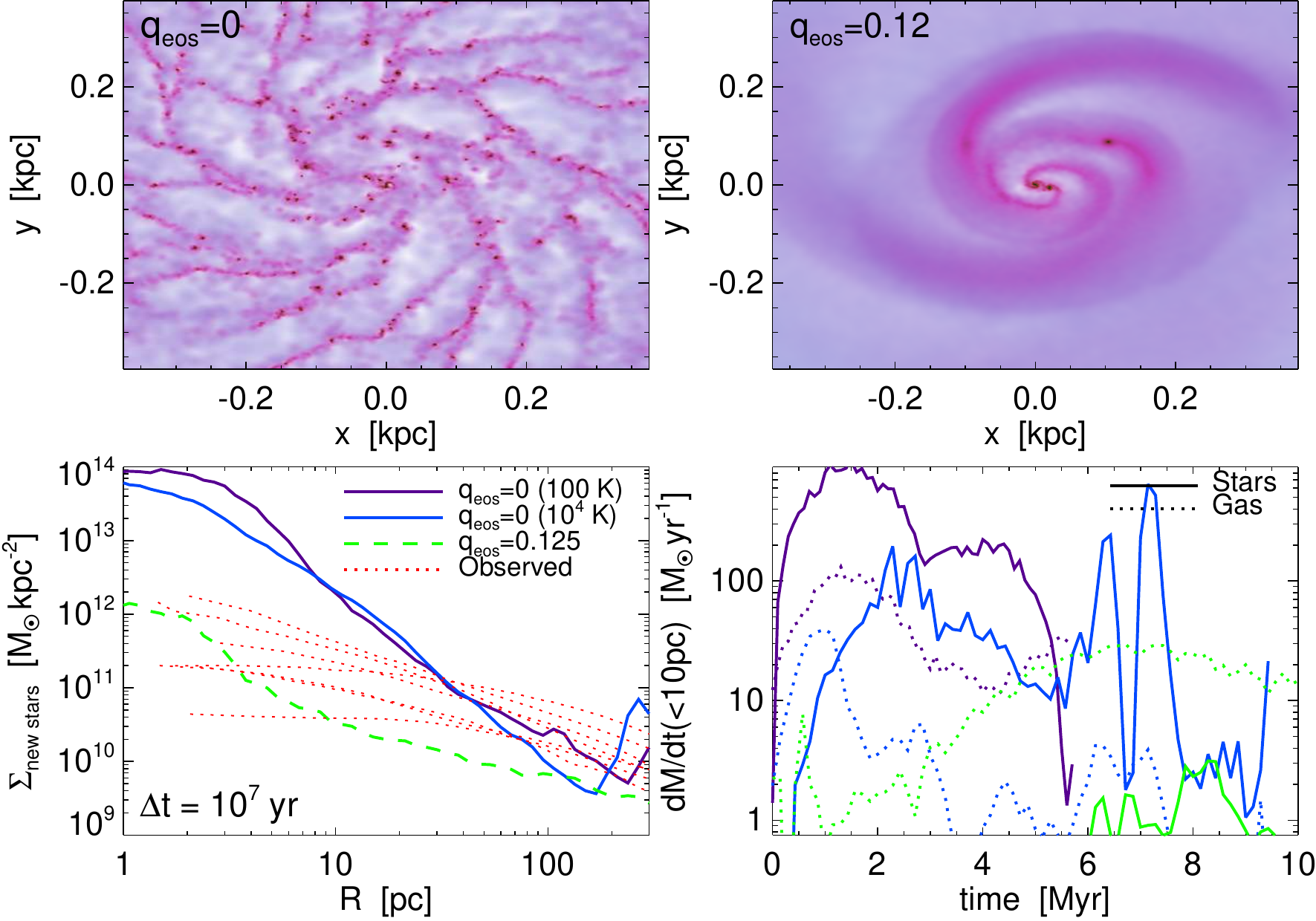}
    \caption{Examples of the problems that occur if systems are
      evolved without some ``feedback'' to prevent runaway cooling and
      fragmentation.  {\em Top Left:} Gas density in such an initially
      smooth re-simulation of the central $\sim$kpc (unlike in the
      main text, here darker areas represent higher-density to
      highlight the fragmentation).  The system here can cool to
      $10^{4}\,$K and has no feedback ($q_{\rm eos}=0$); the image is
      shown after $\sim2 \times 10^{6}$ years (less than the
      global dynamical time).  {\em Top Right:} The same 
      initial conditions but for our fiducial model -- a moderate 
      subgrid (``turbulent'') sound speed $\sim30\,{\rm km\,s^{-1}}$ ($q_{\rm eos}=0.125$) in dense
      star-forming regions.  In the no feedback
      $\qeos = 0$ case, gas clumps at the Jeans scale and rapidly
      turns into stars -- leading to a severe violation of the
      observed {\em global} \citet{kennicutt98} relation.  {\em Bottom
        Left:} Stellar density profiles after $10^{7}\,$yr, for
      simulations with no feedback (two examples shown, with different
      cooling floors as labeled) and our fiducial $\qeos = 0.125$
      model; the initial density is just $10^{10}\,\msun\,{\rm
        kpc^{-2}}$ at small radii.  We also show the observed stellar
      mass density profiles of massive cusp ellipticals in Virgo from
      \citet{jk:profiles}, chosen to have the same mass at
      $>0.5-1\,$kpc.  In the absence of feedback, runaway
      fragmentation and the inability of clumps to dissolve  inevitably leads to the formation of an extremely
      massive nuclear stellar cluster with a mass and surface density at least 
      an order of magnitude larger than any observed.  {\em Bottom
        Right:} The accretion rate into the central $10\,$pc for both
      gas (dotted) and {\em already-formed} stars (solid).  In the
      $\qeos = 0$ models, sinking clumps provide large gas accretion
      rates $\sim10-100\,\msun\,{\rm yr^{-1}}$, but the stellar inflow
      rate typically exceeds the gas inflow rate by a factor of $\sim
      10$; this is inconsistent with the observed stellar densities at small radii.  Our fiducial,
      moderate-feedback case, by contrast, drives primarily gas -- not
      stars -- to small radii.
      \label{fig:nofb.ill}}
\end{figure*}

\begin{figure*}
    \centering
    \plotside{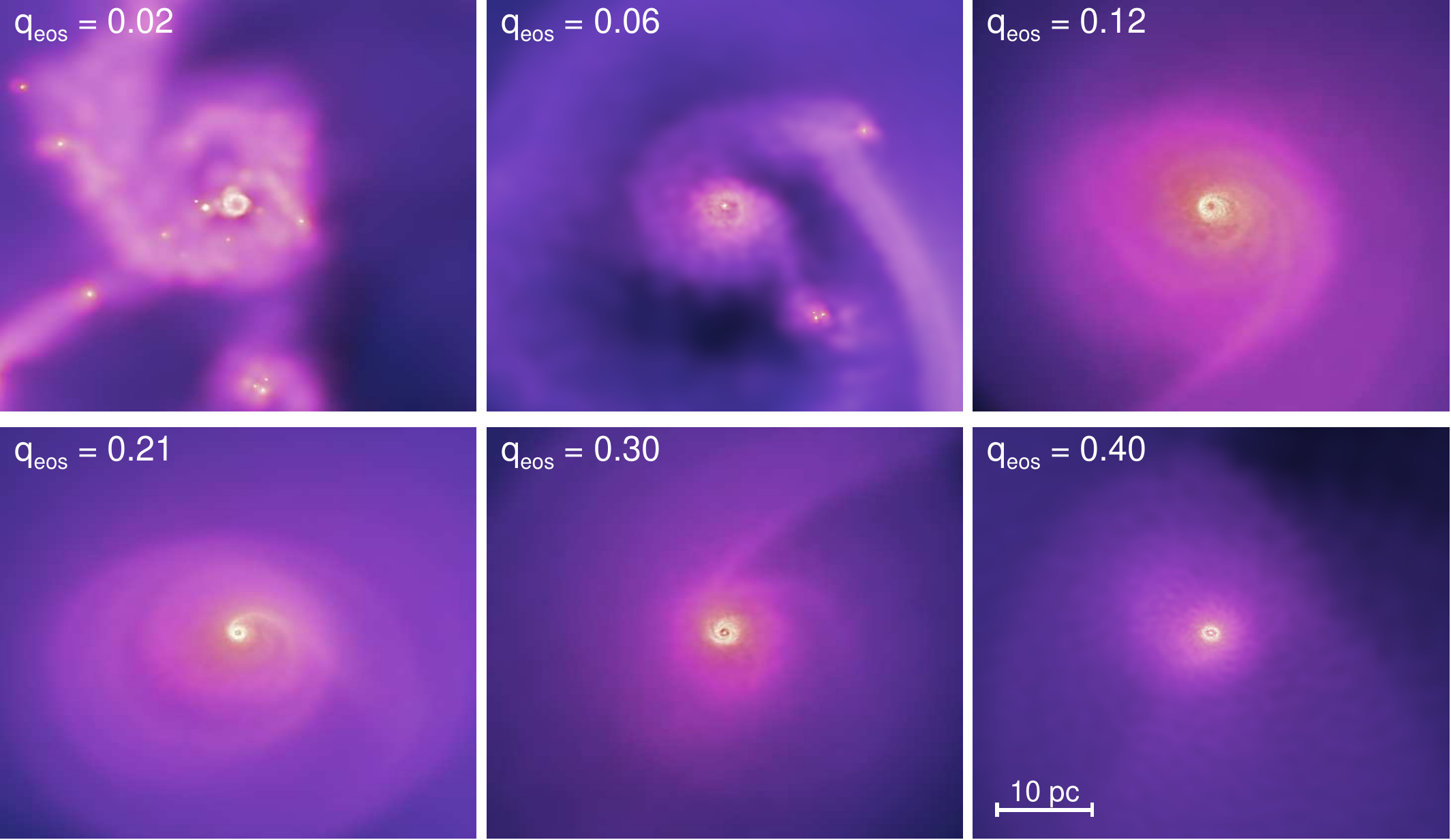}
    \caption{Effects of varying the assumed efficiency of 
     stellar feedback on the resulting 
     nuclear eccentric disk, given identical inflow conditions at $100\,$pc. 
     We show the gas surface density with color 
     indicating specific SFR (as Figure~\ref{fig:seed.test}), 
     for several choices of $q_{\rm eos}$ (from the survey Nf8h2b4q). 
     The range shown corresponds to sub-grid turbulent velocities ranging from 
    $\sim20-100\,{\rm km\,s^{-1}}$ in the high star-formation rate regions of the ISM.
    The efficiency of feedback affects the level of substructure.  However, 
    the general character and appearance of the lopsided disk mode 
    is similar over the entire plausible range. 
    Provided the catastrophic fragmentation in Figure~\ref{fig:nofb.ill} can 
    be avoided, our qualitative conclusions are independent of $q_{\rm eos}$. 
    \label{fig:qtest}}
\end{figure*}

As discussed extensively in \S \ref{sec:sims} \& \S
\ref{sec:discussion}, the largest {\em a priori} uncertainty in our
modeling is the question of what physics should be included to
describe the behavior of the ISM.  Our approach is to include a large
``turbulent'' sound speed in the equation state, as a subgrid model
for the effects of feedback by stellar winds, radiation pressure, and
supernovae on the ISM.  We choose the turbulent velocity largely by
comparison to observed systems (Fig.~\ref{fig:qeos}). The resulting
turbulent velocities $\sim 30-100 \ {\rm km \, s^{-1}}$ are also
reasonably consistent with theoretical estimates of the feedback
required to maintain $Q \sim 1$ and disrupt molecular clouds in
starburst environments
\citep{thompson:rad.pressure,murray:molcloud.disrupt.by.rad.pressure}.

To illustrate the importance of including a subgrid model, in this
Appendix we compare the results of our standard models with a
simulation in which the ISM is isothermal at $\sim 10^4$ K, i.e., $c_s
\simeq 10 \ {\rm km \, s^{-1}}$ ($\qeos = 0$). We have experimented
with a range of isothermal cooling floors from $c_{s}=1-15\,{\rm
  km\,s^{-1}}$ ($100 - 3\times10^{4}\,$K), but the results here are
generic to this range.  Figure~\ref{fig:nofb.ill} shows several of the
key results comparing the $c_s \simeq 10 \ {\rm km \, s^{-1}}$
simulation with our fiducial $\qeos = 0.125$ simulation, for a
re-simulation on $\sim 100\,$pc scales. From the images it is clear
that after just a few local dynamical times in the intermediate-scale
disk, the simulation without feedback has violently fragmented into
large clumps.  It is important to stress that in this simulation we
fully resolve the Jeans and Toomre lengths so the fragmentation is
physical, not numerical.

The difficulty with the $c_s \simeq 10 \ {\rm km \, s^{-1}}$ ($\qeos =
0$) simulation is that, absent a full model for how feedback can
either suppress their formation or suppress their internal star
formation, the only possibility is that the clumps ``run away'' and
turn efficiently into stars. 
The runaway and final consequences of the fragmentation for 
star formation and the IMF are sensitive to details of the cooling 
rates and cooling function, but the end result of 
runaway star formation is inevitable if the 
gas is allowed to be arbitrarily cold and has a small cooling time 
\citep{gammie:2001.cooling.in.keplerian.disks,
nayakshin:sfr.in.clumps.vs.coolingrate,
thompson:rad.pressure}. 
This has several dramatic effects.
First, the global star formation efficiency is significantly enhanced
-- for the same mean gas surface density, most of the gas will turn
into stars in $\sim1-4$\,
 local dynamical times absent feedback.  
 This is a factor of $10-50$ higher than implied by the
Schmidt-Kennicutt laws.  For comparison, our fiducial $\qeos = 0.125$
simulation has a star formation efficiency per dynamical time of a few percent, 
much more consistent with observations.  Because of the rapid and efficient star formation in the $c_s = 10 \
{\rm km \, s^{-1}}$ simulation, the clumps quickly convert a large
fraction of their mass to stars; the resulting massive stellar
clusters sink and coalesce via clump-clump scattering and/or dynamical
friction, to the center of the galaxy.  As the lower-right panel of
Figure \ref{fig:nofb.ill} shows, most of the mass flowing into the
central $\sim 10$\,pc in the $c_s \simeq 10 \ {\rm km \, s^{-1}}$
simulation is in the form of stars, rather than gas.  This results in
the formation of an extremely massive nuclear star cluster, with a
mass of $\sim 10^{9}\,\msun$ inside the central $\sim10-20\,$pc -- a
factor of at least $\sim10\,$ larger than observed nuclear star
clusters
\citep{boker04:nuclei.scalings,cote:virgo,ferrarese:nuclear.cluster.vs.host.mass}.
Indeed, the resulting stellar surface density at small scales (left
panel of Fig.~\ref{fig:nofb.ill}) significantly exceeds the highest
nuclear densities observed in any cusp ellipticals
\citep{ferrarese:profiles,jk:profiles, lauer:bimodal.profiles}, or for
that matter any nuclear star clusters, globular clusters, nuclear
disks, or other high-density stellar systems
\citep{hopkins:maximum.surface.densities}.

In contrast to the constant $c_s \simeq 10 \ {\rm km \, s^{-1}}$
simulation, our fiducial simulation with significant subgrid feedback
does not clump up into many dense star clusters.  This is also not
completely physical since in reality we expect that the ISM should
have significant sub-structure on the scales we model (including,
e.g., molecular clouds and star clusters).  However, when such
self-gravitating gaseous clumps form, observations strongly suggest
that are likely to be short-lived and/or to inefficiently form stars.
Our subgrid model assumes that most of the mass remains in a diffuse
ISM, rather than becoming bound in dense star clusters.  
Not only is the latter physically implausible, but Figure \ref{fig:nofb.ill}
demonstrates that it strongly violates observational constraints.
Note that there are conditions under which even an extremely cold 
gaseous disk could avoid catastrophic fragmentation 
\citep[see e.g.][]{lodato:2004.acc.disk.spiralwaves,rice:maximum.viscous.alpha,
boley:2006.protoplanetary.disk.w.cooling,krumholz:2007.rhd.protostar.modes}, particularly 
the case in which the cooling time is sufficiently long such that continued cooling 
is offset by gravitational heating (giving rise to sustained local structure 
such as tightly-wound spiral arms, but not runaway fragmentation). 
However, the cooling rates under typical conditions in our simulations 
are much too rapid to reach this regime; we could, of course, modify the 
cooling function to suppress the 
cooling rate and/or mimic some continuous heating term -- but this is 
effectively equivalent to adopting a new sub-grid feedback/microphysics 
model, and we find it has the same qualitative effects.

Figure~\ref{fig:qtest} shows the effects of varying the efficiency of
feedback from supernovae and massive stars, encapsulated in the
parameter $q_{\rm eos}$. Specifically, we show results of a set of nuclear scale simulations (Nf8h2b4q in Table 3) having $q_{\rm eos} = 0.02-0.40$
(effective turbulent velocities $\sim20-100\,{\rm km\,s^{-1}}$),
roughly the lower and upper limits allowed by observational
constraints for the systems of interest in Figure~\ref{fig:qeos}.
Figure~\ref{fig:qtest} shows that the choice of sub-resolution model
has a significant effect on the amount of resolved sub-structure in
the simulation.  This is not surprising since larger turbulent
velocities raise the Jeans mass/length, above which gravity is the
dominant force.  The key point, however, is that all of the
simulations show a similar nuclear lopsided disk.  In terms of the
properties of inflow and {\em global} instability discussed throughout 
this paper, the differences produced by changing $q_{\rm eos}$ are 
similar to the differences produced by somewhat different 
galaxy properties.   The fundamental reason for the weak dependence on
the subgrid model is that the torques in our simulations are primarily
determined by gravity, not hydrodynamic forces or viscosity. 
The primary role of the subgrid feedback model is simply to prevent
catastrophic fragmentation of the galactic gas. Provided this can be 
accomplished, our qualitative conclusions do not depend on the details of 
the feedback model.

\end{appendix}

\end{document}